\documentclass[12pt,notitlepage]{article}

\usepackage{amsmath}         
\usepackage{amssymb}         
\usepackage{amsfonts}       
\usepackage{graphicx}        

\usepackage{color}         
\usepackage{slashed}       
\usepackage{framed}        
\usepackage{cite}          
\usepackage{booktabs}      
\usepackage{youngtab}	    
\usepackage{arydshln} 	    
\usepackage{bbm}           
\usepackage{tocloft}		   
\usepackage{setspace}	   
\usepackage{listings}       
\usepackage[utf8]{inputenc}

\usepackage[margin=2cm]{geometry}   
\numberwithin{equation}{section}    

\newcommand{\email}[1]{\href{mailto:#1}{#1}}

\newenvironment{institutions}[1][2em]{\begin{list}{}{\setlength\leftmargin{#1}\setlength\rightmargin{#1}}\item[]}{\end{list}}

\let\oldenumerate\enumerate
\renewcommand{\enumerate}{
  \oldenumerate
  \setlength{\itemsep}{1pt}
  \setlength{\parskip}{0pt}
  \setlength{\parsep}{0pt}
}

\let\olditemize\itemize
\renewcommand{\itemize}{
  \olditemize
  \setlength{\itemsep}{1pt}
  \setlength{\parskip}{0pt}
  \setlength{\parsep}{0pt}
}

\usepackage{fancyhdr}		

\usepackage[
	colorlinks=true,
	citecolor=black,
	linkcolor=black,
	urlcolor=blue,
	hypertexnames=false]{hyperref}

\newcommand{\abs}[1]{\ensuremath{\left|#1\right|}}

\newcommand{\barB}{\ensuremath{\overline{B}}}

\newcommand{\barQ}{\ensuremath{\overline{Q}}}
\newcommand{\barq}{\ensuremath{\overline{q}}}

\newcommand{\Tr}{\ensuremath{\text{Tr\,}}}

\newcommand{\yF}{\ensuremath{\tiny\yng(1)}}
\newcommand{\yFb}{\ensuremath{\tiny \overline{\yng(1)}}}
\newcommand{\yA}{\ensuremath{\tiny\Yvcentermath1 \yng(1,1)}}

\newcommand{\ev}[1]{\ensuremath{ \langle #1 \rangle} }
\renewcommand{\eqref}[1]{Eq.~(\ref{#1})}

\newcommand\Tstrut{\rule{0pt}{2.5ex}}

\begin{document}

\begin{center}

    {\Large \bf Confinement On the Moose Lattice}

    \vskip .5cm

    {\large \bf a.k.a.~Party Trick Confinement}



    \vskip .5cm

    { \bf Benjamin~Lillard} 
    \\ \vspace{-.2em}
    { 
    \footnotesize
    \email{blillard@illinois.edu}
    }
	
    \vspace{-.2cm}

    \begin{institutions}[3.5cm]
    \footnotesize
    \textit{Illinois Center for Advanced Studies of the Universe \& Department of Physics, 
    University of Illinois at Urbana-Champaign, Urbana, IL 61801, USA. }   
    \end{institutions}

\end{center}

\begin{abstract}
\noindent 
In this work we present a new class of $\mathcal N=1$ supersymmetric confining gauge theories, with strikingly simple infrared theories that descend from intricate interconnected networks of product gauge groups.
A diagram of the gauge groups and the charged matter content of the ultraviolet theory has the structure of a triangular lattice, with $SU(N)$ or $SU(3 N)$ gauge groups at each of the vertices, connected by  bifundamental  chiral superfields. 
This structure admits a $U(1)_R$ conserving superpotential with marginal (trilinear) operators.
With the introduction of this superpotential, the $SU(3N)$ and $SU(N)$ gauge groups confine: in the far infrared limit of the supersymmetric theory, the relevant degrees of freedom are gauge invariant ``mesons'' and ``baryons.''
In this paper we show how the properties of the infrared degrees of freedom   depend on the topology and shape of the moose/quiver ``lattice'' of the original gauge theory. We investigate various deformations of the theory, and propose some phenomenological applications for BSM models. 
\end{abstract}

\tableofcontents

\section{Introduction}

Strongly coupled gauge theories are highly relevant to our understanding of the universe, but  challenging to analyze. 
Taking the strong nuclear force (QCD) as an example,  perturbation theory is incapable of deriving the masses and interactions of the various hadrons in the Standard Model from high-energy observables.  A  first-principles calculation in the strongly coupled regime requires nonperturbative methods, such as lattice QCD.
Aspects of supersymmetry make this problem more tractable:  with a sufficiently high degree of symmetry, it may be possible to constrain the form of the low energy theory to the point that its constituents and interactions can be 
more precisely identified.

\medskip

In this article we investigate  product gauge groups of the form 
$(SU(3N) \times SU(N)^3)^k$ in $\mathcal N=1$ supersymmetry (SUSY), with a set  of chiral bifundamental ``quark'' matter fields arranged so that a moose diagram of the theory forms a 
triangular lattice.  With the appropriate superpotential, the theory exhibits two stages of confinement, leading to an infrared theory of ``baryons'' and ``mesons'' that is surprisingly simple.

The ultraviolet phase of our theory is constructed from three ingredients: the $SU(3N)$ and $SU(N)$ gauge groups; chiral matter superfields, transforming in the bifundamental representations of $SU(3N) \times SU(N)$ or $SU(N) \times SU(N)$; and a chiral superpotential $W$, which includes the marginal gauge invariant trace operators of the form $W \supset \lambda \text{Tr}\,(q_1 q_2 q_3)$ for three bifundamentals $q_i$. To analyze the low energy theory we rely on the  Seiberg dualities for supersymmetric QCD (SQCD)~\cite{Seiberg:1994pq, Intriligator:1994sm}.

For the initial discussion, we restrict our attention to the cases where there is some high-energy scale $M_\star$ where all of the gauge couplings are perturbatively small, i.e.~$g_{3N}(\mu), g_N(\mu) \lesssim \mathcal O(1)$ at $\mu = M_\star$. As the $SU(3N)$ and $SU(N)$ gauge couplings run in opposite directions (i.e.~their NSVZ $\beta$ functions~\cite{Novikov:1983uc} have opposite signs), this situation is not entirely generic. We refer to this $\mu \sim M_\star$ regime as the ultraviolet theory (UV), even though the $SU(N)$ gauge groups become strongly coupled in the extreme ultraviolet limit $\mu \gg M_\star$. 

This theory exhibits confinement with chiral symmetry breaking at a scale characterized by $\Lambda_{3N}$, where the one-loop $\beta( g_{3N})$ function diverges. 
By describing the theory as confining, we mean that the $\mu \gg \Lambda_{3N}$ perturbatively coupled $(SU(3N) \times SU(N)^3)^k$ theory with the trilinear superpotential $W$
is Seiberg-dual to 
a theory of singlet ``baryons'' and $SU(N) \times SU(N)$ bifundamental ``mesons'' at scales $\mu \ll \Lambda_{3N}$.
Some of the global symmetries of the UV theory are spontaneously broken by the $\mathcal O(\Lambda_{3N})$ expectation values of the baryon operators. 

A subset of the $SU(N) \times SU(N)$ bifundamentals acquire  vectorlike masses $m_i$ of order $\mathcal O(\lambda_i \Lambda_{3N})$, inherited from the $\lambda_i \text{Tr}\,(q_1 q_2 q_3)$ operators in the UV theory. Like the $\lambda_i \neq 0$ UV superpotential terms, these vectorlike masses lift some of the flat directions that would otherwise be included in the moduli space.
At scales $\mu < m_i$, these massive degrees of freedom can be integrated out. This changes the sign of the $\beta(g_N)$ function, so that the $SU(N)$ gauge groups become strongly coupled at some new $\Lambda_N < m_i$. 
In the far infrared (IR) theory $\mu \ll \Lambda_N$, the only light degrees of freedom are composite baryons and mesons, which can be mapped onto the set of gauge invariant operators of the original UV theory. 
Despite the formidable complexity of the original ultraviolet theory, this phase of the low energy effective theory is remarkably simple. 

\medskip

Sections~\ref{sec:triangular} and~\ref{sec:topology} present our main results, following the evolution of the theory from $\mu \sim M_\star$ down to $\mu \ll \Lambda_N$ step by step, tracking the degrees of freedom and their global symmetries in each regime. Section~\ref{sec:triangular} focuses on the aspects of the calculation that are the least dependent on the actual shape of the ``moose lattice,'' and the easiest to generalize. 
Details about the boundary conditions become much more important in the infrared limit of the theory, as we show in Section~\ref{sec:topology}. For example,  if the moose lattice is given periodic boundary conditions, the $SU(N)$ groups do not confine, but instead have an
 unbroken Coulomb phase in the IR. 
Section~\ref{sec:topology} explores several such variations on the boundary conditions. As an interlude between Sections~\ref{sec:triangular} and~\ref{sec:topology}, Section~\ref{sec:globals} follows the global symmetries of the theory from the UV to the IR. 

Our focus in this work is restricted to four-dimensional spacetime, and the so-called moose lattice is simply a way to keep track of the gauge groups and matter fields --- however, it is highly suggestive of a geometrical interpretation, consistent with the deconstruction~\cite{ArkaniHamed:2001ca, Csaki:2001em,Csaki:2001zx,Skiba:2002nx} of a six-dimensional spacetime with two compact dimensions, as we discuss in our concluding remarks. This view is reinforced by the emergence of some bulk-like and brane-like features in the infrared theory.

In Section~\ref{sec:reviewSUSY} we provide a review of the familiar Seiberg dualities and confinement in SQCD. Section~\ref{sec:reviewProduct} reviews some especially relevant literature on confinement in $SU(N)$ product gauge theories~\cite{Csaki:1997zg, Chang:2002qt, Lillard:2017mon}.

\medskip

\subsection{Review of Confinement in $\mathcal N=1$ Theories} \label{sec:reviewSUSY}

Supersymmetry (SUSY) ameliorates some of the challenges of strongly coupled theories, making it possible to derive some infrared properties of a theory exactly.
The conjectural Seiberg dualities~\cite{Seiberg:1994pq, Intriligator:1994sm} are central to this effort.
Given a gauge group such as $SU(N)$ with some set of matter fields, 
one can sometimes identify a dual theory with a different gauge group, new matter fields, and possibly some superpotential that describes the interactions of the dual matter fields.
In describing the theories as ``dual,'' we mean that the two ultraviolet theories flow to the same infrared behavior, not that this duality is exact at all energy scales.
Seiberg dualities have been identified not only for $SU(N)$ gauge groups, but also $Sp(2N)$, $SO(N)$, and the exceptional Lie groups.

A number of SUSY gauge theories have been shown to confine: that is, rather than being dual to another gauge theory, the dual theory has no gauge interactions.
Well-known examples include  $SU(N_c)$ with $F=N_c$ or $F=N_c+1$ pairs of chiral superfields in the fundamental $(\yF)$ and antifundamental $(\yFb)$ representations of the gauge group (a.k.a.~``SQCD''),
or $SU(N_c)$ with one field in the antisymmetric $(\yA)$ representation and an appropriate number of fundamentals and antifundamentals~\cite{Berkooz:1995km,Pouliot:1995me,Poppitz:1996vh}.

In some cases the infrared theory has a ``quantum deformed moduli space,''  where the classical constraint equations are modified by some terms that depend on the gauge couplings.
We refer to these theories as ``qdms-confining.''
The canonical example is the $F=N_c$ case of SQCD, 
where the infrared behavior is described by the gauge invariant operators
\begin{align}
M_{ij} = Q^\alpha_i \barQ_{\alpha j} ,
&&
B = \det Q \equiv  Q^{N_c} ,
&&
\barB =\det \barQ \equiv \barQ^{N_c},
\label{eq:defBMB}
\end{align}
where $Q^{N_c} \equiv \det Q$ is the completely antisymmetric product of $i=1, 2, \ldots N_c$ distinct fields $Q_i$, each in the fundamental representation of $SU(N_c)$.
Classically, $M$ and $B \barB$ would obey the constraint equation $\det (Q \barQ) = \det Q \det \barQ$, 
\begin{align}
\det M = B \barB,
\end{align}
but this classical constraint is modified quantum mechanically~\cite{Davis:1983mz,Affleck:1984xz,Seiberg:1994bz}
by a term proportional to the holomorphic scale $\Lambda$,
\begin{align}
\det M - B \barB = \Lambda^b,
&&
\Lambda^b = \mu^b \exp\left( -\frac{8\pi^2 }{g^2(\mu) } + i \theta_\text{YM} \right),
&&
b = 3N_c - F.
\label{eq:qdmsN}
\end{align}

In the absence of any superpotential there is an $SU(F)_Q \times SU(F)_{\barQ}$ global flavor symmetry, under which $M$ transforms as a bifundamental and $B$ and $\barB$ are singlets. 
The other global symmetries include a $U(1)_B$ ``baryon number'' symmetry under which $B$ and $\barB$ have opposite charges, and a $U(1)_R$ under which the scalar parts of the $Q$, $\barQ$, $B$, $\barB$, and $M$ superfields are all neutral.\footnote{We use the common shorthand where each superfield and its scalar component are labelled with the same symbol, e.g.~$Q$ or $\barQ$.} This $U(1)_R$ does not commute with the generators of the $\mathcal N = 1$ supersymmetry; we follow the normalization where the chiral superpotential $W$ has charge $+2$ under any conserved $U(1)_R$, while the gauginos have charge $+1$.

The origin of the moduli space $\ev{B} = \ev{\barB} = \ev{M_{ij}} = 0$ is inconsistent with the modified constraint \eqref{eq:qdmsN}, so either $\det M$ or $B \barB$ must acquire some expectation value.
A nonzero $\ev{B \barB}$ spontaneously breaks the $U(1)_B$ baryon number.
If instead $\ev{\det M} \neq 0$, then $SU(F) \times SU(F)$ is broken to a subgroup: in the most symmetric of these vacua, 
$\ev{M_{ij} } \propto \delta_{ij}$, a diagonal subgroup $SU(F)_d \subset SU(F) \times SU(F)$ remains unbroken.
On a more generic point in the moduli space, the $SU(F) \times SU(F) \times U(1)_B$ symmetry is completely spontaneously broken.

\medskip

A second class of theories, known as  ``s-confining''~\cite{Csaki:1996sm, Csaki:1996zb}, confine smoothly without necessarily breaking any of the global symmetries, with the classical constraint equations enforced by a dynamically generated superpotential.
For example, the infrared limit of SQCD with $F=N+1$ flavors is described by the gauge invariants
\begin{align}
B_i &= (Q^N)_i \equiv \frac{\epsilon_{i\, j_1 \ldots j_N}}{N!}  \epsilon_{k_1 \ldots k_N} Q_{j_1}^{k_1} Q_{j_2}^{k_2} \ldots Q_{j_N}^{k_N} ,
\nonumber\\
\barB_i &= (\barQ^N)_i , \\
M_{ij} &= (Q\barQ)_{ij} \equiv Q_{i}^\alpha \barQ_{j}^\alpha, \nonumber
\end{align}
where $k$ refer to $SU(N_c)$ gauge indices, $i$ and $j$ to $SU(F)$ flavor indices, and $\epsilon$ is the completely antisymmetric tensor.
The constraint equations between $M$, $\barB$ and $B$ are not modified quantum mechanically.
Instead, the infrared theory has a dynamically generated superpotential~\cite{Intriligator:1995au}
\begin{align}
W = \frac{1}{\Lambda^{2N - 1} } \left( B M \barB - \det M \right),
\end{align}
which enforces the classical constraint equations
\begin{align}
B_i M_{ij} = 0,
&&
M_{ij} \barB_j = 0,
&&
(M_{ij})^{-1} \det M = B_i \barB_{j}.
\end{align}
It is easy to show that this $W$ has $R$ charge $+2$ under any conserved $U(1)_R$.
The origin of moduli space is now a viable solution to the constraint equations, permitting confinement without chiral symmetry breaking.

\medskip

The Seiberg dualities for SQCD survive a number of nontrivial consistency checks: the dimensionality of the moduli spaces of the UV and IR theories match, the two theories share the same set of global symmetries, and all of the 't~Hooft anomaly matching conditions are satisfied. In Appendix~\ref{sec:reviewAnomaly} we demonstrate this for some $F=N$ examples chosen to highlight some subtleties associated with anomaly matching on the quantum deformed superpotential.

\subsubsection*{Superpotential Deformations} \label{sec:reviewSuperdef}

Each of the models described so far has been derived from a UV theory with no superpotential. In the case of s-confinement, a superpotential is dynamically generated for the IR theory, which respects all the global symmetries and enforces the classical constraints between the operators. 
For qdms~confinement, there is no dynamically generated superpotential: the quantum modified constraint can only be implemented in a superpotential by the use of Lagrange multipliers.

In the SQCD example, one could perturb the theory by including the gauge-invariant superpotential operators
\begin{align}
W \sim m_{ij} (Q \barQ)_{ij} + \frac{(Q\barQ)^2}{M_\star} + \ldots .
\label{eq:Wperturbd}
\end{align}
This $W$ explicitly breaks the $SU(F)_\ell \times SU(F)_r$ symmetry, as well as $U(1)_R$.
Note that if any of the mass terms $m_{ij}$ are larger than $\Lambda$, then it is no longer appropriate to treat the problem as $F=N$ SQCD. After integrating out the heavier quarks with $m_{ij} \gtrsim \Lambda$, the remaining $F' < N$ SQCD theory does not confine: its Seiberg dual has an $SU(N - F')$ gauge group. The infrared theory thus bears no resemblance to the qdms-confining version of SQCD. 
If we are to treat the superpotential \eqref{eq:Wperturbd} as a small perturbation, it should be the case that $m_{ij} \ll \Lambda$. In this case the global symmetries are still approximately conserved, and the infrared effective theory developed in Section~\ref{sec:reviewSUSY} is still applicable at scales large compared to $m_{ij}$ (and small compared to $\Lambda$).

As defined in \eqref{eq:defBMB}, $M$, $B$ and $\barB$ have mass dimensions $2$, $N$ and $N$, respectively. When matching superpotential terms it can be more convenient to normalize these by factors of $\Lambda$ to give the operators canonical mass dimension,\footnote{As the exact form of the K\"ahler potential is not known for the dual theory, the mapping between UV gauge invariants and canonically normalized IR degrees of freedom includes some unknown numeric coefficients. Our $(Q\barQ) \rightarrow \Lambda \mathcal M$ mapping assumes these coefficients are $\mathcal O(1)$. \label{foot:note}}
\begin{align}
M \rightarrow \frac{(Q \barQ)}{\Lambda},
&&
B \rightarrow \frac{(Q^N)}{\Lambda^{N-1}},
&&
\barB \rightarrow \frac{(\barQ^N)}{\Lambda^{N-1}}, 
\end{align} 
so that a generic symmetry-violating superpotential for SQCD includes
\begin{align}
W \sim  A_{ij}\Lambda\, M_{ij} + \alpha_{ijkl} \frac{\Lambda^2}{M_\star}  M_{ij} M_{kl} + \ldots
+ \beta \frac{\Lambda^{N-1}}{M_B^{N-3}} B + \bar{\beta} \frac{\Lambda^{N-1}}{M_{\barB}^{N-3}} \barB + \ldots.
\end{align}
For $W\approx 0$ to be a good approximation in the near ultraviolet as well as the infrared, the mass scales associated with the irrelevant operators should satisfy $M_\star \gg \Lambda$, so that all of the global symmetries are approximately conserved above and below the scale $\Lambda$.
For $G_\text{global}$ to be approximately conserved below $\Lambda$, it must also be the case that $A_{ij} \Lambda \ll \Lambda^2$. In the $N=2$ special case, the same should be true for $M_B \Lambda$ and $M_{\barB} \Lambda$.

Solving the equations of motion for $M_{ij}$, $B$ and $\barB$, we find that the $F^2 + 1$ light degrees of freedom do not remain massless, but instead acquire some potential that lifts various directions of the moduli space.
In the $F=N$ case the addition of $A_{ij}$, $\beta$, and $\bar\beta$ is sufficient to completely break the global symmetry group $SU(F)_\ell \times SU(F)_r \times U(1)_B \times U(1)_R$.
The quadratic terms $M^2$, $M \barB$, $M B$ and $B \barB$ determine the location of the global minimum of the potential on the moduli space, up to corrections from further irrelevant operators that may be included in the superpotential.

\subsection{Confinement in Linear Moose Theories} \label{sec:reviewProduct}

Each of the complicated product gauge group models considered in this paper
uses a collection of alternating $SU(N) \times SU(M) \times SU(N) \times SU(M) \times \ldots$ gauge groups as a building block.
Dubbed the ``linear moose'' model~\cite{Georgi:1985hf}, the matter content of this theory consists of one chiral bifundamental quark for each adjacent pair of $SU(N)\times SU(M)$ or $SU(M) \times SU(N)$ groups.
This type of structure appears in the $k$ site deconstruction of a five-dimensional theory~\cite{ArkaniHamed:2001ca, Csaki:2001em,Csaki:2001zx,Skiba:2002nx}, which in the presence of a $\mathbbm Z_2$ orbifold produces an $SU(N)^k$ chiral gauge theory.

For the ``even'' linear moose with equal numbers of $SU(N)$ and $SU(M)$ gauge groups, $(SU(N) \times SU(M))^k$, the anomaly matching conditions are saturated, indicating that even in a nonsupersymmetric theory the confinement can proceed without breaking chiral symmetry~\cite{Georgi:1985hf},  while for the ``odd'' linear moose
$(SU(N) \times SU(M))^k \times SU(N)$ the chiral symmetry is necessarily spontaneously broken.

Additional information about the low energy behavior can be extracted from supersymmetric moose theories~\cite{Poppitz:1996vh, Csaki:1997zg, Chang:2002qt, Maruyoshi:2009uk, Nii:2016jzi, Brunner:2016uvv, Lillard:2017mon}. For example, with $\mathcal N=1$ supersymmetry, it is often possible to derive the exact form of the chiral superpotential. If an $\mathcal N=1$ theory can be shown to be the limit of $\mathcal N=2$ supersymmetry~\cite{Intriligator:1994sm, Leigh:1995ep, Argyres:1996eh, Hirayama:1997tw, Argyres:1999xu, Maruyoshi:2009uk}, the K\"ahler potential may be similarly constrained based on the form of the holomorphic prepotential.

\begin{figure}
\centering
\includegraphics[width=0.8\textwidth]{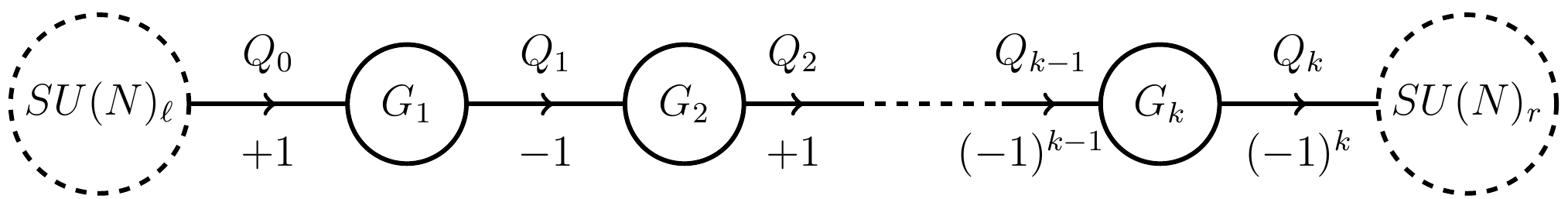}
\caption{The moose diagram for the qdms-confining $SU(N)^k$ model~\cite{Chang:2002qt}, showing the $k$ gauge groups $G_i = SU(N)_i$ and the global $SU(N)_\ell \times SU(N)_r$ symmetry. Each $Q_j$ transforms as a bifundamental $(\overline{N}, N)$ under the adjacent $G_{j} \times  G_{j+1}$. The quark charges under the global $U(1)_B$ symmetry are indicated in the lower row.
Arrows pointing into (out of) a group $G_j$ indicate that a quark transforms in the (anti)fundamental representation of that group. 
}
\label{fig:mooseChang}
\end{figure}

In the special case $N=M$, the Seiberg duality for $F=N$ SQCD can be used to quantify aspects of the infrared theory for the supersymmetric linear mooses~\cite{Chang:2002qt}. 
The moose (a.k.a~``quiver''~\cite{Douglas:1996sw}) diagram for this theory is shown in Figure~\ref{fig:mooseChang}, together with the matter superfield charge assignments under the global symmetries.
Below, we summarize the method and results of Ref.~\cite{Chang:2002qt}, which are utilized several times in Section~\ref{sec:triangular}.

\medskip

Given some large hierarchy between confinement scales, e.g.~$\Lambda_1 \gg \Lambda_2 \gg \ldots$, the gauge groups $G_{2, 3, \ldots}$ in Figure~\ref{fig:mooseChang} can be treated as global symmetries in the regime where only $G_1$ is strongly coupled. In this limit the degrees of freedom at intermediate scales $\Lambda_{2, 3, \ldots} \ll \mu \ll \Lambda_1$ include the baryonic $\det Q_0$ and $\det Q_1$ operators, as well as the mesonic $(Q_0 Q_1)$ that transforms as the bifundamental of $SU(N)_\ell \times G_2$.
The baryon and meson operators satisfy the usual quantum-modified constraint,
\begin{align}
\det(Q_0 Q_1) = \det Q_0\, \det Q_1 - \Lambda_1^b,
\label{eq:detQ01}
\end{align}
with $b=2N$. If $G_2$ were not gauged, then $\det Q_0$, $\det Q_1$ and $(Q_0 Q_1)$ would form the set of gauge-invariant operators that describe the flat directions on the moduli space~\cite{Luty:1995sd}. 

Next, at $\mu \sim \Lambda_2$, the group $G_2 = SU(N)_2$ becomes strongly coupled, thus lifting the pseudo-flat $(Q_0 Q_1)$ directions and selecting the $\ev{\det Q_0}\ev{\det Q_1} = \Lambda^b$ vacuum. 
As the light degrees of freedom $(Q_0 Q_1)$ and $Q_2$ resemble $F=N$ SQCD, confinement of $G_2$ produces the gauge-invariant composite operators $\det(Q_0 Q_1)$, $\det Q_2$, and $(Q_0 Q_1 Q_2)$.
However, given the constraint equation \eqref{eq:detQ01}, the baryonic operator $\det(Q_0 Q_1)$ is not an independent degree of freedom, but is redundant with $\det Q_0$ and $\det Q_1$.

Continuing in this manner for $G_{3, 4, \ldots}$, and replacing the redundant operators $\det(Q_0 Q_1)$ and $\det (Q_0 Q_1 Q_2 \ldots)$ where possible,
the constraint equations have the form~\cite{Chang:2002qt}
\begin{align}
\det(Q_0 Q_1) &= (Q_0)^N (Q_1)^N - \Lambda_1^b, \nonumber\\
\det(Q_0 Q_1 Q_2) &=(Q_0)^N (Q_1)^N (Q_2)^N - \Lambda_1^b (Q_1)^N - (Q_0)^N \Lambda_2^b, \nonumber\\
\det(Q_0 Q_1 Q_2 Q_3) &=Q_0^N Q_1^N Q_2^N Q_3^N - \Lambda_1^b Q_2^N Q_3^N - Q_0^N\Lambda_2^b Q_3^N - Q_0^N Q_1^N \Lambda_3^b + \Lambda_1^b \Lambda_3^b,
\end{align}
where in our shorthand $Q_j^N \equiv \det Q_j$.
Given $k$ copies of the gauge group $SU(N)$, the moduli space is spanned by the reduced set of gauge invariant operators $(Q_0 Q_1 \ldots Q_k)_{ij}$ and $Q_{0, 1, \ldots, k}^N$, where the mesonic operator $(Q_0 \ldots Q_k)$ is a bifundamental of the $SU(N)_\ell \times SU(N)_r$ flavor symmetry,
with the single constraint equation
\begin{align}
\det(Q_0 \ldots Q_k) = \prod_{j=0}^k Q_j^N + \sum_{\substack{\text{all neighbor} \\ \text{contractions}}} \! \! \!  \left\{ Q_0^N \ldots Q_{j-1}^N Q_{j}^N \ldots Q_k^N \right\}.
\label{eq:Q01k}
\end{align}
Following Ref.~\cite{Chang:2002qt}, ``neighbor contraction'' indicates the replacement $Q_{j-1}^N Q_j^N \rightarrow - \Lambda_{j}^{2N}$. The sum in \eqref{eq:Q01k} includes all possible contractions.

Here we see explicitly the difference between ``even'' and ``odd'' moose theories identified in Ref.~\cite{Georgi:1985hf}. If $k$ is even, then the product $(Q_0^N \ldots Q_k^N)$ includes an odd number of terms, and the moduli space includes the origin, $(Q_0 Q_1 \ldots Q_k)_{ij} = 0$, $Q_j^N = 0$, thus permitting confinement without chiral symmetry breaking.
If instead $k$ is odd, then the product $(Q_0^N \ldots Q_k^N)$ can be fully contracted, and the constraint equation includes the constant term $\Lambda_1^b \Lambda_3^b \ldots \Lambda_k^b$, thus forcing at least a subset of the operators to acquire nonzero expectation values.

Even though this analysis used the $\Lambda_1 \gg \Lambda_2 \ldots \gg \Lambda_k$ hierarchy as a simplification, the same conclusion \eqref{eq:Q01k} is reached in any other ordering of scales~\cite{Chang:2002qt}. Furthermore, the model survives a number of consistency checks, including the $\Lambda_{j} \rightarrow 0$ limit where any of the gauge groups is replaced by a global symmetry; the addition of mass terms, where possible; or  spontaneous symmetry breaking, where an $SU(N)_j$ is higgsed to one of its subgroups.

\subsubsection*{Modified Boundary Conditions}

A closely related class of product gauge groups was shown in Ref.~\cite{Lillard:2017mon} to s-confine. In this model the $N$ copies of $Q_0$ in Figure~\ref{fig:mooseChang} were replaced by four quarks $Q$ and one two-component antisymmetric tensor $A$ ($\yA$ in Young tableaux notation).
This product gauge group confines while dynamically generating a superpotential, with the same ``even/odd''  behavior identified in Ref.~\cite{Georgi:1985hf} based on the number of gauged $SU(N)$ groups.

\medskip

In another variation of the linear $SU(N)^k$ theory, the linear moose of  Ref.~\cite{Chang:2002qt} is modified by gauging the diagonal $SU(N)$ subgroup of the global $SU(N)_\ell \times SU(N)_r$, so that the moose diagram forms a closed ring~\cite{Csaki:1997zg}.
To analyze this theory it is easiest to begin with the limit where this gauged $G_0 \subset SU(N)_\ell \times SU(N)_r$ is weakly coupled, with $\Lambda_0 \ll \Lambda_{1, 2, \ldots, k}$. If $G_0$ were not gauged, then the mesonic operator  $M_{ij} = (Q_0 \ldots Q_k)_{ij}$ would be gauge-invariant. With $G_0$ gauged, this bifundamental of $SU(N)_\ell \times SU(N)_r$ decomposes into irreducible representations of $G_0 = SU(N)$: a singlet $\mathcal M_0$, and an adjoint $\mathcal M_\text{Ad}$, defined as
\begin{align}
\mathcal M_0 \equiv \text{Tr}(Q_0 \ldots Q_k)
&&
\mathcal M_\text{Ad} \equiv (Q_0 \ldots Q_k)_{ij} - \frac{1}{N} \text{Tr}(Q_0 \ldots Q_k).
\label{eq:ringadjoint}
\end{align}

The moduli space is spanned by the gauge invariant operators $Q_j^N$; $\mathcal M_0$; and powers of $\mathcal M_\text{Ad}^m$ for $2 \leq m \leq N-1$, with the upper limit on $m$ due to the chiral ring being finitely generated~\cite{Cachazo:2002ry,Witten:2003ye}. 
With the adjoint $\mathcal M_\text{Ad}$ the only $G_0$-charged degree of freedom, it is not possible to completely higgs the gauge group.
By giving $\mathcal M_\text{Ad}$ an arbitrary expectation value, $SU(N)$ can be broken into any of its rank $N-1$ subgroups, leaving at least an unbroken gauged $U(1)^{N-1}$ Coulomb phase.

For the $SU(N)^{k+1}$ ring moose it is possible to identify a holomorphic prepotential that specifies both $W$ and the K\"ahler potential. For example, in the $\Lambda_0 \ll \Lambda_{1\ldots k}$ limit, the chiral matter field $\mathcal M_\text{Ad}$ can be combined with the vector gauge superfield $\lambda_0$ and the antichiral $\mathcal M_\text{Ad}^\dagger$ into an $\mathcal N=2$ supermultiplet, all of which transform in the adjoint representation of $G_0$.
In Ref.~\cite{Csaki:1997zg}, the prepotential hyperelliptic curve   is found by reducing the $SU(N)^k$ product group down to $SU(N)_i\times SU(N)_j$, where $\Lambda_i$ and $\Lambda_j$ are the  two smallest holomorphic scales, in analogy with the $SU(2) \times SU(2)$ theory in Ref.~\cite{Intriligator:1994sm}.

\medskip

Appealing to a geometric interpretation of this theory, we refer to these variants of the $SU(N)^k$  linear moose as having different boundary conditions. In Ref.~\cite{Chang:2002qt}, the linear moose terminates with (anti)fundamental quarks with $SU(N)$ global symmetries at each end; in Ref.~\cite{Lillard:2017mon}, the matter content is altered on one of the boundaries to include the antisymmetric $\yA$ representation; and in Ref.~\cite{Csaki:1997zg} the boundaries are made periodic, turning the moose line into a moose ring.
In Section~\ref{sec:topology} we  investigate similar variations to the boundary conditions on the moose lattice.

\section{Confinement on the Moose Lattice} \label{sec:triangular}

The discretization of  Euclidean space in $n \geq 2$ dimensions is complicated by the fact that there are multiple lattice arrangements that can be said to contain only nearest-neighbor interactions. 
This ambiguity is not present for the $n=1$ dimensional lattice, where each vertex has precisely two neighbors (or one, if the vertex is on the boundary of the lattice). Two dimensional space, on the other hand, can be tiled by squares, triangles, hexagons, or any variety of non-regular polygons.
Our decision to present a triangular (rather than rectangular) lattice is motivated by the fact that the triangular plaquette permits gauge-invariant marginal operators in the superpotential, so that all of the important mass scales in the problem are dynamically generated. This setup leads to a relatively simple analysis, where the confinement proceeds in two stages, at $\Lambda_{3N}$ and $\Lambda_{N}$.

From Section~\ref{sec:triangle-UV} to Section~\ref{sec:triangle-IR} we track the evolution of the theory from the ultraviolet ($\mu \sim M_\star)$ to the infrared ($\mu \ll \Lambda_N$).  This stage of the analysis can be completed without specifying the precise shape of the moose lattice, but for concreteness we will periodically refer to an  $[SU(3N) \times SU(N)^3]^k$ example with $k = 3 \times 3$ as an illustration.
The shape and topology of the moose lattice become much more important in Section~\ref{sec:topology}, where we analyze the different kinds of behavior that can emerge in the far infrared limit of the theory.

\subsection{The Weakly Coupled Regime} \label{sec:triangle-UV}

\begin{figure}
\centering
\includegraphics[width=0.53\textwidth]{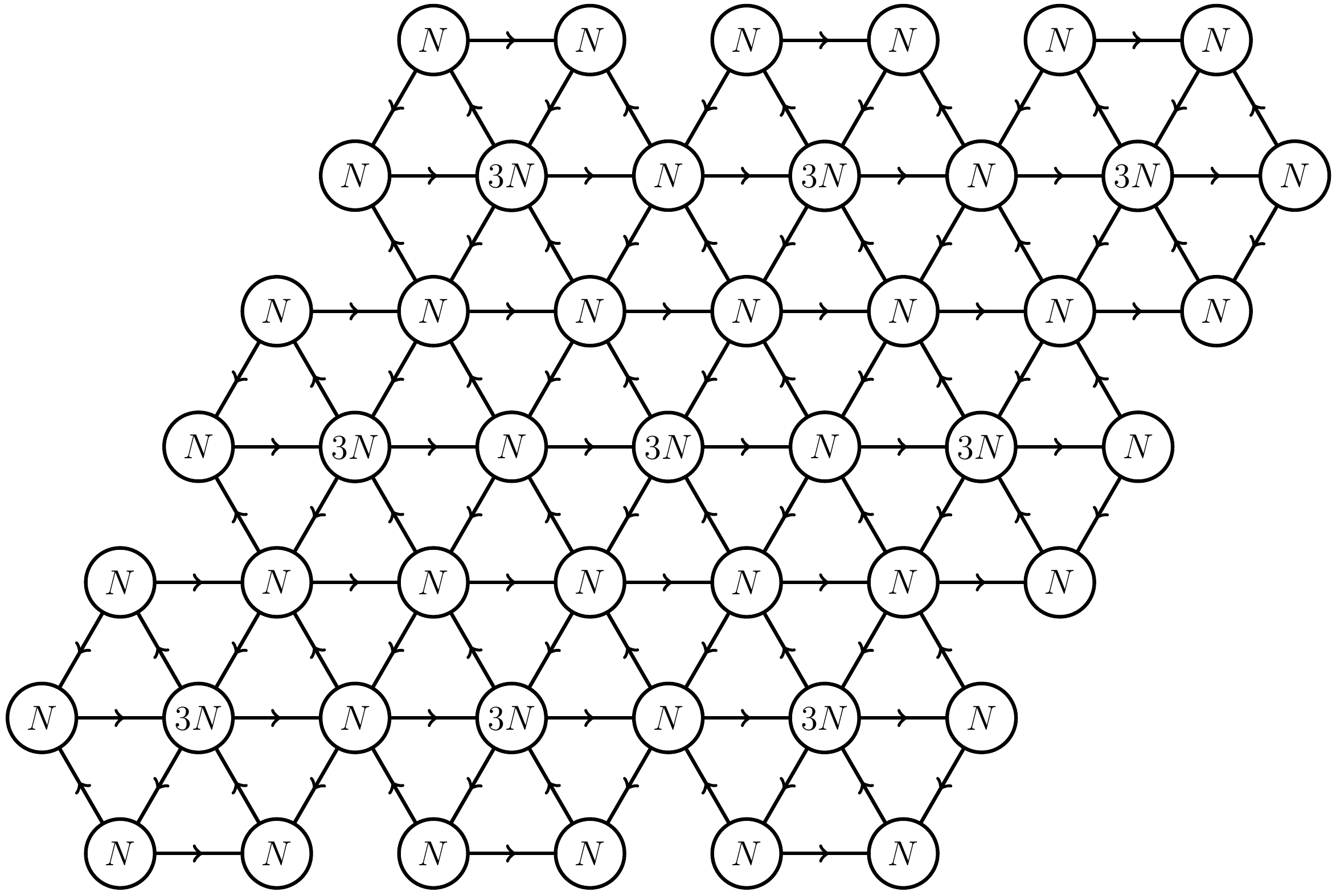}
\hspace{-2cm}
\includegraphics[width=0.53\textwidth]{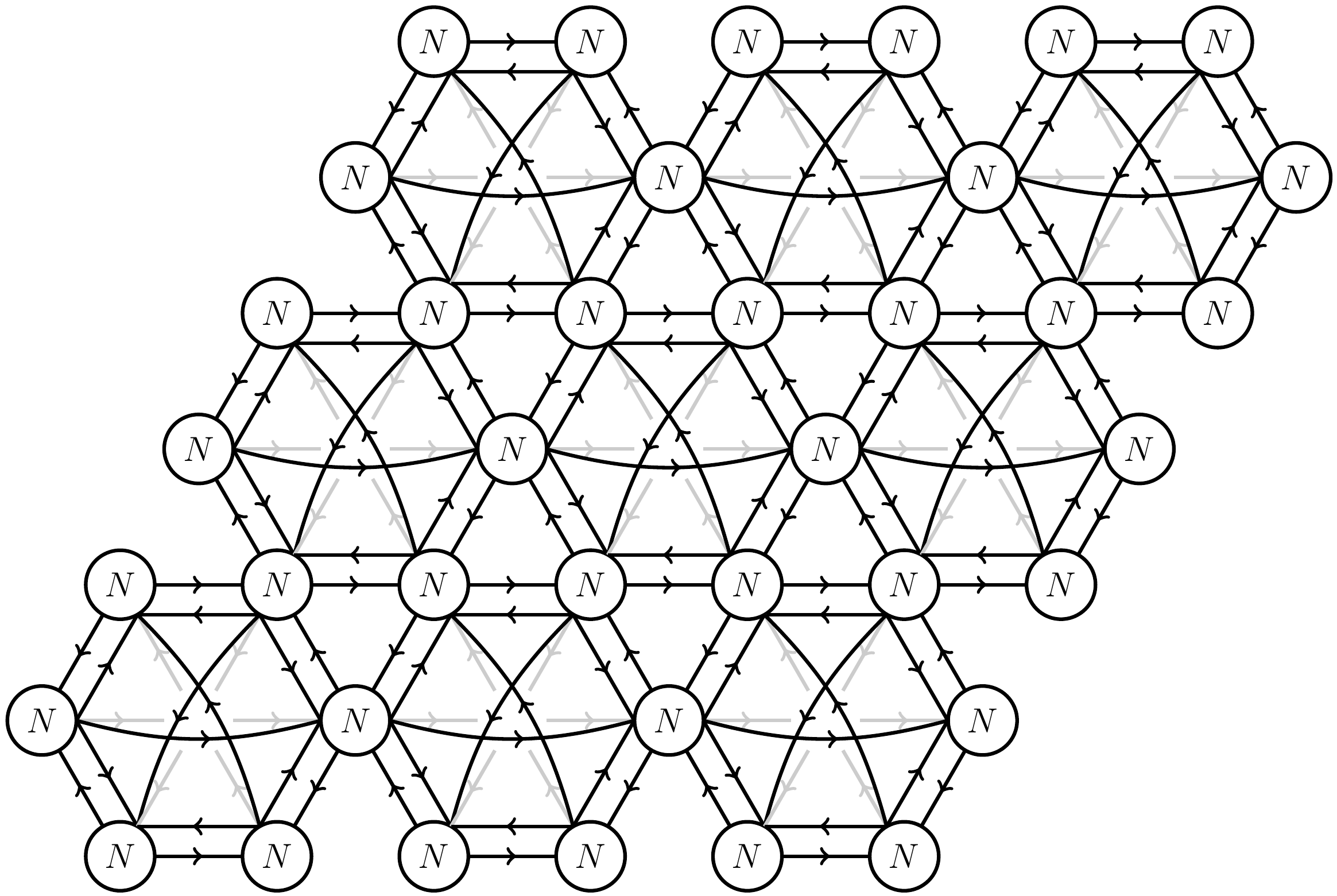}
\caption{ \textbf{Left:} the moose diagram for the $(SU(3N) \times SU(N)^3)^k$ triangular theory, in the regime where the gauged $SU(3N)$ are weakly coupled.
The $SU(N)$ nodes on the boundary of each figure (with nonzero cubic 't~Hooft anomaly coefficients) are global symmetries.
Each triangular plaquette can be encircled by gauge-invariant trace operators of the form $W \sim \lambda \text{Tr}(Q_1 Q_2 Q_3)$, where $Q_{1 \ldots 3}$ are the three bifundamental quarks that lie on the edges of the plaquette.
\textbf{Right:} The moose diagram of the $SU(3N)$ invariant degrees of freedom, in the limit $\mu \ll \Lambda_{3N}$. The paired bifundamental quarks at the edge of each hexagon acquire vectorlike mass terms of order $\mathcal O(\lambda \Lambda_{3N})$ from the trace terms in the superpotential, and are integrated out of the infrared theory.
}
\label{fig:triangle}
\end{figure}

Our discussion begins at the ultraviolet scale $\mu \sim M_\star$ where we take all of the gauge couplings to be perturbatively small, i.e.~$g_{i} \lesssim \mathcal O(1)$ for all of the $SU(3N)$ and $SU(N)$ gauge groups.
A moose diagram of one example is shown in Figure~\ref{fig:triangle}, with non-Abelian groups $SU(N_c)$ represented by circles, and chiral superfields represented by lines.

\medskip

The ``unit cell'' of the lattice consists of a central $SU(3N)$ gauge group surrounded by six $SU(N)$ groups; three pairs of bifundamental quarks $Q + \barQ$, respectively in the fundamental and antifundamental representations of $SU(3N)$; and a bifundamental $q$ for each neighboring  $SU(N) \times SU(N)$ pair on the boundary of the hexagonal unit cell, for a total of six.
Throughout this paper we will borrow lattice-related terms to describe the various components of the model, so that each gauge group is located at a ``site''/``node''/``vertex''; the bifundamentals are ``links'' or edges; and the trilinear gauge invariants $\text{Tr}(q_1 q_2 q_3)$ will be referred to as ``plaquette'' operators.

Figure~\ref{fig:triangle} shows a $k=3\times 3$ example of the $[SU(3N) \times SU(N)^3]^k$ model, 
with nine copies of the unit cell arranged in a parallelogram. 
Each $SU(N_c)$ group on the boundary of the lattice is a global symmetry. At these sites the cubic  anomaly coefficients are nonzero, so these $SU(N)$  groups cannot be gauged without introducing additional matter fields. 

 It is not necessary to include an integer number of unit cells in the lattice. Although Figure~\ref{fig:triangle} shows an example where every $SU(3N)$ group is gauged, and the lattice boundary passes only through $SU(N)$ sites, we could just as well have routed the lattice boundary through some of the $SU(3N)$ groups instead. 
Periodic boundary conditions are more restrictive. A periodic direction should include an integer number of unit cells, so that the gauged $SU(3N)$ and $SU(N)$ sites are all anomaly-free.

\medskip

At the ultraviolet scale $M_\star$,  each of the gauge groups is taken to be weakly coupled. For the $SU(3N)$ groups this is a natural assumption: the matter content of Figure~\ref{fig:triangle} supplies each $SU(3N)$ group with $F = 3N$ pairs of $(Q + \barQ)$ quarks,  and   SQCD with $F = N_c$ exhibits asymptotic freedom.
Given a coupling $g_{3N}(M_\star)$ defined at the short-distance scale $M_\star$, the gauge coupling at another scale $\mu$ is given by the NSVZ $\beta$ function,
\begin{align}
\beta(g_{N_c}) = \frac{d g_{N_c} }{d \log \mu} = - \frac{g_{N_c}^3 b}{16 \pi^2} ,
&&
b = 3 N_c - F,
\end{align}
and $\beta < 0$ for $F = N_c$. Dimensional transmutation defines the holomorphic scale $\Lambda_{3N}$ in terms of the gauge coupling $g$ and the  CP violating Yang-Mills phase $\theta_\text{YM}$,
\begin{align}
\Lambda_{3N}^b = \mu^b \exp\left( -\frac{8\pi^2}{g_{3N}^2(\mu)} + i \theta_\text{YM} \right),
\end{align}
which indicates the scale where the $SU(N_c)$ gauge group becomes strongly coupled, i.e.~$g(\Lambda) \rightarrow \infty$.
If $g_{3N}(M_\star) \ll 4\pi$ is perturbatively small, then there will be a hierarchy $\Lambda_{3N} \ll M_\star$ between the two scales.

In a theory with multiple unit cells, the scales $\Lambda_{3N}^{(i)}$ for the different $SU(3N)$ gauge groups are not necessarily identical, and can even include large hierarchies $\Lambda_{3N}^{(i)} \ll \Lambda_{3N}^{(j)}$, as long as all of the $\Lambda_{3N}^{(j)}$ are smaller than $M_\star$.

Each $SU(N)$ node, on the other hand, has $3N + N + N$ pairs of  quarks in the $\mathbf{N}$ and $\overline{\mathbf{N}}$ representations, for a total of $F = 5 N_c$ flavors. In this case, $b = - 2N$, the $\beta(g_{N})$ function is positive, and there is no asymptotic freedom. However, if we start with a weakly coupled $g_{N} (M_\star) \lesssim 1$ at $\mu = M_\star$, the $SU(N)$ groups remain weakly coupled for all $\mu < M_\star$ down to the phase transition at $\mu \sim \Lambda_{3N}$.
In this work we do not explore the $\mu \gg M_\star$ limit of the theory, where the $SU(N)$ become strongly coupled.

\medskip

The final addition to the theory is a chiral superpotential $W$, composed of the marginal trace operators $W \supset \lambda \text{Tr}\, (Q_1 Q_2 Q_3)$ that encompass each triangular plaquette.
Eventually we anticipate introducing a full set of symmetry-violating marginal and irrelevant operators into $W$, but at this stage of the discussion we restrict $W$ to include only the $U(1)_R$ preserving operators.
 For Figure~\ref{fig:triangle} this includes the plaquettes surrounded by $SU(3N) \times SU(N) \times SU(N)$ groups, but not the $SU(N) \times SU(N) \times SU(N)$ plaquettes.

The moose notation is particularly helpful when constructing gauge invariant operators: one simply follows the arrows on the diagram so as to form closed loops. For example, within the bulk of the lattice, the next set of trace operators are the dimension-6, irrelevant operators $W \supset \alpha (Q_1 Q_2 \ldots Q_6)/ M_\star^3$. 
Another type of gauge invariant operator is formed by open ``Wilson lines,'' which start and end on the boundaries of the lattice. These operators transform as bifundamentals under the $SU(N_f)_L \times SU(N_f)_R$ global symmetries associated with their endpoints, and have the form
\begin{align}
W \supset \frac{\alpha_{ij} }{M_\star^{k-2}}  Q^{(1)}_{i, n_{1} } Q^{(2)}_{n_{1}, n_2 } Q^{(3)}_{n_2, n_3} \ldots Q^{(k)}_{n_{k-1}, n_{k} } Q^{(k+1)}_{n_k, j} ,
\label{eq:supMstarM}
\end{align} 
where $i, j$ are indices for the global symmetries $SU(N_f)_{L, R}$, and the repeated $n_{1 \ldots k}$ indices correspond to the gauged $SU(N_c)$ groups.
Simple examples on Figure~\ref{fig:triangle} include the straight left-to-right lines, which pass either through alternating $SU(N)$ and $SU( 3N)$ nodes, or exclusively though $SU( N)$ nodes.
In addition to the analogous straight lines along the $\varphi = 120^\circ$ and $\varphi = 240^\circ$ directions (where we define $\varphi = 0$ as left to right in the page), generic Wilson line gauge invariants  can include any number of $60^\circ$ corners.
Most of these operators have no relevance in the infrared theory: as we show in Section~\ref{sec:confine3N}, the degrees of freedom associated with the $60^\circ$ corners become massive, so that the low energy theory is dominated by the straight Wilson lines that pass through $SU(3N)$ sites.

An altogether different set of gauge invariant ``baryon'' operators is generated by the  completely antisymmetric products
\begin{align}
W \supset  \frac{(q)^N}{M_\star^{N-3}}  +  \frac{(\barq)^N}{M_\star^{N-3}}  +  \frac{(Q_a)^N (Q_b)^N (Q_c)^N}{M_\star^{3N - 3}} +  \frac{(\barQ_a)^N (\barQ_b)^N (\barQ_c)^N }{M_\star^{3N - 3}} ,
\label{eq:supMstarB}
\end{align}
where $q$ and $\barq$ are any of the $SU(N) \times SU(N)$ bifundamentals, and $Q_{a,b,c}$ and $\barQ_{a,b,c}$ are $SU(3N) \times SU(N)$ bifundamentals in the $(\mathbf{3N}, \overline{\mathbf{N}})$ and $(\overline{\mathbf{3N}}, \mathbf{N})$ representations, respectively.
In the $N=3$ special case, $q^N$ and $\barq^N$ are marginal operators.

\medskip 

By invoking $M_\star$ as the mass scale associated with the irrelevant operators, we complete the promise made earlier in this section, that the theory is weakly coupled at scales $\mu \lesssim M_\star$. This statement now applies to the superpotential couplings as well as the gauge interactions, as long as the dimensionless parameters $\alpha, \lambda$, etc.~are all $\lesssim \mathcal O(1)$.

Theories of gravity are generally expected to violate global symmetries~\cite{Giddings:1987cg,Kamionkowski:1992mf,Barr:1992qq,Kallosh:1995hi,Abbott:1989jw,Coleman:1989zu}, so the flavor symmetry violating superpotential \eqref{eq:supMstarM} and its baryon number violating cousin \eqref{eq:supMstarB} may be generated by Planck-scale effects, i.e.~with $M_\star \rightarrow M_p$.
A lower scale $M_S < M_p$ may be appropriate if this theory is to be embedded within string theory, any other $\mathcal N > 1$ version of supersymmetry, or any more than four continuous spacetime dimensions.

\subsection{First Stage of Confinement} \label{sec:confine3N}

Approaching the scales $\mu \rightarrow \Lambda_{3N}^{(i)}$, the $SU(3N)_i$ gauge groups become strongly coupled.
To understand the behavior of the theory in the infrared, $\mu \ll \Lambda_{3N}$, we refer to the Seiberg duality for $F = N_c = 3N$ SQCD, an $S$-duality that relates the weakly-coupled gauge theory at $\mu \gg \Lambda_{3N}$ to a theory of gauge invariant mesons and baryons at $\mu \ll \Lambda_{3N}$. 
Everything we need to know about the $\mu \lesssim \Lambda_{3N}$ regime of the theory can be deduced from studying a single unit cell of the lattice.

\begin{figure}
\centering
\includegraphics[width=0.95\textwidth]{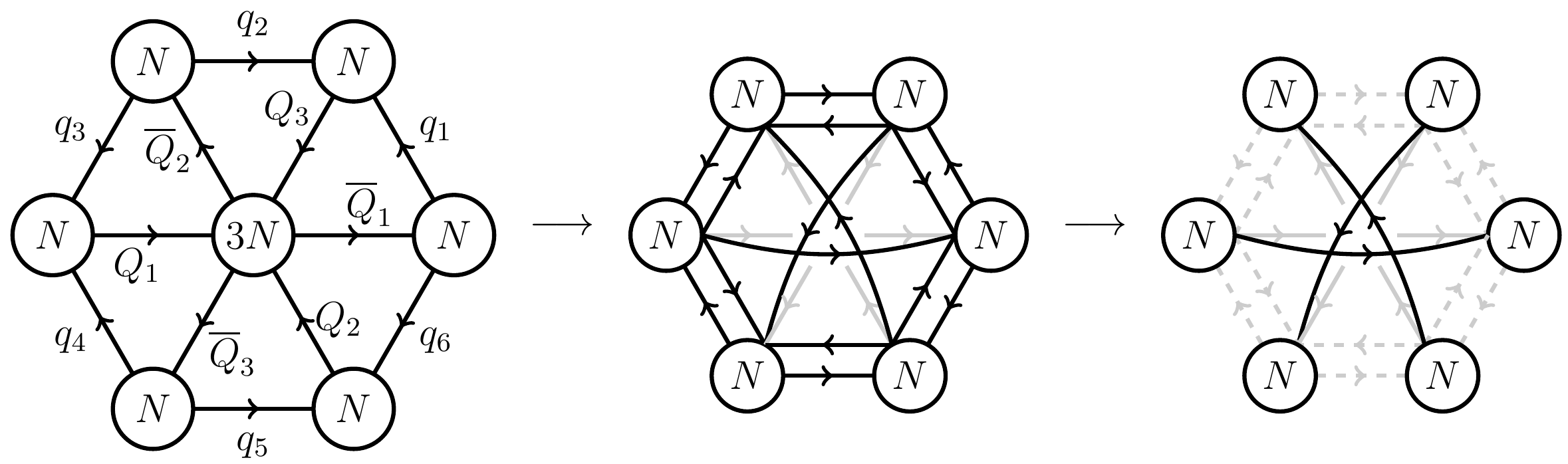}
\caption{When the theory is probed at intermediate scales $\mu < \Lambda_{3N}$, the unit cell evolves towards the moose diagram shown on the right. First, on the left, we show the $\Lambda_{3N} \ll \mu < M_\star$ unit cell for weakly coupled $SU(3N)$. After $SU(3N)$ confines, the light degrees of freedom include the $SU(N) \times SU(N)$ meson operators, shown in the middle diagram either crossing the center of the unit cell or connecting adjacent $SU(N)$ nodes on its boundary. Together with the original $SU(N) \times SU(N)$ bifundamentals $q_i$, the meson operators on the edge of the hexagon acquire $\mathcal O(\lambda_i \Lambda_{3N})$ masses from the plaquette operators $W \sim Q \barQ q \rightarrow (Q \barQ) q$.
The rightmost diagram shows the theory at scales $\mu < \lambda \Lambda_{3N}$, after integrating out the vectorlike quarks. In this infrared limit, the only light $SU(N)$-charged matter fields are the mesons that pass through the center of the unit cell. 
}
\label{fig:3Nsequence}
\end{figure}

Thanks to the positive sign in the $\beta(\mu)$ function for $N_c = N$ with $F = 5N$, the $SU(N)$ gauge coupling $g_N$ --- already perturbative at $\mu = M_\star$ --- becomes even smaller as $\mu$ decreases from $M_\star$ towards $\Lambda_{3N}$. 
As far as the $SU(3N)$ node is concerned, there are $3N$ flavors of quarks $Q$ and antiquarks $\barQ$, with opposite charges under a $U(1)_B$ baryon number, and transforming under approximate $SU(3N)_{Q}$ and $SU(3N)_{\barQ}$ flavor symmetries.
By gauging the $SU(N)$ subgroups, these putative global $SU(3N)_{Q, \barQ}$ flavor symmetries are explicitly broken, but at $\mu \lesssim \Lambda_{3N}$ this effect is a small perturbation.

The gauge groups and chiral matter associated with the unit cell are shown in Figure~\ref{fig:3Nsequence}.
The superpotential includes six $U(1)_R$ conserving plaquette operators per unit cell,
\begin{align}
W \supset \lambda_1 (q_1 Q_3 \barQ_1) + \lambda_2 (q_2 Q_3 \barQ_2) + \lambda_3 (q_3 Q_1 \barQ_2) + \lambda_4 (q_4 Q_1 \barQ_3) + \lambda_5 (q_5 Q_2 \barQ_3) + \lambda_6 (q_6 Q_2 \barQ_1),
\label{eq:plaquettes}
\end{align}
where the trace over gauge indices is implied.
Each $\lambda_{c=1 \ldots 6}$ is a dimensionless complex parameter. 
Under the simplest $U(1)_R$ charge assignment, each $SU(3N)$-charged  $Q_i$ and $\barQ_i$ is neutral, while the $SU(N)\times SU(N)$ bifundamentals $q_i$ have $R$ charges of $+2$.

Following \eqref{eq:defBMB}, the Seiberg dual of the $F = N_c$ SQCD is described by $F^2$ meson operators and two baryon operators, with one constraint equation:
\begin{align}
(M_{mn})_{ij} = (Q_m \barQ_n)_{ij}, 
&&
B = (Q_1^N Q_2^N Q_3^N), 
&&
\barB = (\barQ_1^N \barQ_2^N \barQ_3^N), 
&&
\det M -B \barB =   \Lambda_{3N}^b,
\label{eq:unitW}
\end{align}
with $b=6N$,
for indices $m,n = 1,2,3$ and $i,j = 1, 2, \ldots, N$. 
The determinant $\det M$ is a shorthand for the completely antisymmetric product of $3N$ $(Q \barQ)$ operators, which could also be expressed in terms of determinants of the nine distinct $SU(N)$ charged mesons,  $\det M_{mn}$. 

\medskip

In the $g_N \rightarrow 0$ limit, where the specified $SU(N)$ node becomes a global symmetry, $M_{ab}$, $B$ and $\barB$ all represent flat directions on the moduli space subject to the constraint \eqref{eq:unitW}.
At symmetry-enhanced points on the moduli space, the $SU(F)_Q \times SU(F)_{\barQ} \times U(1)_B \times U(1)_R$ global symmetry may be spontaneously broken to $SU(F)_\text{diag} \times U(1)_B \times U(1)_R$, by the $\ev{M_{ij} } \propto \delta_{ij}$ expectation value (with $i,j = 1 \ldots F$).
Or, the $U(1)_B$ global symmetry can be broken by $\ev{ B \barB} = -\Lambda_{3N}^{b}$.

For gauged $SU(N)$, many of these flat directions are lifted. The true moduli space is spanned by gauge-invariant operators~\cite{Luty:1995sd}, and the $M_{mn}$ are not gauge invariant: they transform as bifundamentals of $SU(N) \times SU(N)$. Each gauged $SU(N)$ introduces a $D$-term potential for the mesons $M_{mn}$, so that the vacuum of the theory lies  on the $\ev{ B \barB} = - \Lambda_{3N}^{b}$ branch, with spontaneously broken baryon number and unbroken $SU(N)$ symmetries.

One linear combination of the $B$ and $\barB$ scalars, the ``$B + \barB$'' direction that changes the value of $\ev{B \barB}$, acquires an $\mathcal O(\Lambda_{3N})$ mass. The other linear combination, the ``$B - \barB$'' or ``$\tan \beta$'' direction tangential to the $\ev{B \barB} = -\Lambda_{3N}^b$ flat direction, remains massless. This flat direction is lifted if $U(1)_B$ is explicitly broken; for example, by the gauge invariant irrelevant operators,
\begin{align}
W & \supset  \frac{(Q_1^N Q_2^N Q_3^N) }{M_\star^{3N-3}} + \frac{(\barQ_1^N \barQ_2^N \barQ_3^N) }{M_\star^{3N-3}} +  \frac{(Q_1^N Q_2^N Q_3^N)(\barQ_1^N \barQ_2^N \barQ_3^N) }{M_\star^{6N-3}} + \ldots
\nonumber\\
W(\mu < \Lambda_{3N}) & \sim \left( \frac{  \Lambda_{3N}^{3N-1}}{M_\star^{3N-3}} \right) \mathcal B +  \left( \frac{  \Lambda_{3N}^{3N-1}}{M_\star^{3N-3}}\right) \overline{\mathcal B}  + \left( \frac{\Lambda_{3N}^{6N - 2} }{M_\star^{6N-3} }\right)  \mathcal B \overline{\mathcal B} + \ldots
\end{align}
which induce small tadpole operators and even smaller baryon mass terms into the superpotential.

Here we have rendered $B$ and $\barB$ as operators with canonical mass dimension $+1$, by extracting the appropriate powers of the confinement scale $\Lambda_{3N}$:
\begin{align}
\mathcal B  &= \frac{(Q_1^N Q_2^N Q_3^N) }{{\Lambda_{3N}}^{3N-1} },
&
\overline{\mathcal B}  &= \frac{(\barQ_1^N \barQ_2^N \barQ_3^N) }{{\Lambda_{3N}}^{3N-1} } ,
&
\mathcal M_{ab} &= \frac{(Q_a \barQ_b) }{{\Lambda_{3N}} } .
\end{align}
Applying the same $(Q_a \barQ_b) \rightarrow \Lambda_{3N}\mathcal  M_{ab}$ mapping to the plaquette superpotential \eqref{eq:unitW}, we see  each of the $a \neq b$ mesons acquires a vectorlike mass pairing with one of the edge quarks $q_i$:
\begin{align}
W(\mu < \Lambda_{3N}) &\supset m_1 q_1 \mathcal M_{31} + m_2 q_2 \mathcal M_{32} + m_3 q_3 \mathcal M_{12} + m_4  q_4 \mathcal M_{13} 
+ m_5 q_5 \mathcal M_{23} + m_6 q_6 \mathcal M_{21},
\label{eq:unitSuperMass}
\end{align}
where $m_a = \lambda_a \Lambda_{3N}$. 

Figure~\ref{fig:3Nsequence} illustrates the transition. The middle diagram shows the nine $\mathcal M_{ab}$ mesons together with the six $q_c$ in one unit cell. This moose diagram describes the theory  at the intermediate scales $m_a < \mu < \Lambda_{3N}$.
All of the mesons shown in Figure~\ref{fig:3Nsequence} are neutral under the spontaneously broken $U(1)_B$. However, as we show in Section~\ref{sec:globals}, there are a number of unbroken $U(1)$ global symmetries under which $B$ and $\barB$ are neutral, and the mesons $\mathcal M_{ab}$ are charged.

\subsection{Integrating Out Heavy Mesons and Quarks} \label{sec:vectorlikeMass}

From the perspective of any of the $SU(N)$ nodes, the confinement at $\Lambda_{3N}$ does not pose a significant change to the theory.
For example, the $3N$ quarks $\barQ_1$ in the $\mathbf{N}$ representation of the $\varphi = 0^\circ$ $SU(N)$ node in Figure~\ref{fig:3Nsequence} are replaced with $N + N +N $ mesons $\mathcal M_{11}$, $\mathcal M_{21}$, and $\mathcal M_{31}$, also in the fundamental representation of $SU(N)$.
For an $SU(N)$ node in the bulk of the moose lattice, the $SU(3N)$ confinement simply replaces $F = 5N$ SQCD with a differently labelled $F=5N$ SQCD model. This is seen explicitly  in the right panel of Figure~\ref{fig:triangle}, for any of the $SU(N)$ sites in the bulk.

The main difference is that the triangle plaquette operators of \eqref{eq:plaquettes} become vectorlike mass terms, \eqref{eq:unitSuperMass}, pairing the mixed mesons $\mathcal M_{a, b\neq a}$ with the $SU(N) \times SU(N)$ bifundamental quarks $q_i$.
Thus, of the $F= 5N$ original flavors, $4N$ of these acquire $\mathcal O(\lambda_i \Lambda_{3N}^{(j)})$ masses, where $j=1,2$ refer to the $SU(3N)$ groups on either side of a given $SU(N)$ node. 
At scales below $\lambda \Lambda_{3N}$ the heavy quarks and mesons are integrated out, leaving only the
$SU(N)$ charged $\mathcal M_{aa}$ mesons, 
as shown in the rightmost diagram of Figure~\ref{fig:3Nsequence}.

\paragraph{Integrating Out:}
On the $\ev{B \barB} =  -\Lambda_{3N}^b$, $\ev{M_{ab} }=0$ branch of the vacuum, all of the $\mathcal M_{ab}$ degrees of freedom correspond to approximately flat directions on the moduli space, at least if we ignore the $D$-term potential from the weakly gauged $SU(N)$ groups.
As can be seen from the supersymmetric Lagrangian, which includes terms of the form $\mathcal L \supset \abs{\partial W/\partial \Phi}^2$ for each of the superfields $\Phi$,
many of the otherwise-flat directions are lifted by \eqref{eq:unitSuperMass}.
To pick one example, the $m_1$ term in $W$ contributes two terms to the so-called $F$-term potential,
\begin{align}
-\mathcal L_F \supset \abs{ \frac{ \partial W}{\partial q_1}}^2 + \abs{ \frac{\partial W}{\partial \mathcal M_{31}}}^2 
\approx m_1^\star m_1 \left( \mathcal M_{31}^\star \mathcal M_{31} + q_1^\star q_1 \right),
\end{align}
In the absence of any other $q_1$ dependent terms in the superpotential, the scalar potential is minimized at  the vacuum solution $\ev{\mathcal M_{31} }= \ev{q_1} = 0$.

Not counting the dimension-6 irrelevant operators, the only other $q_1$ dependent term in the superpotential comes from the triangular plaquette operator involving $q_1$ and the $q_b$ and $q_c$ from the adjacent unit cells, $W \sim \lambda_{bc} (q_1 q_b q_c)$.
Together with the $m_1$ mass term, the vacuum solution nominally shifts to
\begin{align}
m_1 \mathcal M_{31} + \lambda_{bc} q_b q_c = 0,
&&
q_1 = 0.
\end{align}
However, confinement on the $b$ and $c$ unit cells sets $\ev{q_b } = \ev{q_c } = 0$, so that the minimum of the scalar potential remains at $\ev{\mathcal M_{31}} = 0$. 
In principle the dimension-6 operators do have the potential to shift $\ev{\mathcal M_{ab}}$ and $\ev{q_i}$ away from the origin of moduli space; however, thanks to the powers of $M_\star^{3}$ in the denominator of such operators, the resulting shift is small enough that it can be safely ignored.

\paragraph{Matching Holomorphic Scales:}
Before we move on to the strongly coupled regime of the $SU(N)$ gauge theory, let us take a moment to study the  transition at $\mu \sim \lambda \Lambda_{3N}$. This is where the sign of the $SU(N)$ $\beta$ function changes, which is what causes $SU(N)$ to become strongly coupled at $\mu \ll \Lambda_{3N}$ in the first place.
By matching the gauge coupling at the scales $\mu = m_c$, the holomorphic scale $\Lambda_N$ for the $F=N$ theory can be derived from $g_N(\mu = M_\star)$.
Specifically, it is the holomorphic gauge coupling $\tau$ that we match at each threshold,
\begin{align}
2 \pi i \, \tau(\mu) \equiv - \frac{8\pi^2}{g^2(\mu)} + i \theta_\text{YM}.
\end{align} 
We begin with the values of $g_N$ and $\theta_\text{YM}$ evaluated at $\mu = M_\star$, and define a $\Lambda_{F=5N} \gg M_\star$ holomorphic scale: 
\begin{align}
\Lambda_{N, F=5N}^{b}  &= M_\star^{-2N} \exp\left(- \frac{8\pi^2 }{g^2(M_\star)} + i  \theta_{F=5N} \right),
\nonumber\\
\Lambda_{N, F=5N} &= M_\star \exp\left( \frac{+1}{2N} \frac{8\pi^2}{g^2(M_\star)} - \frac{i \theta_{F=5N}}{2N} \right) .
\end{align}
Although the $CP$-odd $\theta_\text{YM}$ parameter is invariant under the RG evolution, it can acquire threshold corrections at $\mu \sim m$, so we specify $\theta_\text{YM}(\mu = M_\star) \equiv \theta_{F = 5N}$.

Allowing the four  superpotential coupling constants $\lambda_{a=1,2,3,4}$ to acquire distinct values, there are generally four distinct mass thresholds, $m_a > m_b > m_c > m_d$.
Matching $\tau(\mu = m_a)$ between the $F=5N$ and $F=4N$ theories, with $b=-2N$ and $b=-N$, respectively, we find
\begin{align}
\frac{ \abs{\Lambda_{F=5N}}^{-2N}  e^{i \theta_{F=5N} } }{m_a^{-2N} }  &= e^{2\pi i\, \tau(m_a) } = \frac{ \abs{\Lambda_{F=4N}}^{-N} e^{i \theta_{F=4N} } }{m_a^{-N} } ,
\nonumber\\
\abs{\Lambda_{F=5N}}^{-2N} m_a^N  e^{i \theta_{F=5N} } &= \abs{ \Lambda_{F=4N}}^{-N}  e^{i \theta_{F=4N} } .
\end{align}
Applying the same matching procedure at $m_{b, c, d}$, we find
\begin{align}
\abs{\Lambda_{F=5N}}^{-2N} (m_a m_b m_c m_d)^N  e^{i \theta_{F=5N} } &= \abs{ \Lambda_{F=N}}^{2N}  e^{i \theta_{F=N} } .
\label{eq:matchABCD}
\end{align}
So, the $\theta_\text{YM}$ phase in the $F=N$ theory is given by
\begin{align}
\theta_{F=N} = \theta_{F=5N} + N \arg\left( m_a m_b m_c m_d\right)  = \theta_{F=5N} + N\arg\left( \lambda_a \lambda_b \lambda_c \lambda_d \right) .
\label{eq:thetaYMmatch}
\end{align}

Inspecting the real part of \eqref{eq:matchABCD}, we find
\begin{align}
\abs{\Lambda_{N, F=N}}  
& =  \abs{\frac{ m_a m_b m_c m_d  }{\Lambda_{N, F=5N}^2  } }^{1/2}
= \frac{ \abs{\lambda_a \lambda_b \lambda_c \lambda_d }^{1/2} \abs{ \Lambda_{3N}}^2 }{\abs{\Lambda_{N, F=5N} } }
= \abs{\lambda_a \lambda_b \lambda_c \lambda_d }^{1/2} \frac{\abs{ \Lambda_{3N}}^2 }{M_\star} \exp\left( - \frac{8\pi^2 /2N }{ g_N^2(M_\star) } \right),
\nonumber\\
\abs{\Lambda_{N, F=N}} &= \abs{\lambda_a \lambda_b \lambda_c \lambda_d }^{1/2} M_\star \exp\left( - \frac{1}{N} \frac{8\pi^2 }{g_{3N}^2(M_\star) }  - \frac{1}{2N} \frac{8\pi^2 }{g_{N}^2(M_\star) } \right).
\label{eq:LambdaFNmatch}
\end{align}
As expected, $\Lambda_{N}$ for the $F=N$ phase of the theory is exponentially small compared to $\Lambda_{3N}$, which is itself exponentially smaller than $M_\star$.
If any of the $\lambda_i$ are much smaller than $\mathcal O(1)$, then $\Lambda_{N, F=N}$ is suppressed by a further factor of $\sqrt{\lambda_i}$.
Note that the ultimate expression for $\Lambda_N$ is unaffected by changes to the assumed ordering, $\lambda_a > \lambda_b > \lambda_c > \lambda_d$.

\begin{figure}[t]
\centering
\includegraphics[width=0.85\textwidth]{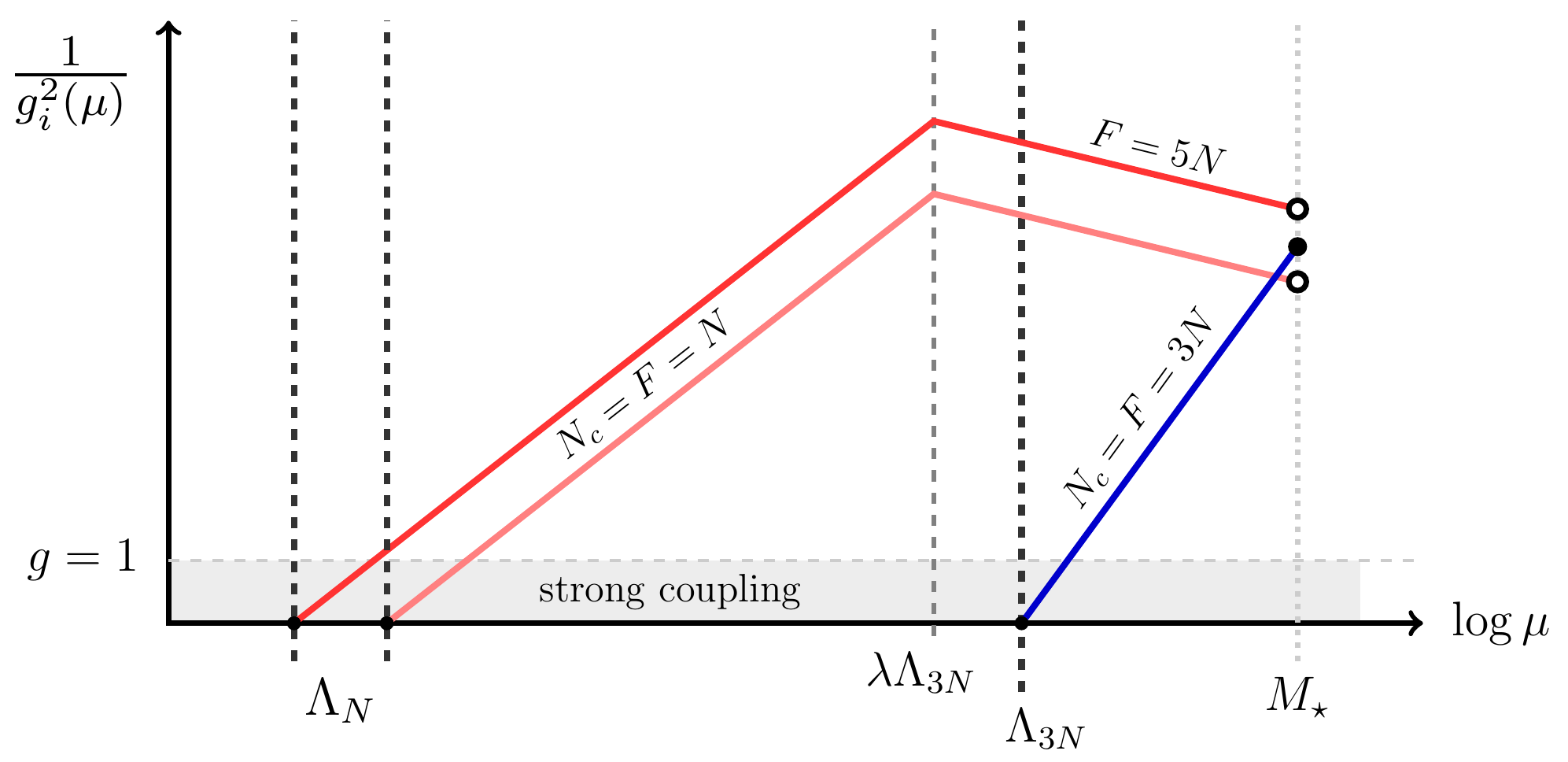}
\caption{An illustration (not exactly to scale) of the running couplings for one $N_c = 3N$ and two $N_c = N$ gauge groups, taken to have similar values near $\mu \sim M_\star$. The narrative in this section follows the plot from right to left, UV to IR. Between $M_\star$ and $\Lambda_{3N}$, $g_{3N}$ (in blue)  runs quickly towards strong coupling, with the slope on the plot given by $b = 3N_c - F = 6N$, while the $g_N$ (in red) become weaker (with $b = 3N-5N= -2N$). After $SU(3N)$ confinement, $4N$ of the $SU(N)$ charged flavors acquire $\mathcal O(\lambda \Lambda_{3N})$ masses, instigating the $F \rightarrow N$ transition. For simplicity we have taken $\lambda_{a,b,c,d} \approx \lambda$ to be nearly equal, while also assuming an approximately uniform value of $\Lambda_{3N}$ for all of the adjacent unit cells. After the last vectorlike pairs are integrated out, the slope has changed from $b=-2N$ to $b=+2N$, and $g_N$ runs to strong coupling as $\mu \rightarrow \Lambda_N$. In this graphic we follow two different $SU(N)$ groups. If their mass thresholds are identical then the ratio between the $g_N^2$ couplings remains fixed, but for generic $\lambda_i \Lambda_{3N}^{(j)}$ this is not so. }
\label{fig:cartoon}
\end{figure}

\medskip

For simplicity, we assumed in \eqref{eq:LambdaFNmatch} that the two unit cells that border the $SU(N)$ node have equal values of $\Lambda_{3N}$.
Very little changes when this assumption is relaxed. If instead the $SU(3N)_L$ and $SU(3N)_R$ gauge groups have couplings $g_{3N,L} \neq g_{3N, R}$ at $\mu = M_\star$, \eqref{eq:LambdaFNmatch} generalizes to
\begin{align}
\abs{\Lambda_{N, F=N}} &= \abs{(\lambda_a \lambda_b) (\lambda_c \lambda_d )}^{1/2} M_\star \exp\left( - \frac{1}{2N} \frac{8\pi^2 }{g_{3N,L}^2(M_\star) } - \frac{1}{2N} \frac{8\pi^2 }{g_{3N,R}^2(M_\star) }  - \frac{1}{2N} \frac{8\pi^2 }{g_{N}^2(M_\star) } \right).
\end{align}
Even if for some reason there is a large hierarchy between $\Lambda_{3N,L}$ and $\Lambda_{3N,R}$, 
it is still true that the $SU(N)$ remains weakly coupled until after both of its neighboring $SU(3N)$ gauge groups have confined.
Once the first of the strongly coupled $SU(3N)$ groups confines at (for example) $\Lambda_{3N, L}$, only two pairs of the bifundamental fields acquire $\mathcal O(\Lambda_{3N, L})$ masses.
After these are integrated out, the $F = 5N$ effective flavors of $SU(N)$ are reduced to $F=3N$; the coefficient of the one-loop $\beta$ function switches from $b = -2N$ to $b=0$; and the gauge coupling $g_N(\mu)$ remains fixed at a perturbatively small value.
Only after the remaining $SU(3N)_R$ group confines does the $\beta(g_N)$ function become negative.

\medskip

After integrating out the vectorlike pairs of quarks $q$ and mesons $M_{ab}$, the $SU(N)$ gauge groups become strongly coupled in the infrared. 
At this stage of the calculation, $\Lambda_{N, F=N}$ is the only relevant version of the $SU(N)$ holomorphic scale, so for the remainder of this section we take $\Lambda_N^b$ to refer exclusively to
\begin{align}
\Lambda_N^b \equiv \Lambda_{N, F=N}^{2N} = \abs{\Lambda_{N, F=N}}^{2N} e^{i \theta_{F=N}},
\end{align}
where $\theta_{F=N}$ and $\abs{\Lambda_{N, F=N}}$ are given in \eqref{eq:thetaYMmatch} and \eqref{eq:LambdaFNmatch}.

In Figure~\ref{fig:cartoon} we show the running gauge couplings for an $SU(3N)$ and two adjacent $SU(N)$ groups. At $\mu = M_\star$ the various $g_i^2$ are taken to be of the same magnitude. In this example we take the simplifying limit where the neighboring $\Lambda_{3N}$ are similar in size, as are  the $\lambda_i$ superpotential coupling constants, so that
the transition between $F=5N$ and $F=N$ for the $SU(N)$ groups occurs sharply at $\mu \approx \lambda \Lambda_{3N}$. 

With $b= 3N_c - F = 6N$ for the $SU(3N)$ gauge group, $g_{3N}$ runs relatively quickly towards strong coupling with decreasing $\mu < M_\star$, while the $SU(N)$ more gradually become more weakly coupled. For $g_N$, the sharp transition from $F=5N$ to $F=N$ at $\mu = \lambda \Lambda$ flips the sign of $b$, implying that 
only at $\mu \approx (\lambda \Lambda_{3N})^2 / M_\star$ has $g_N(\mu)$ returned to its initial value at $\mu = M_\star$. Thus,
$\Lambda_{N} \ll (\lambda \Lambda_{3N})^2 / M_\star$ is generally much smaller than  $\Lambda_{3N}$ and $M_\star$.
If Figure~\ref{fig:cartoon} were drawn to scale, the red lines corresponding to the different $g_N$ would form mirror images in the vicinity of $\mu \sim \lambda \Lambda_{3N}$, and $\Lambda_N$ would be much further to the left on the plot.

\paragraph{Deformed Moduli Space:}

In light of the $\mathcal O(\lambda \Lambda_{3N})$ masses for the mixed mesons, we can simplify the constraint equation \eqref{eq:unitW} by discarding the contributions from the heavy degrees of freedom. Expanding $M$ into its nine components $M_{ab}$, and setting $M_{ab} \rightarrow \ev{M_{ab}} = 0$ for $a \neq b$, the nonzero part of the determinant is
\begin{align}
\det M \longrightarrow \det M_{11} \det M_{22} \det M_{33} .
\end{align}
With this substitution, the moduli space in the limit of weakly coupled $SU(N)$ is given by
\begin{align}
(\det M_{11}) (\det M_{22}) (\det M_{33}) - B \barB = \Lambda_{3N}^b,
\label{eq:M123BB}
\end{align}
for $b = 2N_c = 6N$.

\subsection{Second Stage of Confinement} \label{sec:triangle-IR}

\begin{figure}
\centering
\includegraphics[width=0.53\textwidth]{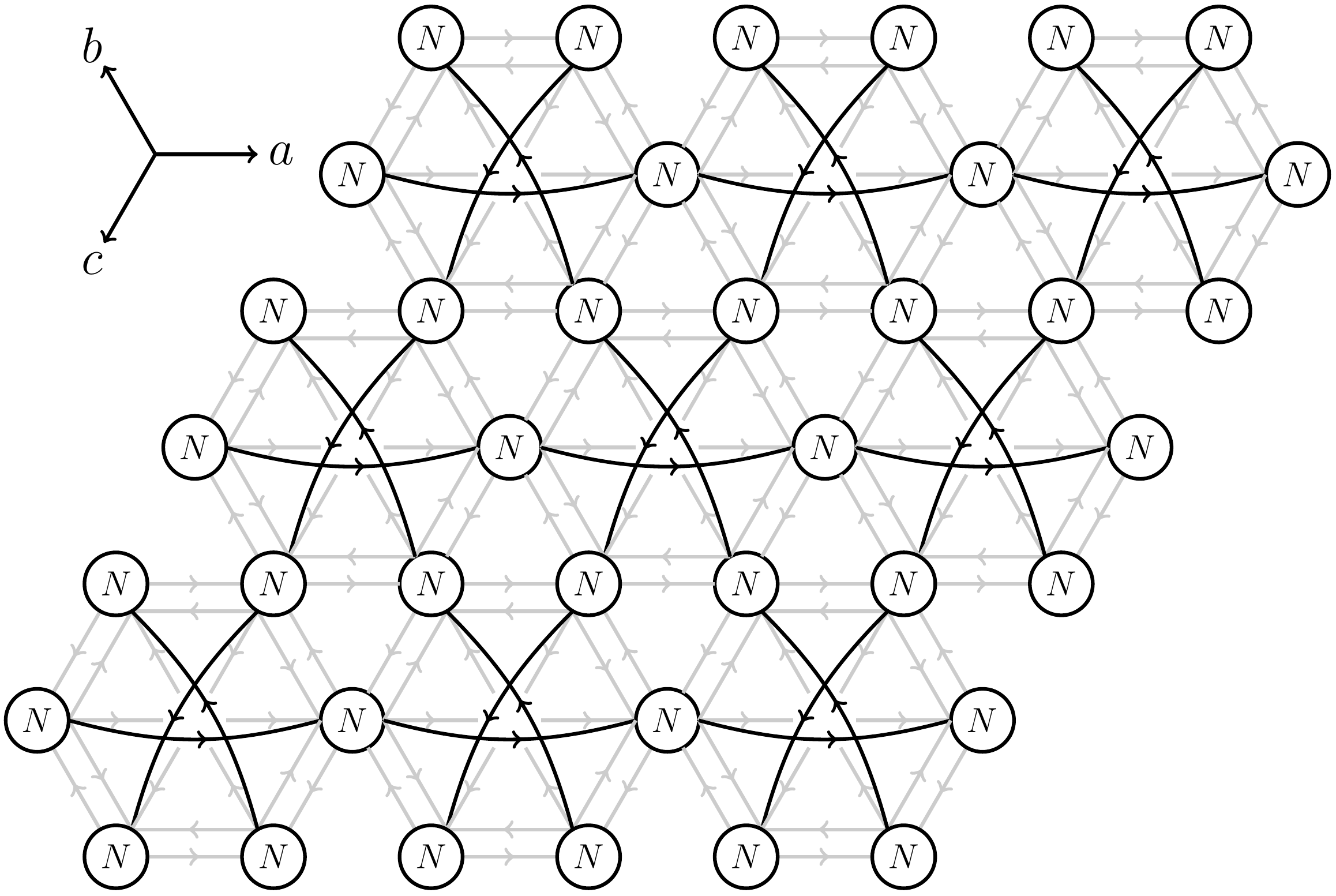}
\hspace{-2cm}
\includegraphics[width=0.53\textwidth]{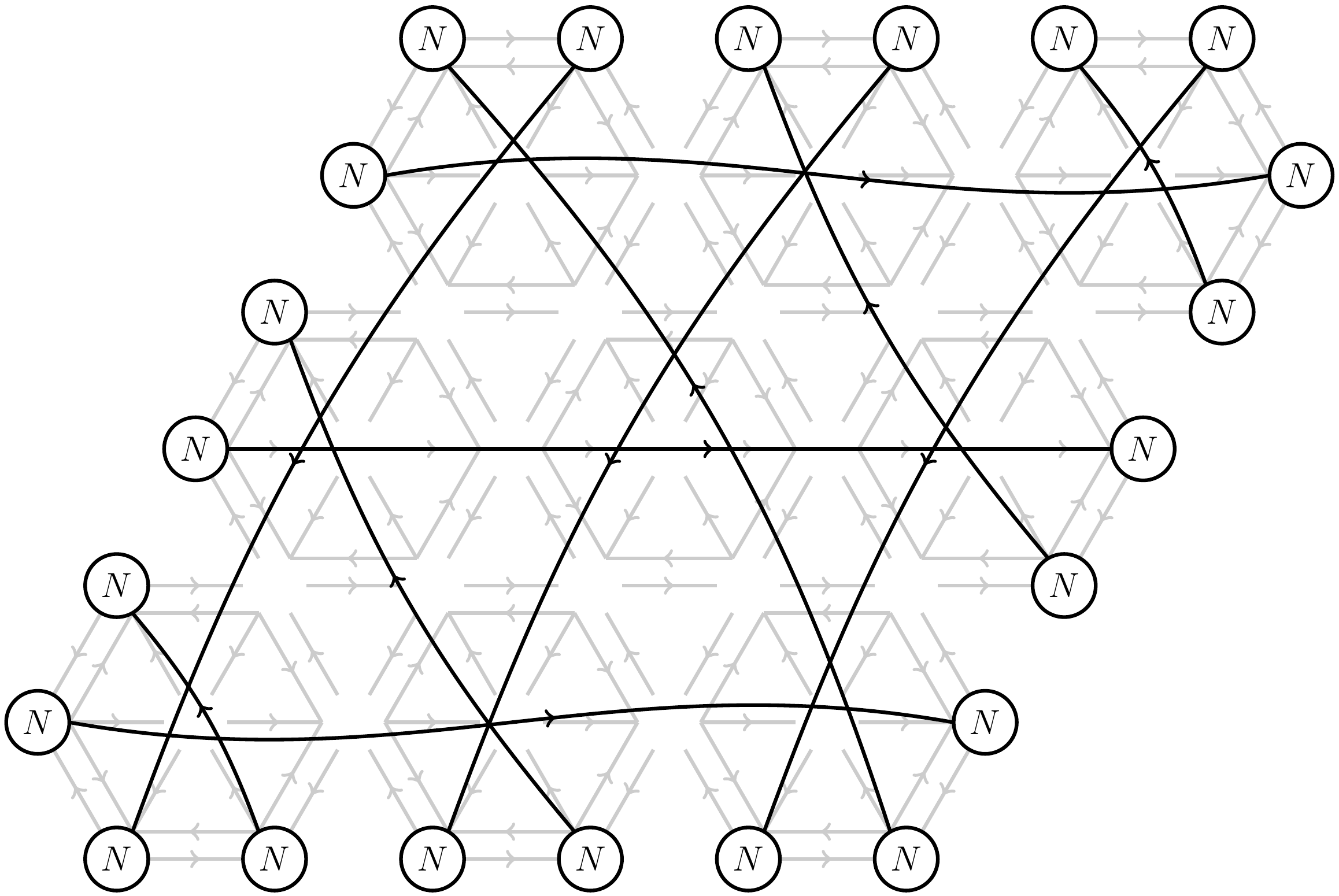}
\caption{ After integrating out the quarks and mesons which have vectorlike masses of $\mathcal O(\lambda \Lambda_{3N})$, the remaining light degrees of freedom are shown in the moose diagrams above. \textbf{Left:} At the intermediate scales $\Lambda_N \ll \mu < \lambda \Lambda_{3N}$, each $SU(N)$ gauge group in the bulk of the lattice is connected by bifundamental mesons to the $SU(N)$ nodes found on the far sides of the two neighboring unit cells. 
\textbf{Right:} In the far infrared of the theory, $\mu \ll \Lambda_N$, the strongly coupled $SU(N)$ product gauge theory is replaced by its Seiberg dual. The degrees of freedom include ``baryon'' operators, $(M_{a}^N)$, and extended ``meson'' operators, shown here as Wilson lines stretching between opposing boundaries of the finite lattice, oriented in the $\varphi = 0^\circ$, $120^\circ$, and $240^\circ$ directions, labelled $a$, $b$, $c$ respectively.
}
\label{fig:triangle2}
\end{figure}

Figure~\ref{fig:triangle2} shows the state of the theory where we left off in Section~\ref{sec:vectorlikeMass}: the six vectorlike pairs of bifundamental quarks and mesons $q_a M_{bc}$ at each unit cell have been integrated out, leaving behind three massless  mesons $M_{aa} = (Q_a \barQ_a)$, which transform as bifundamentals of the $SU(N) \times SU(N)$ nodes on opposite corners of the unit cell.
The three mesons are oriented along the $\varphi = 0^\circ, 120^\circ, 240^\circ$ lattice directions, which largely decouple from each other.
Within the bulk of the lattice, there are no light degrees of freedom that couple mesons in the $\varphi$ direction to any of the $\varphi \pm 120^\circ$ mesons. Each $SU(N)$ gauge group couples exclusively to a single pair of mesons, and there are no couplings between parallel sets of meson operators. 
All of the Wilson lines that changed direction within the lattice incorporated the heavy degrees of freedom $q$ or $(Q_a \barQ_{b\neq a})$.

Ignoring the lattice boundaries, the $SU(N)$ groups and their charged matter fields can be organized into several copies of the $SU(N)^k$ linear moose theory shown in Figure~\ref{fig:mooseChang}.
At this stage of the analysis we may restrict our attention to a single string of unit cells, as follows:
\begin{align}
\begin{aligned}
\includegraphics[width=0.8\textwidth]{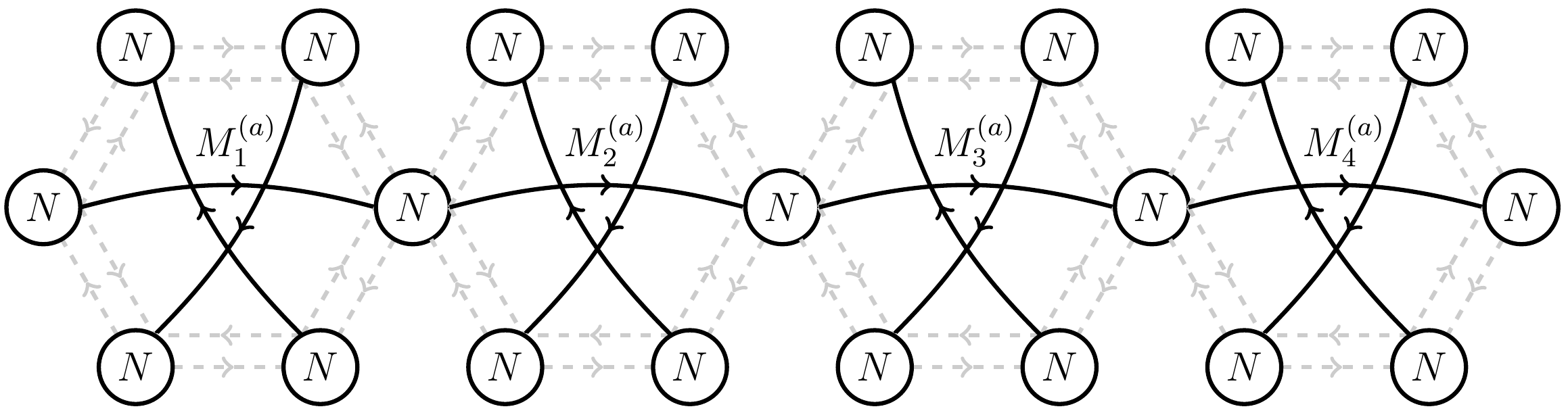} 
\end{aligned}
\label{eq:figureUnits}
\end{align}
Each unit cell has three meson types, $M^{(\gamma)} = M_{\gamma\gamma} = (Q_\gamma \barQ_\gamma)$ for $\gamma = a,b,c$, oriented respectively in the $\varphi = 0^\circ, 120^\circ, 240^\circ$ directions. 
Given $m=4$ unit cells, there are $m-1$ gauged $SU(N)$ groups.

This theory is described in detail by Chang and Georgi~\cite{Chang:2002qt}, and summarized in Section~\ref{sec:reviewProduct}, so there is relatively little work left to do here.
The mesons $M^{(a)}$ charged under the strongly coupled $SU(N)$ gauge groups form gauge-invariant operators,
\begin{align}
\mathcal M_a = (M_1^{(a)} M_2^{(a)} \ldots M_k^{(a)} ) ,
&&
B_1^{(a)} = \det M_1^{(a)}, 
&&
B_2^{(a)} = \det M_2^{(a)}, 
&&
\ldots
&&
B_m^{(a)} &= \det M_m^{(a)}, 
\label{eq:B1k}
\end{align}
which satisfy a constraint equation
\begin{align}
\det \mathcal M_a &= \prod_{j=1}^m B_j^{(a)} + \sum_{\substack{\text{all neighbor} \\ \text{contractions}}} \! \! \!  \left\{ B_1^{(a)} \ldots B^{(a)}_{j-1} B^{(a)}_{j} \ldots B^{(a)}_m \right\}.
\label{eq:modulispacechang}
\end{align}
Following Ref.~\cite{Chang:2002qt}, the term ``neighbor contractions'' refers to the replacement of $(B_{j-1} B_j)$ by a factor of the holomorphic scale $-\Lambda_N^b$ associated with the $SU(N)$ node that connects the $M^{(a)}_{j-1}$ and $M^{(a)}_{j}$ bifundamentals.
The superpotential coupling constants $\lambda_i$ appear indirectly in \eqref{eq:modulispacechang}: as established in Section~\ref{sec:vectorlikeMass}, the holomorphic scales $\Lambda_N$ depend on both $\lambda_i$ and $g_{N_c}(M_\star)$.

Note that none of these mesons $M_i^{(a)}$ is charged under ``the'' $U(1)_B$ symmetry; but, it should be possible to identify some other $B'$ symmetry under which the $M_i$ mesons have $\pm 1$ charges, as in the $SU(N)^k$ model of Ref.~\cite{Chang:2002qt}. There are indeed many such $U(1)$ symmetries, each with some nonzero $SU(N)^2 U(1)$ anomaly coefficients with the global $SU(N)$s of the top and bottom edges of \eqref{eq:figureUnits}. A complete discussion of this ``meson's baryon number'' is postponed to Section~\ref{sec:globalbrane}.

If the number of unit cells in the row is even, then the product $B_1 B_2 \ldots B_m$ can be fully contracted. In the $m=4$ example of \eqref{eq:figureUnits}, the constraint equation is expanded as
\begin{align}
\det \mathcal M_a &= B^{(a)}_1 B^{(a)}_2 B^{(a)}_3 B^{(a)}_4 - \Lambda_1^b B^{(a)}_3 B^{(a)}_4 - B^{(a)}_1 \Lambda_2^b B^{(a)}_4 - B^{(a)}_1 B^{(a)}_2 \Lambda_3^b + \Lambda_1^b \Lambda_3^b,
\end{align}
 where $\Lambda_{j=1,2,3}^b$ refer to the $j$th gauged $SU(N)$ group, which lies on the border between the $j$th and $(j+1)$th unit cells. The constant term $\Lambda_1 \Lambda_3$ removes the origin from the moduli space: either $\det \mathcal M_a$ or some product of baryon operators must acquire an expectation value.
If $k$ is odd, then the product $(B_1 \ldots B_k)$ cannot be fully contracted: instead, every term in the sum contains an odd number of $B_j$ factors. In this case, $\ev{\mathcal M_a} = \ev{B_i^{(a)}} = 0$ is included on the moduli space.

\medskip

A memory of $\Lambda_{3N}$ is preserved in the baryonic operators through the spontaneous symmetry breaking.
Recall from \eqref{eq:unitW} that $SU(3N)$ confinement produced two baryon operators $B$ and $\barB$, and that one linear combination of these remains light.
As demonstrated in Section~\ref{sec:vectorlikeMass}, the $M_{ab}$ degrees of freedom for $a \neq b$ acquire $\mathcal O(\lambda \Lambda_{3N})$ masses, and expectation values $\ev{M_{ab} } = 0$.
This simplifies the $\det( Q \barQ)$ determinant that appears in the constraint equation for $B$ and $\barB$:
after replacing $M_{a \neq b}$ with their expectation values, the only remaining term in $\det(Q \barQ)$ is
\begin{align}
B \barB + \Lambda_{3N}^b = \det (Q \barQ) = \det M_{aa} \det M_{bb} \det M_{cc}  .
\end{align}
In the language of \eqref{eq:B1k}, the $SU(3N)$ constraint equation from the $j$th unit cell becomes
\begin{align}
(B\barB+ \Lambda_{3N}^{6N})_j = B_j^{(a)} B_j^{(b)} B_j^{(c)},
\label{eq:BBbbb}
\end{align}
thus coupling the ``$B- \barB$'' flat direction to the more recently formed baryons $B^{(\gamma)} = \det M_{\gamma\gamma}$, for $\gamma=a,b,c$.

\medskip 

For the $k = 3\times 3$ arrangement shown in Figures~\ref{fig:triangle} and~\ref{fig:triangle2}, the deformed moduli space includes a copy of \eqref{eq:modulispacechang} for each of the 11 strings of meson operators shown in Figure~\ref{fig:triangle2}, combined with one copy of \eqref{eq:BBbbb} for each of the nine $SU(3N)$ unit cells.
Along the $\varphi = 0^\circ$ and $\varphi = 240^\circ$ directions, the constraint equations all follow the $m=3$ form of \eqref{eq:modulispacechang}, with $SU(N) \times SU(N)$ gauge groups.
In the $\varphi = 120^\circ$ direction, on the other hand, the width of the moose lattice is not constant. 
At the lower left and upper right corners of Figure~\ref{fig:triangle2}, there are no gauged $SU(N)$ groups, and only a single meson $M^{(b)} = (\barQ_2 Q_2)$ in each cell, following the labelling convention of Figure~\ref{fig:3Nsequence}.
The off-center $\varphi = 120^\circ$ strings have a single gauged $SU(N)$, so that the effective infrared theory is simply $F=N$ SQCD, while the string passing through the central $SU(3N)$ has two gauged $SU(N)$ groups, like the $\varphi = 0^\circ$ and $\varphi = 240^\circ$ lines.

This is the stage of the calculation where the boundary conditions, i.e.~the shape and topology of the moose lattice, begin to have significant effects on the low energy theory. In this section we have taken the $SU(N)$ nodes at the boundary of the lattice to be conserved global symmetries. This kind of boundary is the easiest to analyze: unit cells can be added to or deleted from the $k = 3\times 3$ example without any change to the analysis above.
With different boundary conditions come significantly altered infrared behaviors. In one example,  diagonal subgroups of the global $SU(N)_\ell \times SU(N)_r$ groups can be gauged: depending on the resulting topology of the moose lattice, this alteration may prevent the $SU(N)$ gauge groups from confining. We explore these kinds of possibilities in Section~\ref{sec:topology}.

\paragraph{Symmetry-Breaking Superpotentials:}

At the corners where the lattice is only one unit cell wide (in the $\varphi = 120^\circ$ direction), our imposition of $SU(N)_\ell \times SU(N)_r$ global symmetry conservation is important for the consistency of the analysis. These cells admit the gauge invariant mass terms of the form $\barQ_2 Q_2$, e.g.~$W \supset \alpha_{ij} M_\star (\barQ_2 Q_2)_{ij}$. Unless the dimensionless $\alpha_{ij}$ is exponentially small, $\alpha \lesssim \Lambda_{3N}/M_\star$, then these quarks should be integrated out. The theory left behind,  $SU(3N)$ SQCD with $F < N_c$, is not described by the methods of Section~\ref{sec:confine3N}.

Where the moose lattice is wider ($m > 1$), the analogous gauge invariant operators are irrelevant, with mass dimension $2m > 3$.
In the $m=2$ case, the perturbation to the $SU(3N)$ theory is small. After $SU(3N) \times SU(3N)$ confinement, the operator $W \sim \alpha (\barQ^{(1)} Q^{(1)})(\barQ^{(2)} Q^{(2)}) / M_\star$ induces a mass for the mesons $M_1$ and $M_2$, which is of order $\alpha \Lambda_{3N}^{(1)} \Lambda_{3N}^{(2)} / M_\star$. By assumption, both $\Lambda_{3N}^{(i)}$ are small compared to $M_\star$, so $\Lambda_{3N}^2/M_\star \ll \Lambda_{3N}$.
However, these meson masses can in principle be large compared to the scale of  $SU(N)$ confinement, $\Lambda_N$, because $\Lambda_N \ll \Lambda_{3N}$ is also exponentially small.

In order for the $SU(N)$ confinement to proceed as in Section~\ref{sec:triangle-IR} for the $m=2$ wide moose lattice, even in the face of global symmetry violation, the associated meson mass scale $\alpha \Lambda_{3N}^2 / M_\star$ should be small compared to $\Lambda_N$.

\subsection{Conclusion}
It is not unseemly to pause here in a spirit of celebration at the simplicity of the low-energy theory. In the $\mu \sim M_\star$ theory depicted in Figure~\ref{fig:triangle}, we started with 108 matter fields; nine $SU(3N)$ and sixteen $SU(N)$ gauge groups; a superpotential consisting of 62 plaquette operators; and a global symmetry group that includes  22 copies of $SU(N)$ and a similar, as-yet-uncounted number of $U(1)$ symmetries.
In the $\mu < \Lambda_N$ limit shown in Figure~\ref{fig:triangle2}, on the other hand, there are 11 mesons, each transforming as a bifundamental of a global $SU(N) \times SU(N)$, and a collection of light baryon operators associated with the spontaneously broken $U(1)$ symmetries, subject to constraint equations of the form \eqref{eq:modulispacechang} and \eqref{eq:BBbbb}.

\section{Tracking the Global Symmetries} \label{sec:globals}

There are several good reasons why we should keep track of the global symmetries. 
Matching the anomaly coefficients of the global symmetries in the UV and IR limits provides a consistency check for the Seiberg duality, for example.
We may also want to embed the Standard Model within the moose lattice, or to test whether the moose theory descends from some higher dimensional QFT; tracking the global symmetries is important to either effort. 
A number of the (approximate) global symmetries  are spontaneously broken during the various stages of confinement, generating (pseudo-) Nambu-Goldstone bosons and their superpartners. For phenomenological applications these details are important: the approximately massless degrees of freedom may be desirable, e.g.~for QCD axion models, or they may be harmful, if for example their presence can be ruled out by cosmological data.
Some of the global symmetries may be gauged, changing the low-energy degrees of freedom in the theory.

\medskip

For the moose lattice, the global symmetries can be split into two types. The first class consists of ``localized'' $U(1)$ symmetries that are associated with a single unit cell. These symmetries have mixed $SU(N_c)^2 U(1)$ anomalies that are cancelled using only the fields from that unit cell. 
Due to this independence from the neighboring cells and the lattice boundary, we associate these symmetries with the interior ``bulk'' of the moose lattice.

The second type of global symmetry is associated with the lattice boundary. For these symmetries, the gauge anomalies generated by the matter fields on one boundary are cancelled by matter fields on another part of the boundary, flowing through some set of charged matter fields in the lattice bulk that generally span multiple unit cells.

\subsection{Global Symmetries In the Bulk} \label{sec:globalbulk}

A complete accounting of the anomaly-free global symmetries depends on the shape of the moose lattice. However, it is possible to identify some non-$R$ $U(1)$ symmetries that are properties of the unit cell: that is, where the only matter fields with $U(1)$ charges are the 12 bifundamentals shown in Figure~\ref{fig:3Nsequence}. 

A simple counting exercise shows why this should be possible. Starting with 12 $U(1)$ phases (one for each matter field), six of the linear combinations are broken by the $R$-conserving plaquette superpotential; $W$ of \eqref{eq:plaquettes} is invariant under the remaining 6 linear combinations. The mixed $SU(3N)^2 U(1)$ anomaly breaks another linear combination. 
At a generic point within the bulk of the lattice (the central $SU(3N)$ in Figure~\ref{fig:triangle}, for example), each $SU(N)$ node at the boundary of the unit cell is shared with one neighboring cell: so, although there are six $SU(N)$ groups appearing in Figure~\ref{fig:3Nsequence}, \emph{on average} only three linear combinations of $U(1)$ phases are broken by the mixed $SU(N)^2 U(1)$ anomalies.
So, for each unit cell within the bulk of the moose lattice, we can identify
\begin{align}
12 - 6 - 1 - 3 = 2
\label{eq:goodmath}
\end{align}
new non-$R$ $U(1)$ symmetries, which are anomaly-free and not broken by the plaquette superpotential.

\begin{figure}
\centering
\includegraphics[width=0.85\textwidth]{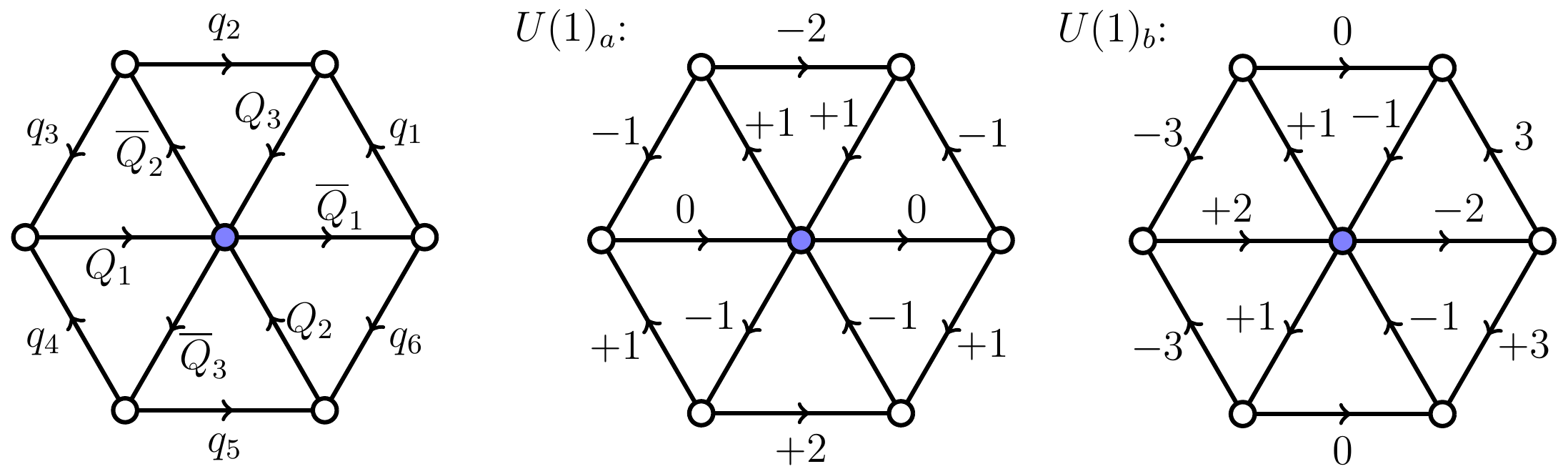}
\caption{In the left diagram we label the matter superfields associated with the unit cell according to Figure~\ref{fig:3Nsequence}. The $SU(3N)$ and $SU(N)$ labels are suppressed, replaced by filled and open circles (respectively) at the nodes of the diagram. 
In the central and right diagrams we list the charges of the superfields under the $U(1)_a$ and $U(1)_b$ symmetries, respectively. 
}
\label{fig:cellUab}
\end{figure}

To make this explicit, consider the following $U(1)_a \times U(1)_b$ charge assignment for the quarks in Figure~\ref{fig:cellUab}:
\begin{align}
U(1)_a:~~~& \hat a(Q_1) = \hat a(\barQ_1) = 0,
&
 \hat a(\barQ_2) &= \hat a(Q_3) = \hat a(q_4) = \hat a (q_6) = +1, \nonumber\\
& \hat a(q_5) = - \hat a(q_2) = + 2,
&
 \hat a (Q_2) &=  \hat a(\barQ_3) = \hat a(q_1) = \hat a(q_3) = -1.
\label{eq:U1a} \\
U(1)_b:~~~& \hat b(Q_1) = - \hat b(\barQ_1) = +2,
&
 \hat b(\barQ_2) &  =  \hat b(\barQ_3) = - \hat b (Q_2) = -\hat b(Q_3) = +1, \nonumber\\
& \hat b(q_5) =  \hat b(q_2) = 0,
&
  \hat b(q_1) &=\hat b(q_6) =  - \hat b (q_3) = -\hat b(q_4)   = +3.
  \label{eq:U1b}
\end{align}
Here we use a new notation, $\hat x(\Phi)$, to concisely report the $U(1)_x$ charge of a superfield $\Phi$, or the charge of its scalar component if the $U(1)$ is an $R$ symmetry.

Each unit cell has its own $U(1)_a \times U(1)_b$ global symmetry, acting only on the quarks associated with that cell. Transformations of this kind can be called ``localized,'' to distinguish them from global symmetries like $U(1)_B$ that act on matter fields from multiple unit cells. This is distinct from (but reminiscent of) a truly local (a.k.a gauged) symmetry.

\medskip
After $SU(3N)$ confinement, \eqref{eq:plaquettes} induces $U(1)_a \times U(1)_b$ conserving  mass terms for the mixed mesons $(Q_i \barQ_{j \neq i})$  and the quarks $q_i$. 
The light degrees of freedom are the mesons $M_{ii} = (Q_i \barQ_i)$ and the baryons $B$ and $\barB$, subject to the constraint  \eqref{eq:M123BB}.
It is easy to see from Eqs.~(\ref{eq:U1a}) and~(\ref{eq:U1b}) that $B$, $\barB$, and $M_{ii}$ are all neutral under $U(1)_a \times U(1)_b$: the only superfields with nontrivial charges are the ones that acquired vectorlike masses, $q_i$ and $M_{i,j\neq i}$.
Once these fields are integrated out, the light degrees of freedom are decoupled from $a$ and $b$. 

It is possible to gauge $U(1)_a \times U(1)_b$ or any of its $U(1)$ subgroups without adding any additional matter fields, thanks to the cancellation of all of the mixed gauge anomalies involving $U(1)_{a,b}$ and the various gauged $SU(N_c)$ groups.
In this case the localized nature of $U(1)_a \times U(1)_b$ in the moose lattice is now \emph{highly} reminiscent of a gauged $U(1)\times U(1)$ symmetry from a 6d spacetime, where the discretization of two compact dimensions causes the local transformation to be realized as a separate $U(1)_a \times U(1)_b$ gauge group for each unit cell.

\paragraph{Plaquette Operators and $R$:}

If $U(1)_a$ and $U(1)_b$ are not gauged, then the superpotential \eqref{eq:plaquettes} may be expanded to include trace operators from the $SU(N) \times SU(N) \times SU(N)$ plaquettes. For any three mutually adjacent unit cells $r, s, t$, these plaquette operators take the form
\begin{align}
W \supset \Tr\!\left( q_1^{(r)} q_3^{(s)} q_5^{(t)} \right) 
&& \text{or} && 
W \supset  \Tr\!\left( q_2^{(r)} q_4^{(s)} q_6^{(t)} \right)  . \label{eq:newplaquettes}
\end{align}
Adding all such operators to the superpotential generally breaks each individual localized $U(1)_{a,b}$ symmetry. 
On average, each unit cell in the bulk of the moose lattice comes with two new $\Tr(q_a q_b q_c)$ plaquette operators, so that the number of $U(1)$ symmetries in the moose lattice no longer scales as the number of unit cells. In terms of \eqref{eq:goodmath}, 
\begin{align}
12 - 6 - 1 - 3 - 2 = 0.
\label{eq:bettermath}
\end{align}
In Section~\ref{sec:globalbrane} we show that once \eqref{eq:newplaquettes} is added to $W$, the number of $U(1)$s scales with the size of the perimeter of the lattice, rather than its area.

A version of $U(1)_a \times U(1)_b$ remains unbroken by  \eqref{eq:newplaquettes}: the symmetric linear combinations of the localized $U(1)$ symmetries from every unit cell in the moose lattice,
$a' = a^{(r)} + a^{(s)} + a^{(t)} + \ldots$ and $b' = b^{(r)} + b^{(s)} + b^{(t)} + \ldots$.
This is easily seen from \eqref{eq:U1a} and \eqref{eq:U1b}, by assigning the same charge to all $q_i^{(r)}$ independently of the unit cell label $r$.

So, the addition of the $(q_1 q_3 q_5)$ and $(q_2 q_4 q_6)$ type plaquettes to $W$ has replaced the $k$ independent localized $U(1)_a \times U(1)_b$ with a single global $U(1)_{a'} \times U(1)_{b'}$. 
This picture further reinforces the notion that the moose lattice reconstructs two extra dimensions: either $U(1)_a \times U(1)_b$ is locally conserved, \eqref{eq:newplaquettes} is forbidden, and there is an independent $U(1)_a \times U(1)_b$ at every unit cell; or, $U(1)_a \times U(1)_b$ is only globally conserved, and there is a single copy of it acting on the whole moose lattice.

Under the $U(1)_R$ identified in \eqref{eq:plaquettes}, all quarks $q_i$ have $R$ charge $+2$. \eqref{eq:newplaquettes} breaks this particular symmetry. More precisely, the $SU(N)^3$ plaquette operators have nontrivial charges under $U(1)_R \times \prod_r U(1)_a^{(r)} \times U(1)_b^{(r)}$, and there is some linear combination of $R$, $a^{(r)}$ and $b^{(r)}$ under which \eqref{eq:newplaquettes} has charge $+2$.

\subsection{Global Symmetries From the Boundary} \label{sec:globalbrane}

The $U(1)_{a, b}$ type symmetries  are distinct from $U(1)_R$ and  the baryon number $U(1)_B$ that is spontaneously broken in the $\ev{B \barB} \neq 0$ vacuum. Indeed, if we restrict our view to a single unit cell, we find that both $U(1)_B$ and $U(1)_R$ have nonzero mixed $SU(N)^2 U(1)$ anomaly coefficients. This means that the anomaly-free versions of $U(1)_B$ and $U(1)_R$ on the moose lattice must involve matter fields from multiple unit cells.
These are the simplest examples of global symmetries associated with the boundaries of the lattice: with the correct charge assignment for a conserved $U(1)_B$, all of the mixed $SU(N)^2 U(1)_B$ anomalies cancel for the gauged $SU(N)$ groups, but not for the $SU(N)_{\ell, r}$ on the boundary of the moose lattice.

\begin{figure}
\centering
\includegraphics[width=0.85\textwidth]{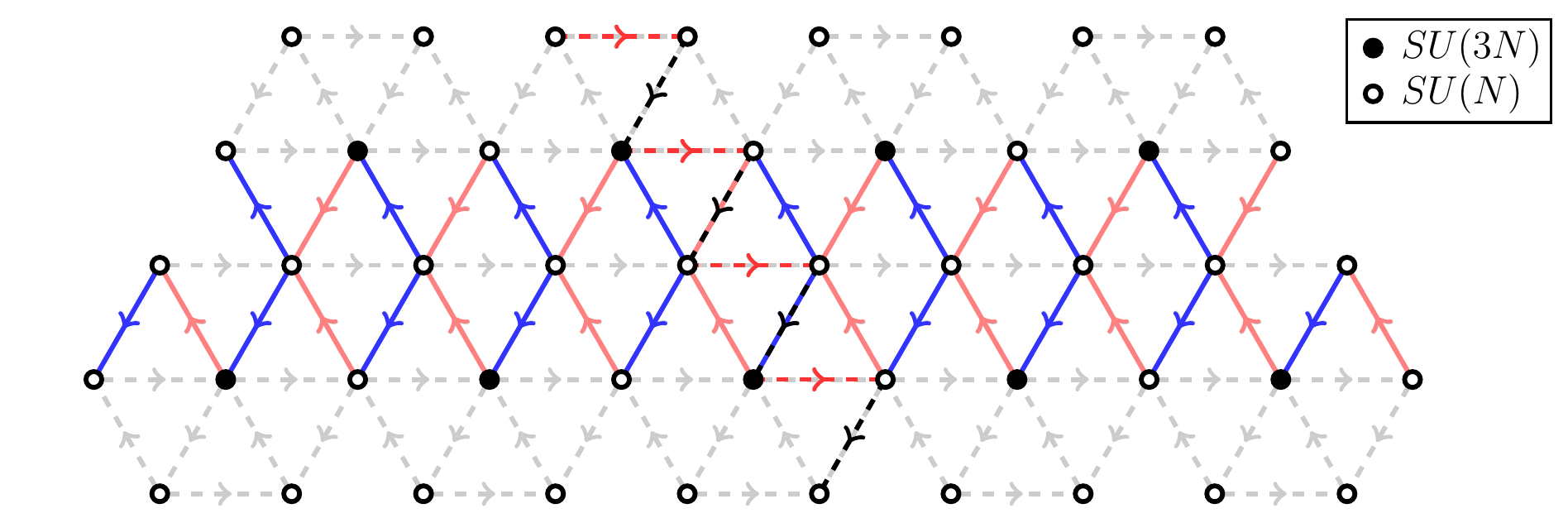}
\caption{An illustration of the quark charge assignments under the boundary-associated conserved global symmetries, using the chevron basis referred to in the text. In the main example the quark charges are indicated by the line color, blue and red for $\pm 1$ charges. Quarks not charged under this $U(1)$ are drawn with faint gray dashed lines. Solid and open circles at the nodes represent $SU(3N)$ and $SU(N)$ groups. Every plaquette operator is neutral under this $U(1)$, and all of the $SU(N_c)^2 U(1)$ mixed anomalies cancel for the gauge groups. 
As a counterexample, we also show an anomalous $U(1)_A$ as the zigzag  line along the $\varphi = 120^\circ$ direction, with black and red dashed lines for $\pm1$ charges. This $U(1)_A$ is unbroken by $W$ and the $SU(3N)^2 U(1)_A$ gauge anomaly; but, some of the $SU(N)^2 U(1)_A$ anomaly coefficients are nonzero, so this $U(1)_A$ is  broken explicitly by $\Lambda_N$ scale effects.
}
\label{fig:globalexample}
\end{figure}

In this section we describe a systematic method for enumerating the other boundary-associated global $U(1)$ symmetries. 
Unlike Section~\ref{sec:globalbulk}, we restrict our attention to $U(1)$ rotations that leave \eqref{eq:newplaquettes} as well as \eqref{eq:plaquettes} invariant.
A relatively simple charge basis can be constructed from strings of adjacent quarks with alternating $\pm1$ charges, traversing the bulk of the moose lattice in a zigzag pattern. All plaquette operators are neutral under such a charge assignment. 
For the mixed $SU(N_c)^2 U(1)$ anomalies to cancel for all gauged $SU(N_c)$, this type of global symmetry needs to involve adjacent rows of unit cells, as shown in Figure~\ref{fig:globalexample}. The result is a chevron-like charge assignment. 
The only nonzero $SU(N)^2 U(1)$ anomaly coefficients involve the global symmetries at the boundaries of the lattice, pairing two $SU(N)_\ell$ groups with their $SU(N)_r$ counterparts on the opposite edge.

There are $U(1)$ symmetries of this type along each of the $\varphi = 0, 120^\circ, 240^\circ$ directions. In the (arbitrarily chosen) moose lattice of Figure~\ref{fig:globalexample}, there are $3 + 5 + 5 = 13$ distinct chevron charge assignments. This accounting includes the single-row zigzag versions that run along the edges of the moose lattice: for example, the dashed $\varphi = 120^\circ$ line in Figure~\ref{fig:globalexample}, but pushed all the way to the right edge of the lattice.
In the $k = 3\times 3$ model of Figure~\ref{fig:triangle}, there are $4+4+6 = 14$ of these $U(1)$ symmetries. More generally, for an arbitrarily shaped moose, the number of chevron global symmetries parallel to $\varphi$ depends on the number of distinct rows of unit cells, i.e.:
\begin{align}
N_\text{chev.} = \sum_{\varphi = 0, 120^\circ, 240^\circ} (N_\text{rows}(\varphi) + 1).
\end{align}
Because the number of these $U(1)$ symmetries is proportional to the number of rows in the moose lattice, it scales with the length of the perimeter of the lattice, rather than as the area (number of unit cells) of the lattice.

These chevron $U(1)$ symmetries can be thought of as partially localized, in analogy with Section~\ref{sec:globalbulk}. Rather than being confined to a single unit cell, these $U(1)$s act on the quarks within a specific horizontal band (or a band parallel to $\varphi = \pm 120^\circ$) while ignoring the rest of the moose lattice.
Unlike the $U(1)_a \times U(1)_b$ symmetries, these chevron $U(1)$s mix with the $SU(3N)$ baryon number: that is, for any participating unit cell, the operators $B = (Q_1^N Q_2^N Q_3^N)$ and $\barB = (\barQ_1^N \barQ_2^N \barQ_3^N)$ have nonzero charges under the chevron $U(1)$. 
When $\ev{B \barB}$ acquires its $\mathcal O(\Lambda_{3N})$ expectation value, it breaks the $U(1)_B$ under which $B$ and $\barQ$ are charged. In the infrared limit of the theory, well below $\Lambda_{3N}$, the $U(1)_B$ is not manifest as a global symmetry of the composite degrees of freedom, but shows up as a collection of Nambu-Goldstone superfields, one for each gauged $SU(3N)$.

There is a global $U(1)_B$ with a particularly simple charge assignment, uniform with respect to the various unit cells:
\begin{align}
\includegraphics[width=0.85\textwidth]{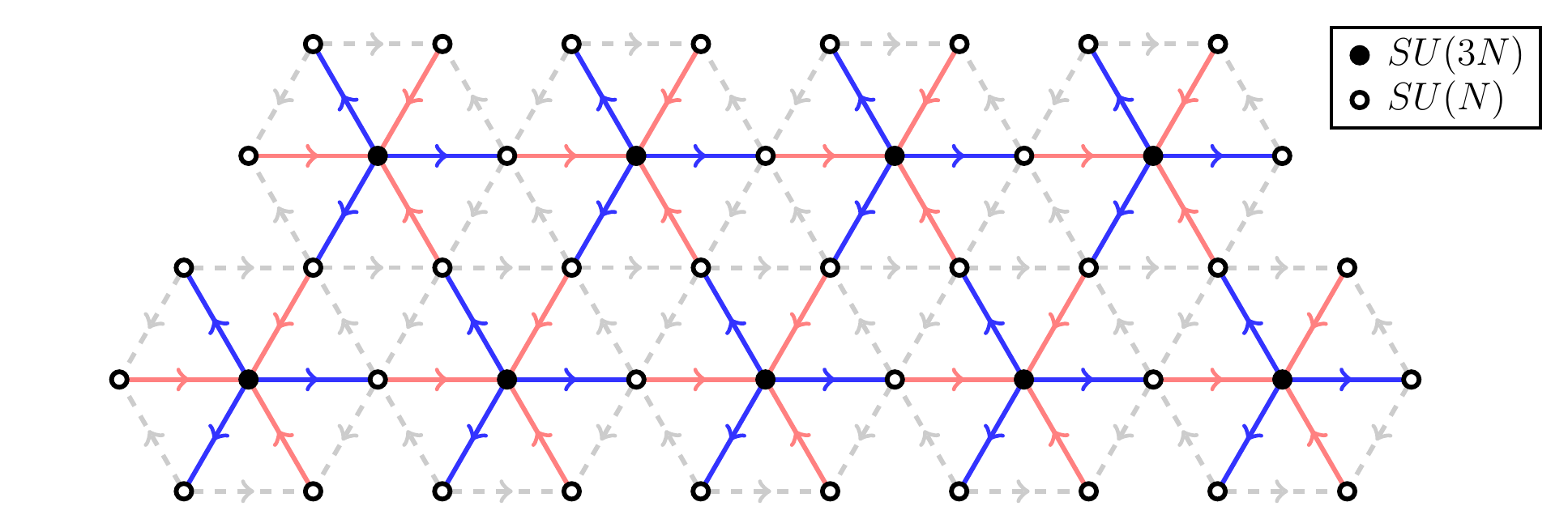},
\label{eq:globaryon}
\end{align}
where as in Figure~\ref{fig:globalexample} we use blue and red to indicate $\pm 1$ charges for the quarks. All of the $q_i$ are neutral. The plaquette operators are manifestly invariant under $U(1)_B$, and all the $SU(N_c)^2 U(1)_B$ gauge anomalies cancel.

In addition to $B$, there are three related conserved $U(1)$ symmetries, obtained by shifting the $U(1)_B$ charge assignment in the $\varphi = 0, 120^\circ, 240^\circ$ directions. Rather than converging at the $SU(3N)$ nodes, these other $B'$ symmetries follow a triangular lattice with vertices at alternating $SU(N)$ nodes.

\medskip

After $SU(3N)$ confinement, $U(1)_B$ and most linear combinations of chevron $U(1)$s are spontaneously broken. 
An alternating combination of chevron $U(1)$s is left unbroken: given the full set of chevron symmetries $U(1)_{j}$, $j = 0, 1, 2, \ldots, N_\text{rows}$, for a fixed value of $\varphi$, the combination $(0) - (1) + (2) - (3) + \ldots$ leaves all of the $B$ and $\barB$ operators invariant.
To give an explicit example, the charge assignments under the unbroken $\varphi = 0$ chevron $U(1)$ are:
\begin{align}
\includegraphics[width=0.7\textwidth]{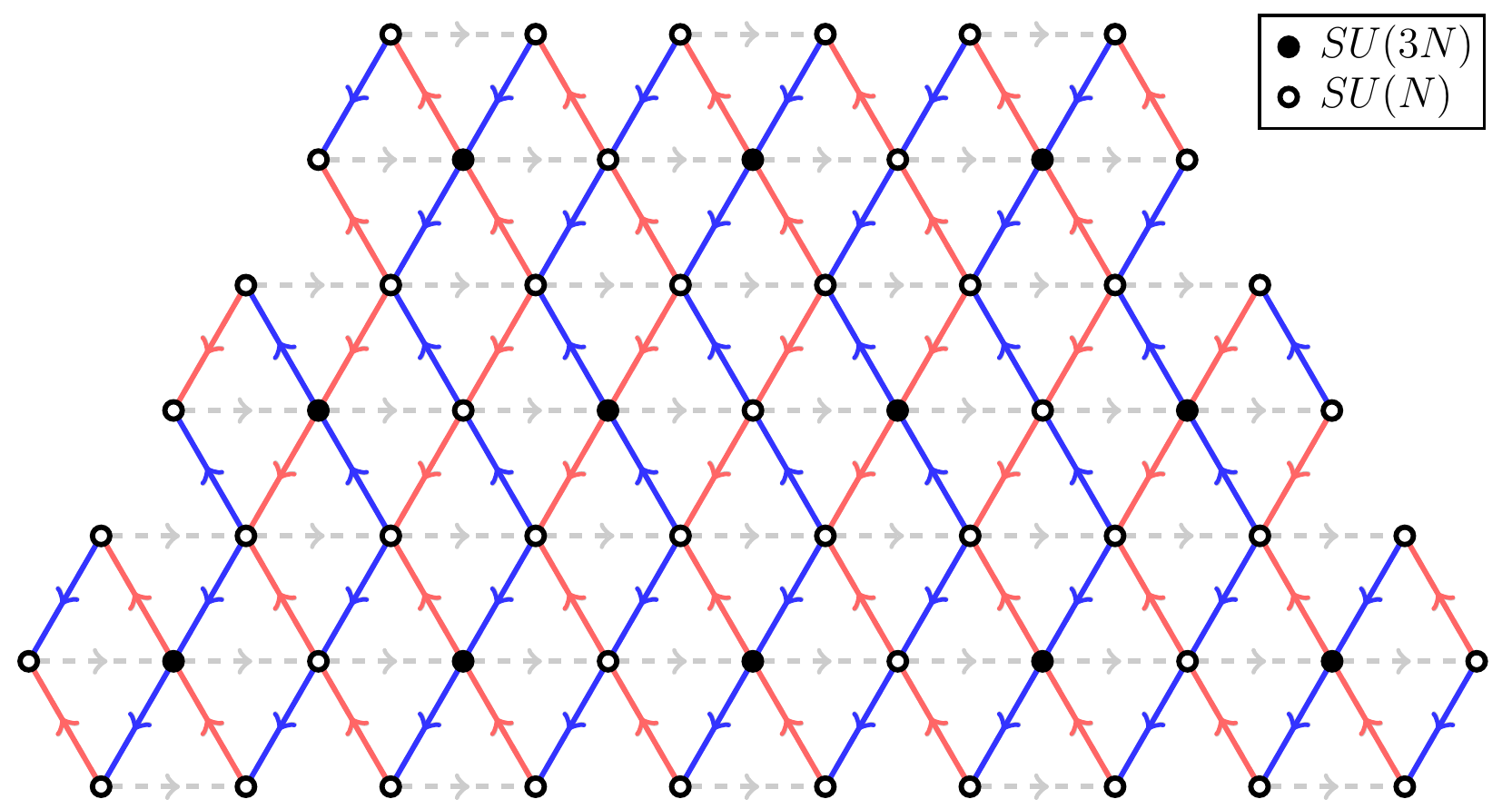}.
\label{eq:glochev}
\end{align}
In this example we added a third row of cells to illustrate the alternating pattern. 
Note that the charge assignment is symmetric with respect to shifts in the $\varphi$ direction, and that the only $SU(N)$ sites with nonzero $SU(N)^2 U(1)$ anomaly coefficients are those on the top and bottom edges, parallel to $\varphi$.
This $U(1)$ and its $\varphi = \pm 120^\circ$ analogues are particularly relevant to the phase of the theory outlined in Section~\ref{sec:triangle-IR}: although $B$ and $\barB$ are neutral under this $U(1)$ and its $\varphi = \pm 120^\circ$ analogues,
the meson operators $M_{2,3}^{(j)}$ have nontrivial charges of $\pm 2$.

In the last stage of confinement, where the $SU(N)$ gauge groups become strongly coupled, the chevron $U(1)$ groups may or may not be spontaneously broken. Following \eqref{eq:figureUnits}, the effective $SU(N)_\ell \times SU(N)_r \times U(1)_{B'}$ symmetry of a particular row in the moose lattice could be broken to the diagonal $SU(N)_d \times U(1)_{B'}$ by a vev for the meson line operator $(M_1^{(a)} M_2^{(a)} \ldots M_{k}^{(a)})$; or, a baryon vev $\ev{(M_{i}^{(a)} )^N}$ could spontaneously break $U(1)_{B'}$ to a discrete Abelian subgroup. The $B'$ in this scenario is a linear combination of the chevron $U(1)$ symmetries.
Alternatively, if the number of unit cells in each row is consistently an even number, then the deformed moduli space \eqref{eq:modulispacechang} may yet include the origin, $\ev{\mathcal M} = \ev{B^{(a)}} = \ldots = 0$, and the global symmetry group need not be broken.
Most moose lattices will include some odd-length rows in one direction or another, forcing some spontaneous symmetry breaking, but an all-even moose lattice can be constructed from doubly periodic boundary conditions. As we discuss in Section~\ref{sec:topology}, periodic boundary conditions lead to Coulomb phases rather than confinement.

\subsection{Sources of Explicit Symmetry Breaking}

Exactly conserved $U(1)$ symmetries can pose phenomenological problems, especially in contexts (such as spontaneous symmetry breaking) where they correspond to exactly massless particles. 
For this reason alone, we should parameterize the sources of $U(1)$ breaking which could introduce mass terms for otherwise massless Nambu-Goldstone bosons.

In this context, despite the fact that $\Lambda_N$ is much smaller than $\Lambda_{3N}$, it is still (by definition) large compared to the ultimate IR limit of the theory, $\mu \ll \Lambda_N$. In particular, we have assumed that SUSY is preserved during the $SU(N)$ confinements. If SUSY is broken (which it must be at some point, if the theory is to describe anything resembling our universe) then the scale of soft SUSY breaking should satisfy $m_s \ll \Lambda_N$.

Though we can be glad to ignore any degrees of freedom with $\mathcal O(\Lambda_N)$ masses, this does \emph{not} necessarily extend to the approximate global symmetries that are explicitly broken by their mixed anomalies with $SU(N)$ gauge groups. 
Many of these anomalous $U(1)_A$ are broken spontaneously by the $\ev{B \barB}$ vevs from $SU(3N)$ confinement at a much higher scale, $\Lambda_{3N}$, where $U(1)_A$ can be treated as approximately conserved. 
The ratio between $\Lambda_N$ and $\Lambda_{3N}$ suppresses the particle masses introduced by the triangle anomaly: indeed, this is exactly the setup of a typical axion model~\cite{Peccei:1977ur, Peccei:1977hh,Wilczek:1977pj,Weinberg:1977ma,Kim:1979if,Shifman:1979if,Dine:1981rt,Zhitnitsky:1980tq}, where the axion mass $m_a$ is related to $f_A$,  the scale of spontaneous $U(1)_A$ breaking, and $\Lambda_\text{QCD}$, the source of explicit $U(1)_A$ violation, via
\begin{align}
m_a^2 \sim \frac{\Lambda_\text{QCD}^4}{f_A^2} \equiv \frac{m_\pi^2 f_\pi^2}{f_A^2}.
\end{align}
In our examples $f_A$ is proportional to $\Lambda_{3N}$, while $\Lambda_N$ stands in for $\Lambda_\text{QCD}$, i.e.~$m_a \sim \Lambda_N^2/\Lambda_{3N}$. Incidentally, the fact that all of these mass scales are dynamically generated makes the moose lattice an ideal playground for model building. For example, Refs.~\cite{Lillard:2017cwx,Lillard:2018fdt} use similar features in simpler SUSY product gauge theories to construct composite QCD axion models.

Global symmetries can also be broken by superpotential operators. We have seen this when adding  plaquette operators to the superpotential; for example, the localized $U(1)_a \times U(1)_b$ global symmetries of \eqref{eq:plaquettes} are broken by the addition of \eqref{eq:newplaquettes} to $W$.
Other possible gauge-invariant operators include the moose lattice analogue of Wilson lines,
\begin{align}
W_\text{line} \supset \frac{\alpha_{ij} }{ M_p^{2k-3} }(Q_1 \barQ_1 Q_2 \barQ_2 \ldots Q_k \barQ_k)_{ij},
\label{eq:Wline}
\end{align}
where the Wilson line runs through $k$ different unit cells.  
The $j$th unit cell in the product should be adjacent to its $(j\pm1)$th neighbors, though the path through the lattice does not need to be in a straight line. We write \eqref{eq:Wline} in terms of the Planck scale $M_p$, though depending on the context a lower UV scale (e.g.~$M_\star$) may be more appropriate.
Operators of this form are charged under the global $SU(N)_\ell$ and $SU(N)_r$ where the Wilson line begins and ends on the lattice boundary, and the addition of $W_\text{line}$ to the superpotential explicitly breaks the non-Abelian global symmetries. If $k$ is large, however, the effect on the IR theory is generally small. After $SU(3N)$ confinement, the effective operator is suppressed by $k$ factors of $\Lambda_{3N} / M_p$, so that the theory of $SU(N)$ charged mesons $M_{i} = (Q_i \barQ_i)$ takes the form:\footnote{See comment in footnote~\ref{foot:note} about matching IR degrees of freedom and UV gauge invariants.} 
\begin{align}
W_\text{line} \longrightarrow \frac{\alpha_{ij} \prod_{i=1}^{k} \Lambda_{3N}^{(i)}}{ M_p^{k} } \frac{(M_1 M_2 \ldots M_k)_{ij} }{M_p^{k-3}}.
\end{align}
Especially for $k \geq 3$, or $\Lambda_{3N} \lll M_p$, this explicit breaking of the non-Abelian global symmetry may be quite small.
In the furthest infrared limit, the straight line operators $\mathcal M_{ij} \sim (M_1 M_2 \ldots M_k)_{ij}$ may acquire an expectation value, subject to a constraint equation of the form \eqref{eq:modulispacechang}, which includes vacua with $\det \mathcal M \neq 0$. At an arbitrary point on the moduli space, the $SU(N)_\ell \times SU(N)_r$ global symmetry is entirely (spontaneously) broken, and operators of the form \eqref{eq:Wline} can generate masses for the pNGBs.

The other type of symmetry violation comes from baryonic superpotential operators, e.g.
\begin{align}
W_\text{bary} \supset \frac{\beta}{M_p^{3N-3}} (Q_1^N Q_2^N Q_3^N) + \frac{\bar\beta}{M_p^{3N-3}} (\barQ_1^N \barQ_2^N \barQ_3^N) + \frac{\gamma_i}{M_p^{N-3}} (q_i^N) + \ldots .
\end{align}
Each unit cell supplies a $B$ and $\barB$ type operator, as well as six $q_i^N = \det q_i$ from the $SU(N) \times SU(N)$ bifundamentals.
Applying the same mapping between canonically normalized degrees of freedom across the $SU(3N)$ transition, 
\begin{align}
W_\text{bary} \supset  \left(\frac{ \Lambda_{3N}}{M_p} \right)^{3N-3} \left( \beta \Lambda_{3N}^2 \mathcal B + \bar\beta \Lambda_{3N}^2 \overline{\mathcal B} \right)  + \sum_{i=1}^6 \frac{\gamma_i}{M_p^{N-3}} (q_i^N)  .
\label{eq:Wbaryb}
\end{align}
As detailed in Section~\ref{sec:vectorlikeMass}, the quarks $q_i$ acquire $\mathcal O(\lambda \Lambda_{3N})$ masses and are integrated out of the theory. 
The baryons are light degrees of freedom, subject to the constraint equation $\det  M -  B \overline{B} = \Lambda_{3N}^{6N}$. Without \eqref{eq:Wbaryb}, the $\ev{B \barB} \neq 0$ vacuum would create massless Nambu-Goldstone bosons for the various $U(1)_B$ type global symmetries, but the $\beta$ and $\bar\beta$ terms introduce masses for these fields.

\paragraph{Gauged Abelian Symmetries:}

If any of the spontaneously broken $U(1)$s are  locally conserved, then it is the (super) Higgs mechanism that saves us from phenomenologically unfriendly massless fields. This would be the case if a subgroup of $U(1)_a \times U(1)_b$ is gauged, for example. Although the baryons $B$ and $\barB$ are neutral under $a$ and $b$, the mesons $M_i \sim (Q_1 \barQ_1)$ are not. After $SU(N)$ confinement, the baryon operator $B^{(i)} \sim (M_i)^N$ can acquire a vev, spontaneously breaking $U(1)_a \times U(1)_b$ to a subgroup and endowing the gauge boson with a mass.

In principle any of the conserved $U(1)$ symmetries can be gauged, as long as the anomaly cancellation conditions are satisfied. Take $U(1)_B$ of \eqref{eq:globaryon} for example. With an integer number of unit cells included in the moose lattice, the $U(1)_B^3$ and mixed gauge anomalies all cancel. From the perspective of the 4d theory, gauging $U(1)_B$ is straightforward.
Unlike $U(1)_a \times U(1)_b$, however, there is no local $U(1)_B$ conservation in the cells of the moose lattice: a $U(1)_B$ rotation acting  only on the $Q_i$ and $\barQ_i$ of a single unit cell has nonzero $SU(N)^2 U(1)$ anomaly coefficients.
For this 4d product gauge theory to descend from a higher dimensional theory with a locally conserved $U(1)_B$,  the local $U(1)$ transformations acting on the compact space need to be  spontaneously broken as part of the 6d$\,\rightarrow\,$4d discretization scheme.

\section{Boundary Conditions} \label{sec:topology}

For most of Section~\ref{sec:triangular}, the shape of the moose lattice mattered very little to the analysis. 
In Section~\ref{sec:triangle-IR}, where the $SU(N)$ groups become strongly coupled, we did make the assumption that the moose lattice was not periodic.  
Our discussion of global symmetries in Section~\ref{sec:globals} is more directly dependent on the shape of the lattice boundary, but our methods remain generic enough that we could switch between the $k=3\times 3$ example and a $k=5+4$ or  $k = 5+4+3$ version with impunity, as in~(\ref{eq:globaryon}) and~(\ref{eq:glochev}).
Every example was constructed in the same basic way: by connecting an integer number of unit cells, gauging the $SU(N)$ nodes that connect adjacent unit cells, while leaving the $SU(N)$ nodes on the boundary of the moose lattice as global symmetries.

By altering the boundary of the moose lattice, we can construct several new types of gauge theories from this template, often by finding ways to gauge the $SU(N)$ boundary nodes.
For example, we can add new matter fields charged under a single $SU(N)$ so as to cancel its cubic $SU(N)^3$ anomaly; we can add bifundamentals of $SU(N)_\ell \times SU(N)_r$ to gauge a pair of boundary nodes; or, we could gauge the diagonal subgroup $SU(N)_d \subset SU(N)_\ell \times SU(N)_r$ of two boundary nodes. None of these perturbations have much effect on $SU(3N)$ confinement, but can considerably alter the behavior of the $SU(N)^m$ gauge theory.

\subsection{Reflective Boundaries} \label{sec:top-reflect}

The reader has probably noticed that the moose lattices depicted in Figures~\ref{fig:triangle} and~\ref{fig:triangle2} have notches missing from the edges. Considering the matter further, the reader may have decided that the missing $SU(N)_i \times SU(N)_j$ bifundamentals have a minimal impact on the theory after all: if $SU(N)_{i,j}$ are global symmetries, each of these edge fields is just $N^2$ chiral fields with no gauge charges. Naturally, if the edge fields are  charged under some gauged $U(1)$, or if they are coupled to the other quarks in the superpotential, they are not entirely irrelevant, but they are not especially interesting either. 

\begin{figure}
\centering
\includegraphics[width=0.5\textwidth]{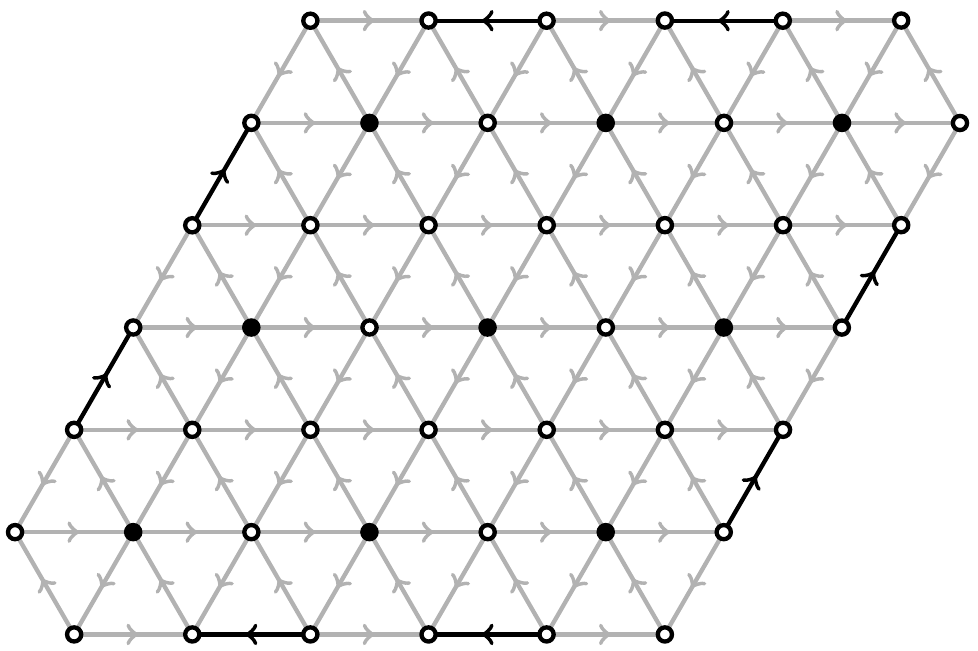}
\hspace{-1cm}
\includegraphics[width=0.5\textwidth]{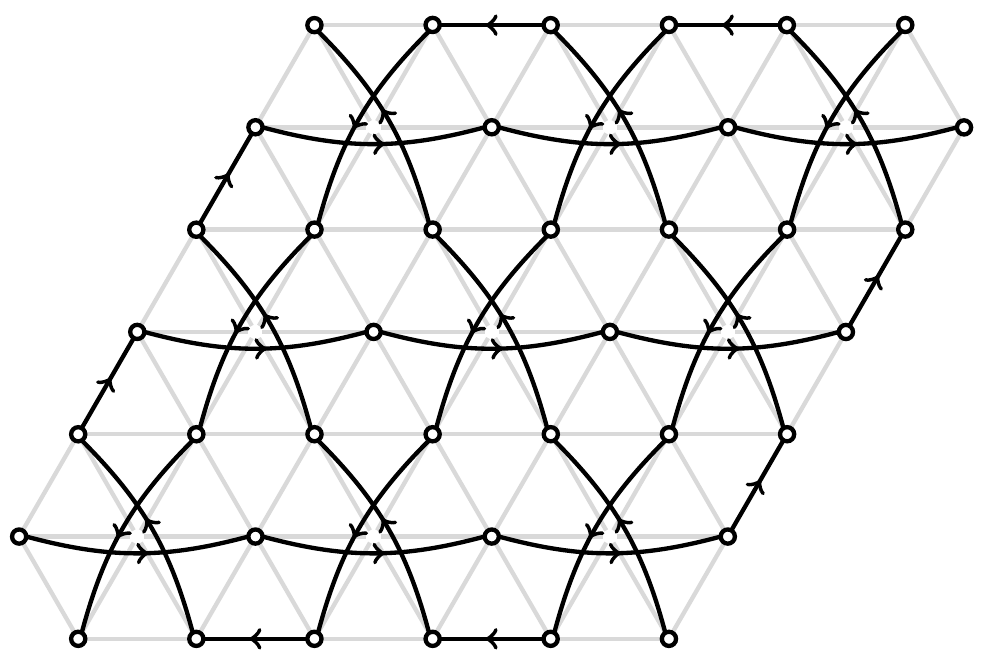}
\caption{Reflective boundary conditions.
\textbf{Left:} The $k=3\times3$ moose lattice from Figure~\ref{fig:triangle} is modified by the addition of eight ``wrong-way''  bifundamental quarks on the boundaries of the lattice. The wrong-way quarks are depicted in black; the matter fields already present in the Section~\ref{sec:triangular} are shown in gray.  Following the concise notation of the figures in Section~\ref{sec:globals}, the $SU(N)$ and $SU(3N)$ sites are represented (respectively) by white and black nodes. The only $SU(N)$ sites with nonzero $SU(N)^3$ anomaly coefficients are those at the corners of the parallelogram.
\textbf{Right:}~After the $SU(3N)$ groups confine, and after the mesons and quarks with vectorlike $\mathcal O(\Lambda_{3N})$ masses are integrated out, the theory includes the $SU(N)$ charged quarks and mesons shown here. 
Thanks to the new edge quarks, the fields can be organized into three disjoint $SU(N)^{m-1}$ gauge theories, two with $m=10$ and one with $m=3$.
}
\label{fig:refl}
\end{figure}

If the addition of edge quarks allows the boundary $SU(N)$ nodes to be gauged, their impact on the theory can be much more interesting. Take for example the modified $k=3\times 3$ theory shown in Figure~\ref{fig:refl}. Here we have added eight bifundamentals to fill in the notches; however, these edge quarks have the opposite $SU(N)_\ell \times SU(N)_r$ charges, i.e.~in the $(\overline{\mathbf{N}}, \mathbf{N})$ representation rather than $(\mathbf{N}, \overline{\mathbf{N}})$, or vice versa.
The arrows on the moose diagram for these edge quarks appear to be pointing in the wrong way: that is, unlike every other matter field in the moose lattice, the arrows of the new edge quark point in the $\varphi' = \pm 60^\circ, 180^\circ$ directions.

From the perspective of the new gauged boundary nodes, there are $3N$ fundamentals (e.g.~from an $SU(3N)$ node) and $3N$ antifundamentals (from the three adjacent $SU(N)$ nodes). This is $F = 3N_c$ SQCD, for which the NSVZ $\beta$ function vanishes, and asymptotic freedom is lost. With our working assumption that $g_N(M_\star) \lesssim \mathcal O(1)$ is perturbatively small, the edge $SU(N)$ gauge couplings begin to run below $\mu = \lambda_i \Lambda_{3N}$, where the vectorlike mass terms for the mesons and $q_i$ quarks become relevant. The right panel of Figure~\ref{fig:refl} shows the theory after these fields have been integrated out. This is the same scale at which the $\beta$ function for the $SU(N)$ nodes in the bulk switches sign. 
If the edge and bulk $SU(N)$ groups have similarly strong couplings at $M_\star$, i.e.~$g_N^{(i)}(M_\star) \sim g_N^{(j)}(M_\star)$, then the running of the different gauge couplings will tend to make the bulk $SU(N)$ groups more weakly coupled than the edge nodes at $\mu \sim  \Lambda_{3N}$, so that the edge nodes are the first to confine.

\medskip

Previously, the $\mu < \lambda \Lambda_{3N}$ phase of the $k=3\times 3$ theory involved eleven disjointed strings of $SU(N)$ charged mesons, as depicted in the right panel of Figure~\ref{fig:triangle2}: three sets each in the $\varphi = 0^\circ, 240^\circ$ directions, and five in the $\varphi = 120^\circ$ direction. With the wrong-way edge quarks and gauged $SU(N)$ nodes on the boundaries,  previously decoupled $SU(N)^m$ sets are joined together.
In Figure~\ref{fig:refl} there are now only three sets of gauge invariant Wilson line operators. One is a simple example of the form \eqref{eq:Wline}, running from corner to corner along the $\varphi = 120^\circ$ direction. The other two appear to bounce off of the lattice boundaries: one originates and terminates in the lower left corner, the other begins and ends in the upper right corner.
This is why we refer to the boundary conditions as ``reflective.''

Aside from this detail, the $SU(N)^{m-1}$ groups confine in the manner described in Section~\ref{sec:triangle-IR} and Ref.~\cite{Chang:2002qt}, now with $m =3$ or $m=10$.
In the far infrared limit the degrees of freedom include three composite mesons of the form $\mathcal M \sim \prod_i^m M_i$, with each $\mathcal M$ charged under a different $SU(N)_\ell \times SU(N)_r$.

Although we have added eight edge quarks to the $k=3\times 3$ model, the reflective theory has fewer global $U(1)$ symmetries. There is no gauge invariant plaquette superpotential associated with the edge quarks $\chi_i$, because products of the form $(\chi_i q_a q_b)$ are charged under the boundary $SU(N)$ nodes. However, by gauging two $SU(N)$ nodes for each new edge quark $\chi$, some of the $U(1)$ global symmetries acquire $SU(N)^2 U(1)$ anomalies that cannot be canceled by assigning $\chi$ a charge under $U(1)$.

\begin{figure}
\centering
\includegraphics[width=0.53\textwidth]{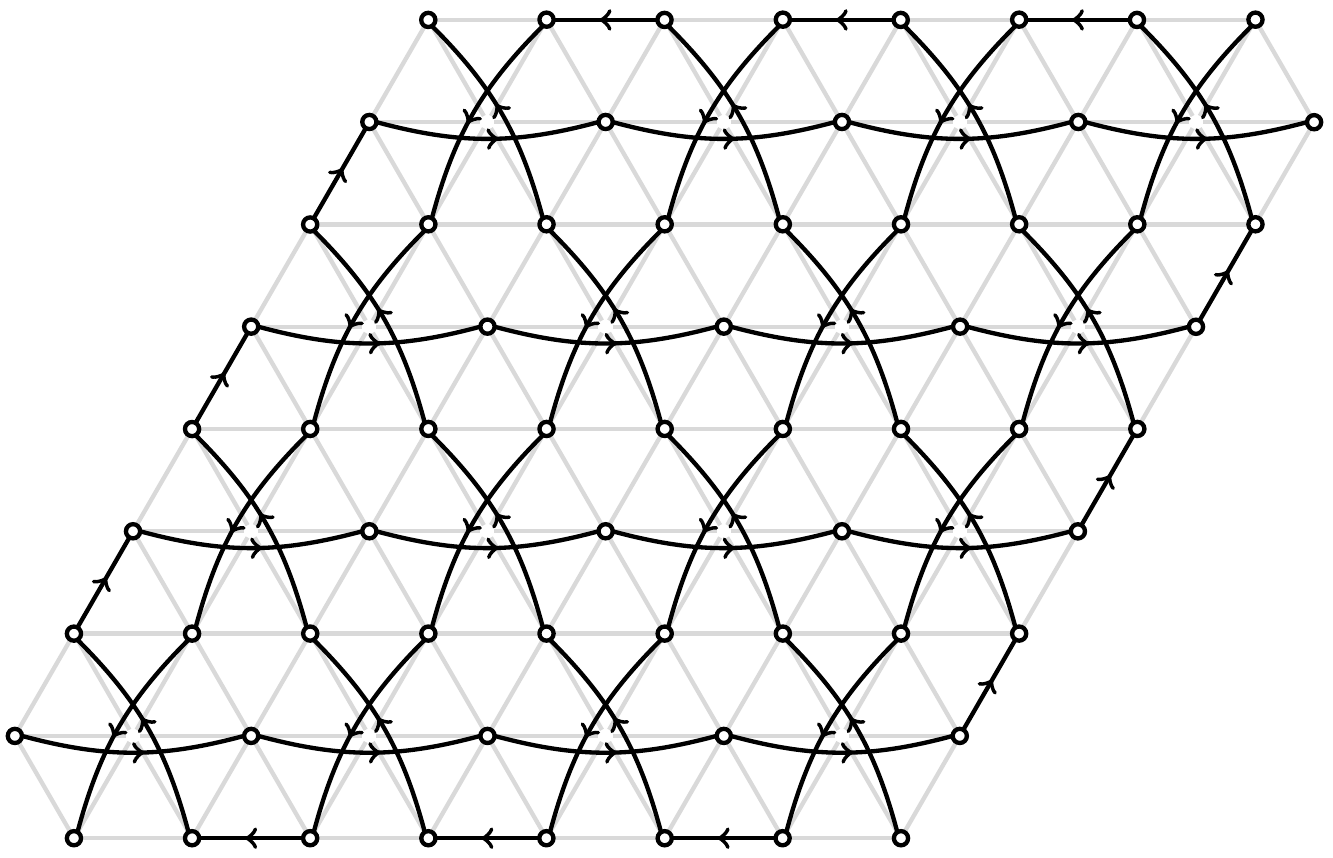}
\hspace{-1.5cm}
\includegraphics[width=0.53\textwidth]{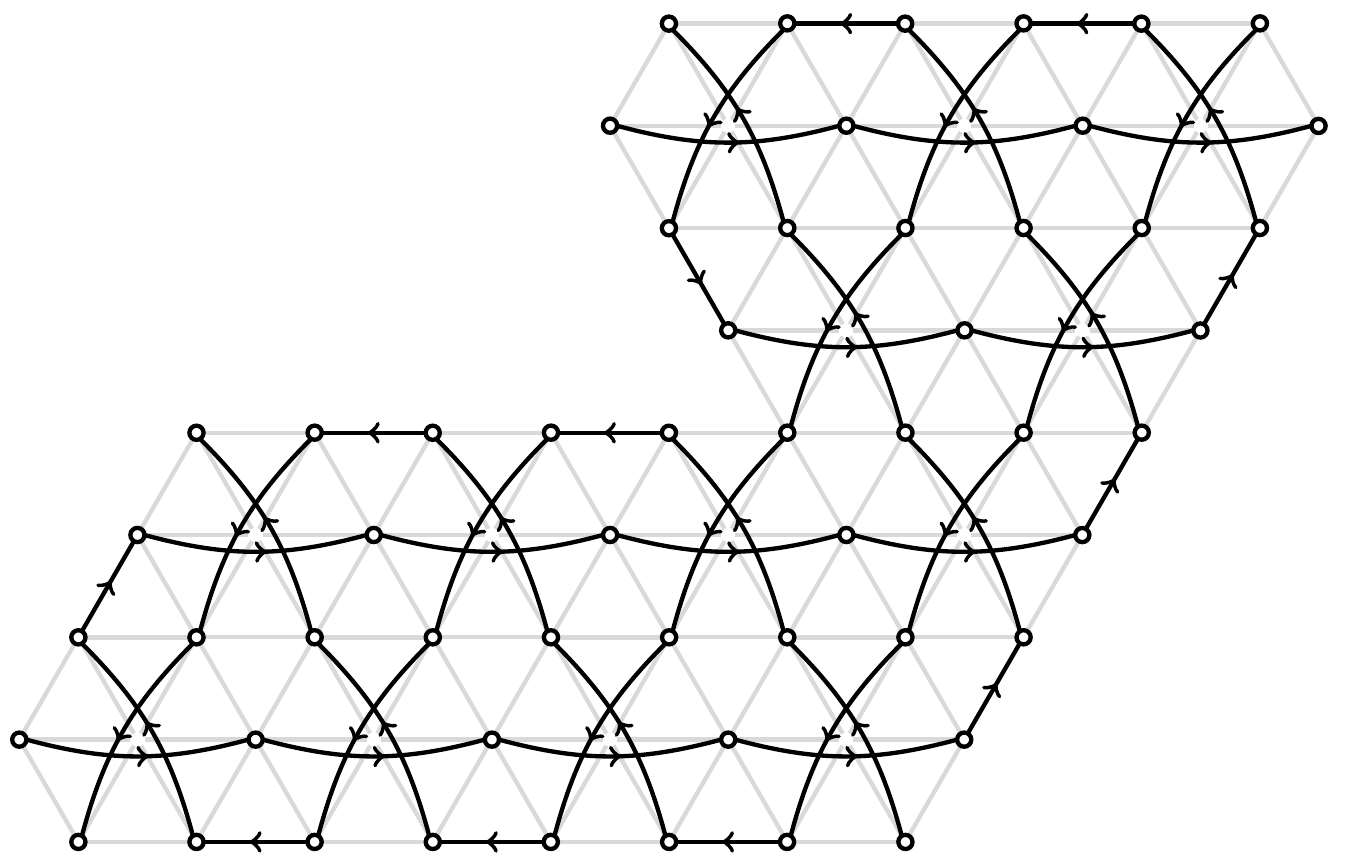}
\caption{
\textbf{Left:} In the $k=4\times 4$ parallelogram moose lattice with reflective boundary conditions, there are two sets of closed $SU(N)^m$ loops, and three open lines with $SU(N)_\ell \times SU(N)_r$ global symmetries. The open ended $SU(N)^{m-1}$ theories confine, following Ref.~\cite{Chang:2002qt}, while the closed loop $SU(N)^m$ theories have Coulomb phases, as in Ref.~\cite{Csaki:1997zg}.
\textbf{Right:}  Here we show an asymmetric moose lattice with reflective boundaries and an arbitrarily chosen shape. In this example there are no closed loops: the gauge invariant Wilson lines transform under the $SU(N)_\ell \times SU(N)_r$ global symmetries of the nodes on the corners of the lattice. 
}
\label{fig:ref4}
\end{figure}

\paragraph{Coulomb Phases:} 

In the reflective $k = 3\times 3$ example, all of the IR mesons transform as the bifundamental of a different global $SU(N)_\ell \times SU(N)_r$.
This feature is not generic, and is not even generic for $k=n \times n$ parallelogram arrangements. In the $k = 4\times 4$ parallelogram of Figure~\ref{fig:ref4}, for example, 
some of the strings of connected $SU(N)$ gauge groups form closed loops, with no $SU(N)_{\ell, r}$ endpoints on the lattice boundary. 
This arrangement is studied in Ref.~\cite{Csaki:1997zg}, which we review in Section~\ref{sec:reviewProduct}. These closed loop product groups appear especially frequently in the examples with periodic  boundary conditions.

At a generic point on the moduli space, each $SU(N)_1 \times SU(N)_2 \times \ldots SU(N)_m$ gauge theory is spontaneously broken to $U(1)^{N-1}$, a Coulomb phase with $N-1$ massless photons.
There are two disjoint closed loops of this form in the $k= 4\times 4$ theory, providing a total of $2(N-1)$ unbroken $U(1)$ gauge groups at an arbitrary point on the moduli space.
As noted in Ref.~\cite{Intriligator:1994sm}, for theories in the Coulomb phase one can describe the Lagrangian using a holomorphic prepotential, borrowing methods from $\mathcal N = 2$ supersymmetry.
The hyperelliptic curves for the $SU(N)^m$ theory are given in Ref.~\cite{Csaki:1997zg}.
%
%

\paragraph{Irregular Boundaries:}

There is no requirement that the moose lattice should have a symmetric shape. We emphasize this point with the example shown on the right hand side of Figure~\ref{fig:ref4}. This example happens to have no closed loops, and a total of four open-ended Wilson lines of varying length. The shortest line is the $\varphi = 0$ gauged $SU(N) \times SU(N)$ in the topmost row; the longest begins in the lower right corner, passes through 29 gauged $SU(N)$ sites before finding the middle-left corner.

Naturally, this program of adding  charged  edge quarks and gauging $SU(N)_\ell \times SU(N)_r$ can be extended to any $(\ell, r)$ pair of boundary nodes, even non-adjacent ones.
In principle every global $SU(N)_\ell \times SU(N)_r$ can be contracted in this way, until all of the $SU(N)$ nodes in the lattice are gauged.


\subsection{Cylindrical Moose} \label{sec:top-cylinder}

All of the examples discussed so far have involved topologically trivial moose lattices, aside from the possibility floated in Section~\ref{sec:top-reflect} of connecting non-adjacent $SU(N)_\ell \times SU(N)_r$ boundary nodes.
By making one of the dimensions in the moose lattice periodic, we can construct cylindrical lattices with $S^1$ topology. 

Periodic lattices have additional discrete symmetries associated with reflections or translations. If the coupling constants $\lambda_i$, $g_N$ and $g_{3N}$ also respect these symmetries, e.g.~$g_{3N}^{(i)} = g_{3N}^{(i+1)} = g_{3N}^{(i+2)} = \ldots$ along the periodic direction, then the discrete symmetries should be manifest in the low energy degrees of freedom.

\begin{figure}
\centering
\includegraphics[width=0.65\textwidth]{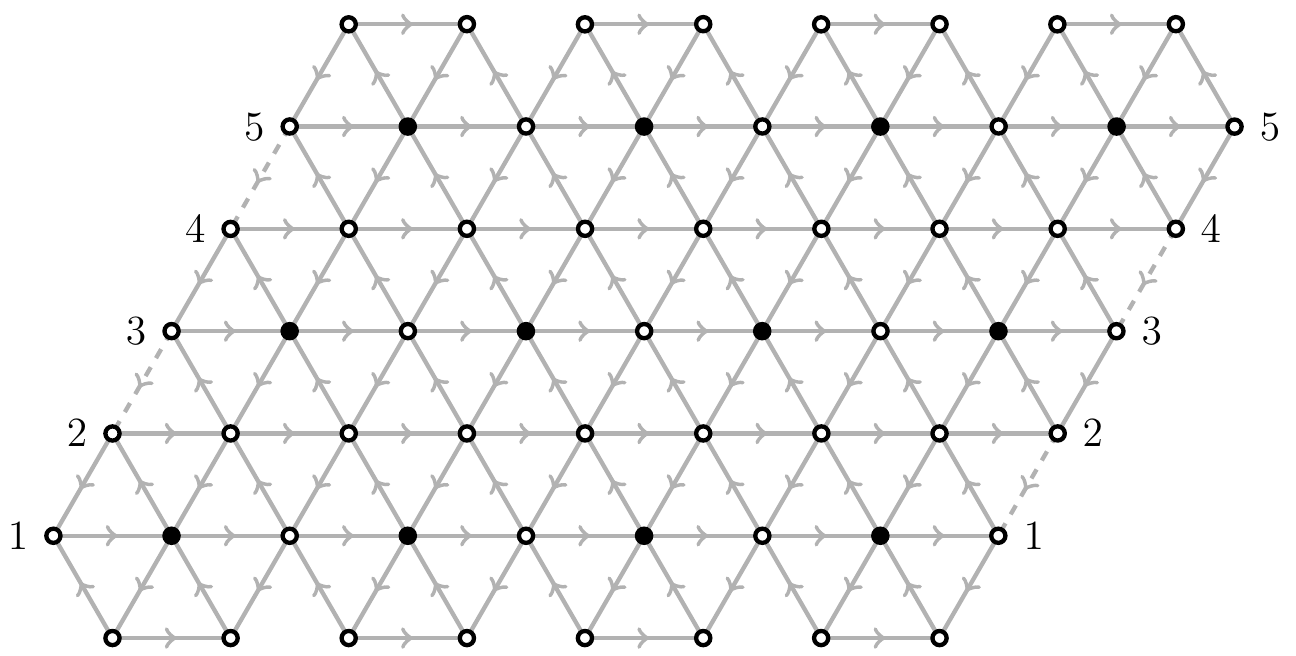}
\caption{A cylindrical moose lattice with $k = 4\times 3$ unit cells, periodic in the horizontal direction. The labels $j = 1 \ldots 5$ indicate that the $SU(N)_j$ nodes on the left and right boundaries are identical. Dashed lines indicate that a quark (for example, the bifundamental of $SU(N)_{1} \times SU(N)_2$) appears twice on the diagram. 
}
\label{fig:cylinder}
\end{figure}

\subsubsection{Symmetric Cylindrical Moose}

Figure~\ref{fig:cylinder} shows a $k = 4\times 3$ example, periodic in the $\varphi=0^\circ$ direction. The gauge group is $SU(3N)^{12} \times SU(N)^{28}$, and the non-Abelian global symmetry from the nodes on the top and bottom edges of the diagram is $SU(N)_\ell^8 \times SU(N)_r^8$. Each pair of $SU(N)_{j}$ nodes ($j = 1 \ldots 5$) on the left and right edges corresponds to a single gauged $SU(N)$ group.
 The dashed lines in the moose diagram depict quarks that would otherwise be double counted; for example, the ``$2 \rightarrow 1$'' edge quark is shown as a solid line on the left edge, and dashed on the right. 

Thanks to the periodicity of the moose lattice, it 
is invariant under shifts of an integer number of unit cells in the $\varphi = 0^\circ$ direction. 
This symmetry of the lattice becomes a symmetry of the theory if the coupling constants within the unit cells of each row are identical.
Each row in the bulk can still have three distinct $SU(N)$ coupling constants $g_N^{(i)}$ and eight plaquette coupling constants $\lambda_i$ (including the $SU(N)^3$ plaquettes), without spoiling the translation symmetry.
If additional constraints are imposed on the $\lambda_i$ and $g_{N}^{(i)}$, then the theory may also inherit the reflection symmetries of the lattice.
More precisely, the theory can be made invariant with respect to the combination of global charge conjugation ($\mathbf{N_c} \leftrightarrow \mathbf{\overline{N}_c}$ at all nodes) with reflections across any vertical ($\varphi' = 90^\circ$) plane aligned with the lattice nodes.

After $SU(3N)$ confinement and the subsequent integrating-out of the massive quarks, the theory includes three sets of $SU(N)^4$ ring product gauge theories, with charged mesons from the $1 \rightarrow 1$, $3\rightarrow 3$ and $5\rightarrow 5$ rows.
These $SU(N)$ groups do not confine, but each have an unbroken $U(1)^{N-1}$ at a generic point on the moduli space.
Following Ref.~\cite{Csaki:1997zg}, the gauge-invariant degrees of freedom can be written in terms of the $\mathcal M_\text{Ad}$ operators of \eqref{eq:ringadjoint},
\begin{align}
\mathcal M_\text{Ad} \equiv (M_a^{(1)} \ldots M_a^{(4)})_{ij} - \frac{1}{N} \text{Tr}(M_a^{(1)} \ldots M_a^{(4)} ),
\end{align}
where $M_a^{(i)} = (Q_a^{(i)} \barQ_a^{(i)})$ is the $\varphi = 0^\circ$ meson of the $i$th unit cell. 
This $\mathcal M_\text{Ad}$ is an adjoint of $SU(N)_1$, and neutral under the other $SU(N)_i$.
The moduli space is spanned by the gauge invariant products of $\mathcal M_\text{Ad}$, 
\begin{align}
u_k \equiv \frac{1}{k} \text{Tr}\left( \mathcal M_\text{Ad}^k \right), ~~~k \geq 2,
\label{eq:uk}
\end{align}
as well as the usual baryon operators, $(M_a^{(i)})^N$.
Note that the values of $u_k$ are insensitive to the choice of which $SU(N)$ group should correspond to $i=1$. The cyclicality of the trace operator ensures that $u_k$ is invariant under shifts of the form $i \rightarrow i+1$.

At $k=N$ this $u_k$ is related to the other degrees of freedom by a classical constraint, 
\begin{align}
 \mathcal M^N \sim (M_a^{(1)})^N (M_a^{(2)})^N (M_a^{(3)})^N (M_a^{(4)})^N .
\end{align}
This classical constraint receives quantum corrections of the form $B_a^{(i)} B_a^{(i+1)} \rightarrow - \Lambda_{i+1}^b$, including a complete set of nearest neighbor replacements as in \eqref{eq:Q01k}~\cite{Csaki:1997zg,Hailu:2002bh}.

At scales $\mu$ where the $SU(N)$ groups are weakly coupled (e.g.~$\mu \sim \lambda \Lambda_{3N}$), 
an arbitrary point on the moduli spaces sees the gauge invariant operators $u_k$  acquire expectation values that spontaneously break $SU(N)^4$ to its maximal Abelian subgroup, $U(1)^{N-1}$.
Ref.~\cite{Csaki:1997zg} uses methods from $\mathcal N=2$ supersymmetry to derive the kinetic terms of the supersymmetric Lagrangian in the infrared limit of the theory.

\medskip

Along the $\varphi=240^\circ$ direction we find four copies of $SU(N)^2$ gauge theories with open boundaries, as in Figure~\ref{fig:mooseChang} or Ref.~\cite{Chang:2002qt}. As described in Section~\ref{sec:triangle-IR}, these $SU(N)$ groups confine, so that in the far infrared the theory is described by baryons $B_a^{(i)} \sim (Q_a^{(i)}\barQ_a^{(i)})^N$ and mesons $\mathcal M_i \sim (Q_i^{(1)} \barQ_i^{(1)} \ldots Q_i^{(3)} \barQ_i^{(3)} )$, where each $\mathcal M_i$ transforms as a bifundamental of an $SU(N)_\ell \times SU(N)_r$. These gauge invariants are related to each other by the modified constraint equation \eqref{eq:Q01k}.

In the $\varphi = 120^\circ$ direction we find exactly the same behavior, except that two of the $SU(N)^2$ sets include the $SU(N)_2$ or $SU(N)_4$ depicted on the boundaries of Figure~\ref{fig:cylinder}.
The $SU(N)$ groups confine, and the light degrees of freedom include four sets of $\mathcal M$ mesons, each charged under its own $SU(N)_\ell \times SU(N)_r$ global symmetry.

Under the $i \rightarrow i+1$ shift symmetry, the four $\varphi = 120^\circ$ mesons are permuted cyclically with each other, as are their associated baryons. The same is true for the four $\varphi = 240^\circ$ mesons. 
There are three sets of $u_k^{(j)}$ operators, one for each row in Figure~\ref{fig:cylinder}. Each $u_k^{(j)}$ transforms trivially under the shift operation, i.e.~$u_k^{(j)} \rightarrow u_k^{(j)}$.
Under reflection-plus-conjugation operations, on the other hand, each $\varphi = 120^\circ$ meson is interchanged with one of the  $\varphi = 240^\circ$ mesons, while again the $u_k^{(j)}$ transform trivially.

Each discrete symmetry of the lattice is only a symmetry of the particle model if the coupling constants cooperate. Unequal values of $\lambda_i$ between different unit cells, for example, tend to break the translation or reflection symmetries.

\subsubsection{Alternative Cylinder Boundaries} \label{sec:altcyl}

In Figure~\ref{fig:cylinder} we chose an example where the aperiodic boundaries (on the top and bottom rows) respected the shift symmetry in the $\varphi = 0^\circ$ direction. A more generic example can have cylindrical topology without this ``azimuthal'' symmetry. For example, one could add or delete unit cells from the aperiodic edges. The periodic edge can be twisted, for example by matching the ``1'' and ``3'' on the right edge of Figure~\ref{fig:cylinder} to the ``3'' and ``5'' nodes (respectively) on the left edge.
Other modifications preserve the shift symmetry: one could add wrong-way quark fields to the cylinder ends to recreate the reflective boundary conditions of Section~\ref{sec:top-reflect}.

\paragraph{Reflective:}
If the $4\times 3$ example of Figure~\ref{fig:cylinder} is given reflective boundary conditions, it is possible for all 16 of the boundary $SU(N)$ nodes to be gauged. After $SU(3N)$ confinement, the charged matter content is given by:
\begin{align}
\includegraphics[width=0.7\textwidth]{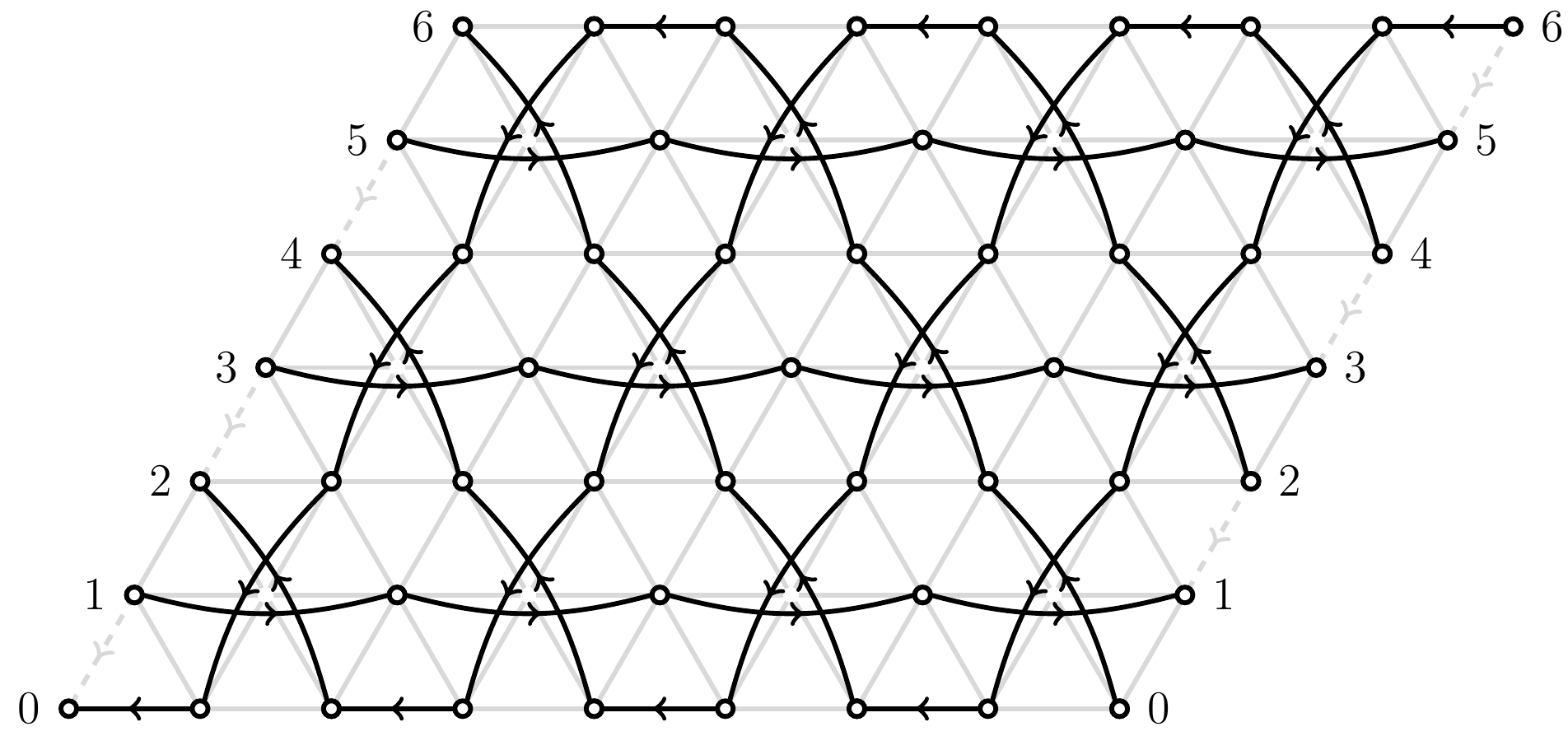}
\label{eq:periodreflect}
\end{align}
The eight sets of $\varphi = \pm 120^\circ$ mesons merge into four rings of $SU(N)^{8}$, following $0 \rightarrow 0$, $2 \rightarrow 2$, $4 \rightarrow 4$, and $6 \rightarrow 6$ (respectively). Horizontal shifts on the lattice induce cyclic permutations to these operators, and now none of the $SU(N)$ groups confine. Instead, there is a Coulomb phase with $7(N-1)$ unbroken $U(1)$ gauge groups; three from the three horizontal rows of $SU(N)^4$, four from the four $SU(N)^8$ loops.
The $\varphi = 0^\circ$ mesons are unaffected by the modification to the boundary conditions.

\paragraph{Barbershop:}
The shift symmetry of the cylinder can be broken by adding an apparent rotation or twist to the periodic moose lattice. 
In Figure~\ref{fig:cylinder}, the moose lattice is exactly periodic in the $\varphi = 0^\circ$ direction. We could instead have rolled up the lattice vertically in the page, rather than horizontally: for example,
\begin{align}
\includegraphics[width=0.7\textwidth]{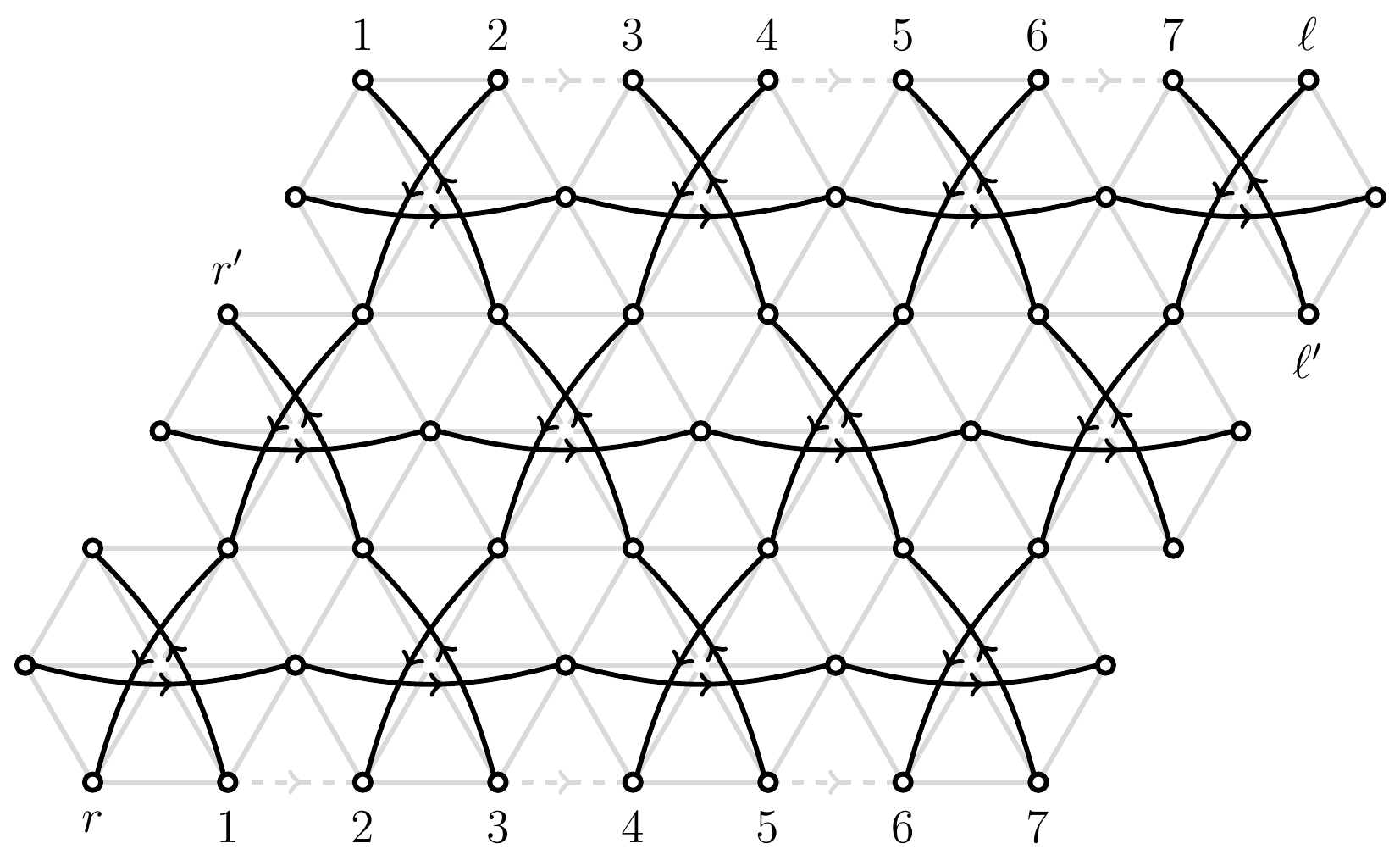}.
\label{eq:barber}
\end{align}
The seven numbered edge nodes are gauged, so that the lattice is approximately periodic in the $\varphi' \approx 90^\circ$ direction.
Unlike the Figure~\ref{fig:cylinder} example, this twisted cylinder does \emph{not} generate any closed rings of $SU(N)$ gauge groups after $SU(3N)$ confinement. 
There are three disjoint open meson lines in the $\varphi=0$ direction, one for each horizontal row, each with an $SU(N)_\ell \times SU(N)_r$ type global symmetry.
In the $\varphi = 240^\circ$ direction there is a single string of $SU(N)$ charged mesons, following $\ell \rightarrow 6 \rightarrow 4 \rightarrow 2 \rightarrow r$ through a total of 11 gauged $SU(N)$ sites. 
The $\varphi = 120^\circ$ direction hosts two open strings with $SU(N)^5$ gauge groups: one passes through $\ell' \rightarrow 7 \rightarrow 3 \rightarrow r'$, the other through $SU(N)_5$ and $SU(N)_1$.

Like the stripes on a barbershop pole, the gauge invariant  line operators in the $\varphi = \pm 120^\circ$ directions wrap around the $S^1$ direction while traveling horizontally along the cylinder.  The number of distinct line operators depends on the size of the cylinder, and on the degree to which it is twisted.

In the example of~(\ref{eq:periodreflect}), the combination of periodic and reflective boundaries removed all of the global $SU(N)_{\ell,r}$ symmetries, ensuring that the IR limit of the theory exhibits a Coulomb phase for each of the $\varphi$.
With the twisted  cylindrical moose lattice of~(\ref{eq:barber}), we encounter the opposite behavior: there are no closed $SU(N)$ rings or Coulomb phases in any direction, but instead all of the $SU(N)$ sites confine as in Section~\ref{sec:triangle-IR}.

\paragraph{M\"obius:}

As a final $S^1$ related example, we can further modify the $k=4\times 3$ cylinder by adding a $180^\circ$ twist about the central row, to construct a M\"obius strip rather than a simple cylinder.
Taking reflective boundary conditions on the top and bottom edges (or rather, the single ``$\text{top} = \text{bottom}$'' edge), the moose diagram after $SU(3N)$ confinement takes the form:
\begin{align}
\includegraphics[width=0.7\textwidth]{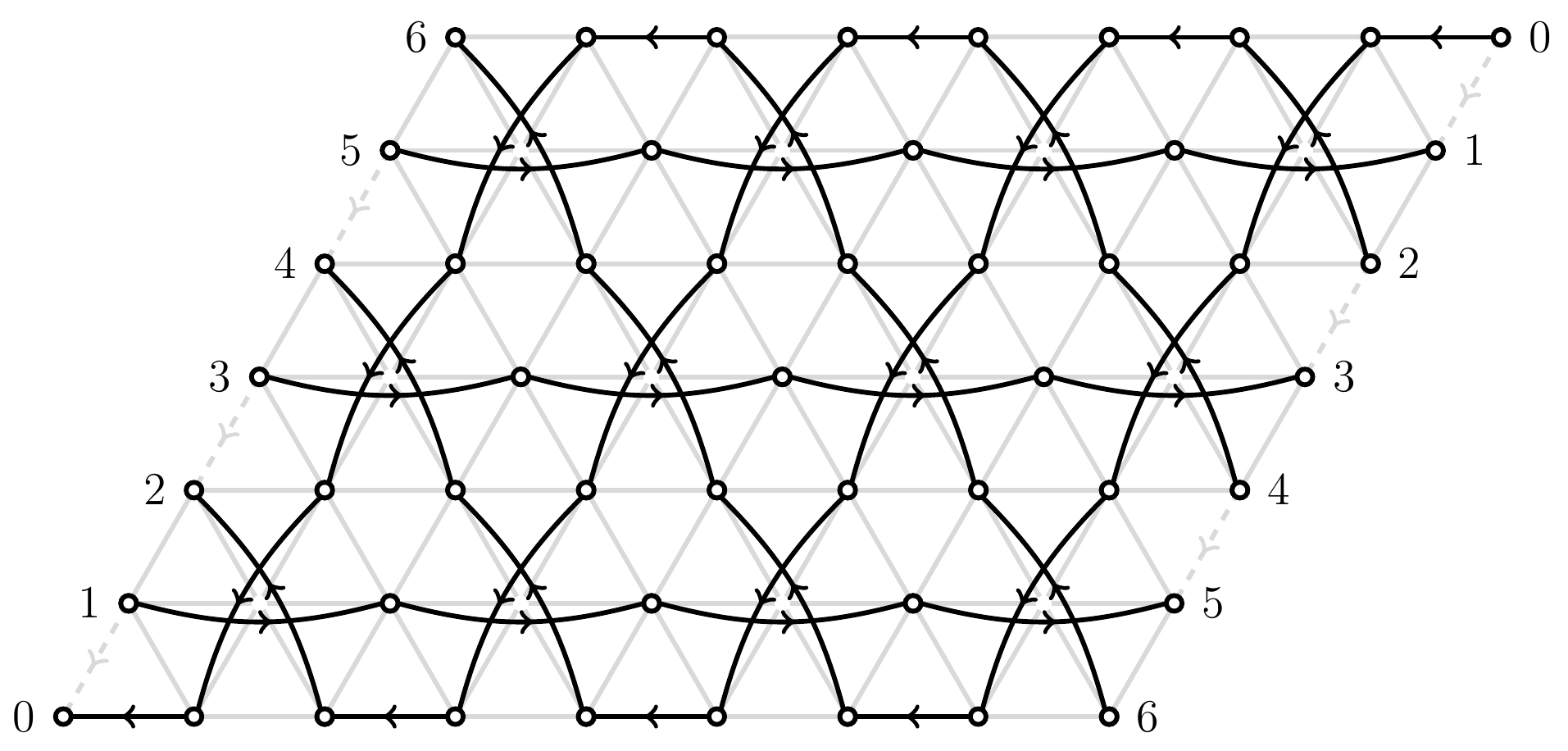} .
\label{eq:mobius}
\end{align}
Compared to~(\ref{eq:periodreflect}), there are even fewer $\varphi = \pm 120^\circ$ distinct gauge invariant line operators: one along $0 \rightarrow 6 \rightarrow 0$, another following $2 \rightarrow 4 \rightarrow 2$, each of which encounters 16 gauged $SU(N)$ groups.
Also, the ``$5$'' and ``$1$'' $\varphi = 0$ meson lines now join together into a single $1 \rightarrow 5 \rightarrow 1$.
So, the $SU(N)$ charged mesons can be organized into four sets of  $SU(N)^m$ rings, with $m=4, 8$ for $\varphi = 0$, and $m=16, 16$ for $\varphi = \pm 120^\circ$.
The infrared theory exhibits a Coulomb phase with just four  copies  of $U(1)^{N-1}$, rather than the $U(1)^{7(N-1)}$ of~(\ref{eq:periodreflect}).

\subsection{Toroidal Moose} \label{sec:top-toroid}

Finally, let us discuss the toroidal topologies ($T^2 = S^1 \times S^1$) that arise when periodic boundary conditions are imposed on all edges of the moose lattice.
By construction, these geometries have no $SU(N)$ sites that are not gauged. 
A generic periodic flat torus can be represented by a $k = m \times n$ parallelogram, with some scheme for matching the nodes on opposite edges. That is, the boundaries can be twisted in the manner of~(\ref{eq:barber}), while preserving the two-dimensional shift symmetries of the lattice.

\begin{figure}
\centering
\includegraphics[width=0.56\textwidth]{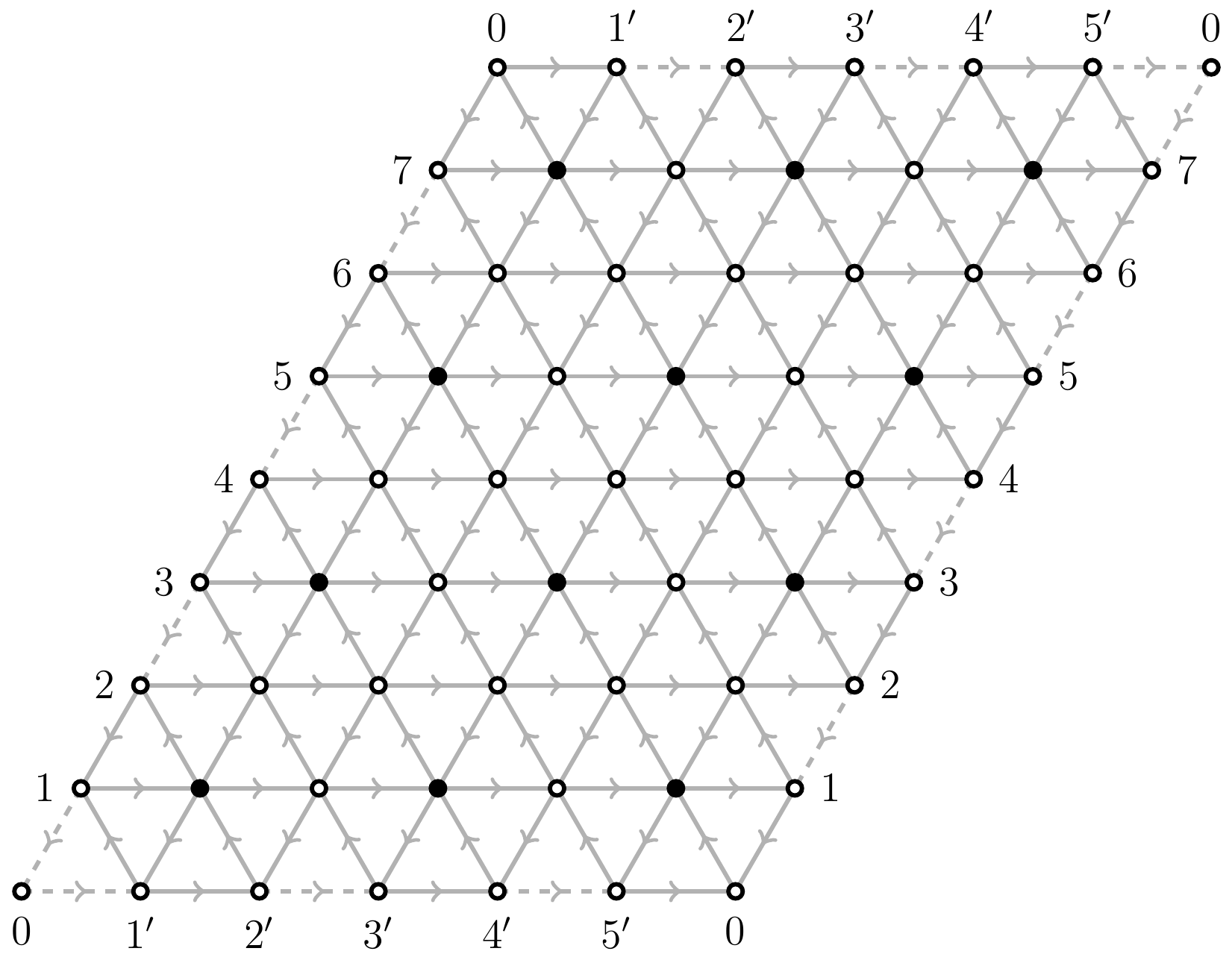}
\hspace{-2.8cm}
\includegraphics[width=0.56\textwidth]{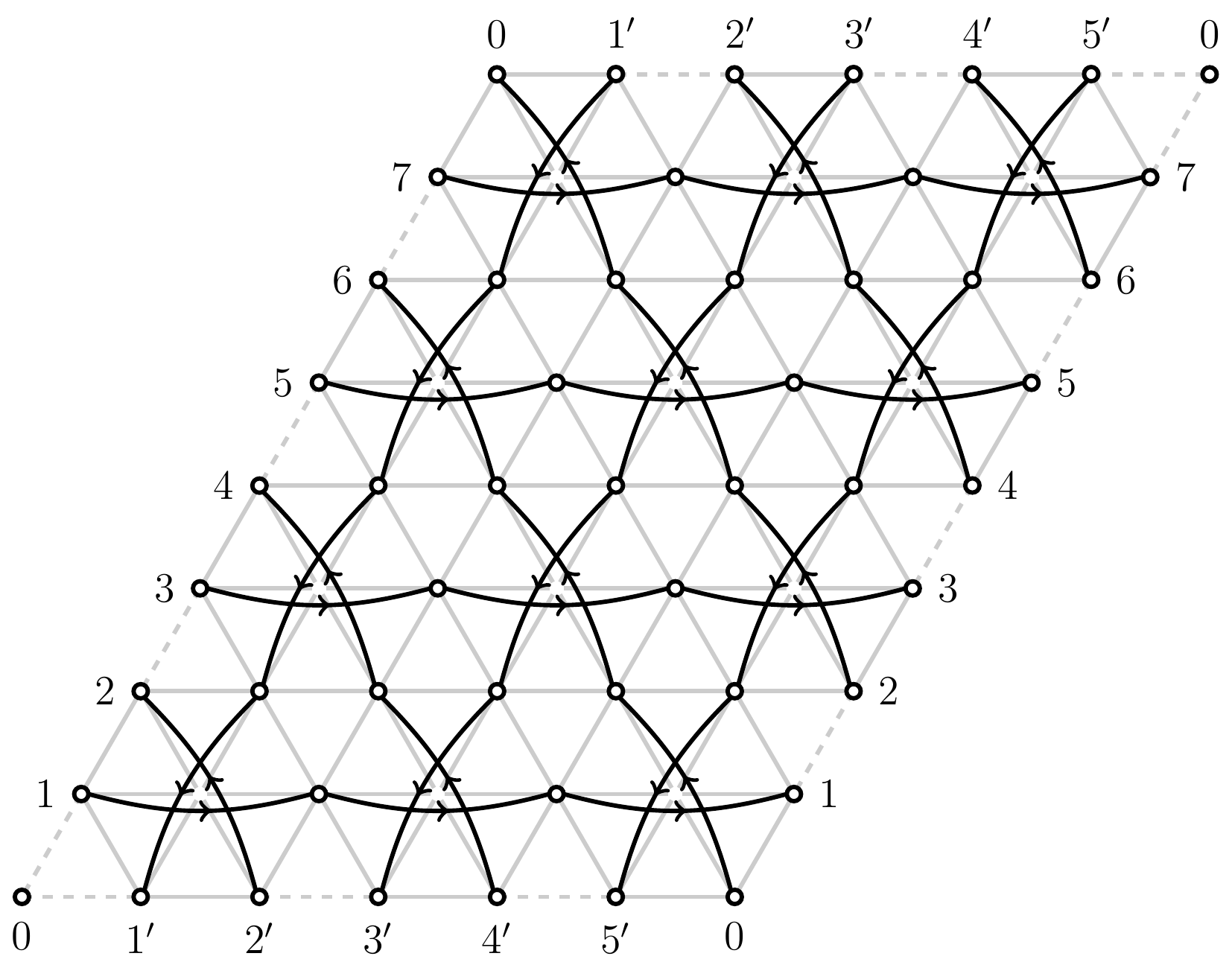}
\caption{\textbf{Left:} A $T^2$ torus, based on a $k = 3\times 4$ rectangular arrangement with periodic boundaries, is shown here in the UV limit where the gauged $SU(3N)$ are asymptotically free. The labels $j=1 \ldots 7$ and $j' = 1' \ldots 5'$ indicate the connections between the boundary nodes. Each pair of  $SU(N)_j$ or $SU(N)_{j'}$ nodes on the diagram corresponds to a single gauged $SU(N)$; likewise, the four ``$0$'' nodes represent a single gauged $SU(N)$.
\textbf{Right:} After the $SU(3N)$ groups confine and the vectorlike pairs are integrated out, these $SU(N) \times SU(N)$ bifundamental mesons are the only charged degrees of freedom. Along the $\varphi = 0^\circ$ and $240^\circ$ directions they form closed loops, with $SU(N)^3$ and $SU(N)^4$ product groups respectively. 
For this rectangular $k = m \times n$ torus with $m \neq n$, the $\varphi = 120^\circ$ wraps around the torus in a single $SU(N)^{12}$ ring, following $0 \rightarrow 6 \rightarrow 4' \rightarrow 4 \rightarrow 2' \rightarrow 2 \rightarrow 0$.
}
\label{fig:toroid}
\end{figure}

As a first concrete example, we return to the $k=4\times 3$ parallelogram, depicted in Figure~\ref{fig:toroid} before and after $SU(3N)$ confinement. With this particular choice for the periodic boundaries, straight lines in the $\varphi = 0$ and $\varphi = 240^\circ$ directions  wrap exactly once about each $S^1$, e.g.~$3 \rightarrow 3$ or $5' \rightarrow 5'$.
Each node number $j=1\ldots 7$ or $j'=1'\ldots 5'$ corresponds to a single gauged $SU(N)$, as usual; similarly, although the $SU(N)_0$ group appears at each of the four corners on the lattice, it is a single gauge group.
So, the UV theory (perturbatively coupled at $\mu \sim M_\star$) is composed of an $SU(3N)^{12} \times SU(N)^{36}$ gauge group, with  $12 \times 6$ bifundamentals of $SU(3N) \times SU(N)$ and an equal number of $SU(N) \times SU(N)$ quarks. 
Assuming no additional $U(1)$ factors are gauged, the superpotential admits $12 \times 8$ trilinear plaquette operators.

Using the technology from Section~\ref{sec:triangular}, it is straightforward to follow the theory from $\mu \sim M_\star$ towards the infrared, past $SU(3N)$ confinement and the generation of  masses for the vectorlike pairs of mesons and quarks.
The remaining light $SU(N)$ charged degrees of freedom are shown in the right side of Figure~\ref{fig:toroid}. The $\varphi = 0$ mesons form four sets of $SU(N)^3$, following $j \rightarrow j$ for $j = 1, 3, 5, 7$. Similarly, the $\varphi = 240^\circ$ mesons provide another three $SU(N)^4$ rings, with $j' \rightarrow j'$ for $j' = 1', 3', 5'$. 
Lastly, the $\varphi = 120^\circ$ mesons form a single closed loop encompassing an $SU(N)^{12}$ gauge group,  following $0 \rightarrow 6 \rightarrow 4' \rightarrow 4 \rightarrow 2' \rightarrow 2 \rightarrow 0$.

The doubly periodic lattice has a two dimensional shift symmetry, spanned by unit cell translations in the $\varphi = 0^\circ$ and $\varphi = 240^\circ$ directions. Respectively, these operations cyclically permute the sets of $\varphi = 240^\circ$ and $\varphi = 0^\circ$ mesons. The single $\varphi = 120^\circ$ line transforms as the identity under both kinds of translation. The moduli space is spanned by the $u_k$ type operators of \eqref{eq:uk}, together with $\Tr(M_a^{(1)} \ldots M_a^{(m)})$  and the various baryonic operators.
At an arbitrary point on the moduli space the $SU(N)^{36}$ group is broken to $U(1)^{8(N-1)}$.

\begin{figure}
\centering
\includegraphics[width=0.64\textwidth]{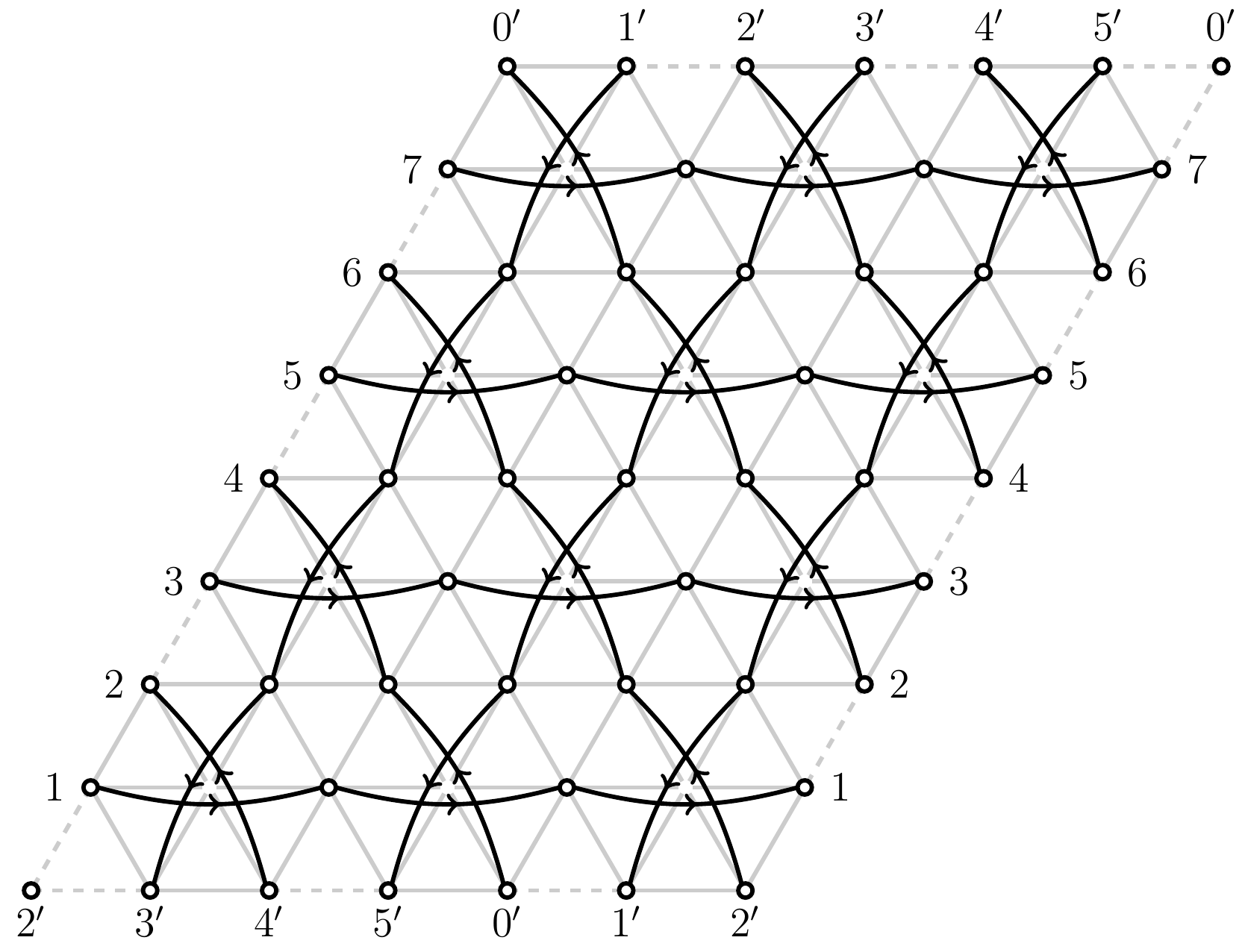}
\caption{Here we show another symmetric $T^2$ torus, periodic in the $\varphi = 0^\circ$ and $\varphi' = 90^\circ$ directions.  The $j \rightarrow j$ lines ($j=1,3,5,7$) are unaffected by the altered boundary conditions (compared to Figure~\ref{fig:toroid}), but now the $\varphi = \pm 120^\circ$ mesons each form one closed ring of $SU(N)^{12}$. 
Reflections of the lattice in the $\varphi = 0^\circ$ or $\varphi' = 90^\circ$ directions exchange the $\varphi = 120^\circ$ and $\varphi = 240^\circ$ operators with each other.
}
\label{fig:toroid90}
\end{figure}

Either type of edge can be twisted, by acting on the node labels with cyclic permutations.
As a simple example, we could shift the bottom row of labels four spaces to the right, $0 \rightarrow 4'$, so that the top and bottom rows now match along the $\varphi' = 90^\circ$ vertical direction. 
The $k=3\times 4$ version is shown in Figure~\ref{fig:toroid90}, after $SU(3N)$ confinement.
In this example the lattice is symmetric with respect to reflections about the vertical axis: this was not true of Figure~\ref{fig:toroid}.
Now, both sets of  $\varphi = \pm120^\circ$ mesons form rings of $SU(N)^{12}$. The $\varphi = 120^\circ$ example follows $0' \rightarrow 4 \rightarrow 2' \rightarrow 6 \rightarrow 4' \rightarrow 2 \rightarrow 0'$; for $\varphi = 240^\circ$, the line operator follows $1' \rightarrow 3' \rightarrow 5' \rightarrow 1'$ instead.
The $\varphi = 0^\circ$ operators are not impacted by the twist: they still form four sets of $SU(N)^3$ rings.
At a generic point on the moduli space, the $SU(N)$ groups are spontaneously broken to $U(1)^{6(N-1)}$.

\medskip

In the present work we are content to restrict ourselves to $T^2$ and $S^1$ topologies for the moose lattice. More complicated topologies can of course be constructed by folding  and connecting lattices of different shapes, but the methods for determining the low energy behavior remain the same.

\section{Conclusion} \label{sec:conclusion}

This paper is dedicated to the $\mathcal N = 1$ supersymmetric triangular moose lattice in four spacetime dimensions, with the $[SU(3N) \times SU(N)]^k$ style product gauge group. Assuming that there is a high energy scale $M_\star$ at which the coupling constants are perturbatively small, we have shown that the $SU(3N)$ gauge groups confine, and that some of the $SU(N)$ charged quarks and mesons subsequently acquire vectorlike masses. Depending on the lattice boundary conditions, the $SU(N)$ gauge groups may either confine or form Coulomb phases.

Aspects of the infrared theory are highly suggestive of a higher dimensional interpretation, where the effectively two-dimensional moose lattice is associated with two compact extra dimensions.
Some of the degrees of freedom, such as the baryon operators, are localized to specific segments of the moose lattice. Each gauged $SU(3N)$, for example, is associated with the $B$ and $\barB$ operators defined in \eqref{eq:unitW}, as if $B$ and $\barB$ were composite fields propagating in the bulk of the extra-dimensional theory. As we show in Section~\ref{sec:globalbulk},  some global symmetries can act in the same way, especially if some subgroup of the $U(1)_a \times U(1)_b$ is gauged. Other degrees of freedom, like the $\mathcal M$ of \eqref{eq:B1k}, are associated more closely with the boundaries of the lattice, as are the chevron-type global $U(1)$ symmetries of Section~\ref{sec:globalbrane}.

Especially for the examples with periodic boundary conditions presented in Section~\ref{sec:topology}, we see that the meson and baryon operators in the farthest infrared limit possess shift symmetries that recall a discretized version of translation invariance.
An approximate version of this translation invariance appears in the bulk of the moose lattice even in non-periodic topologies.
All of these signs point towards a geometric physical interpretation of the moose lattice gauge theory, providing a clear direction for future research on this topic. 

The Coulomb phases associated with the periodic or reflective boundary conditions provide another area for future study. Aside from noting that Ref.~\cite{Csaki:1997zg} provides expressions for the holomorphic prepotential of the $SU(N)^m$ ring theories, we have not yet taken advantage of the approximate $\mathcal N=2$ supersymmetry to constrain the Lagrangian for the infrared degrees of freedom.

\medskip

The structure of the model also provides many opportunities for model building. The strong $CP$ problem, and the associated axion quality problem, supply one such well-motivated target. The QCD axion provides an elegant mechanism to explain the otherwise confoundingly tiny value of the $SU(3)_c$ $CP$ violating $\theta$ parameter, where an approximate $U(1)_\text{PQ}$ with nonzero $SU(3)_c^2 U(1)_\text{PQ}$ anomaly coefficient is spontaneously broken at a high scale, $f_a \gg \Lambda_\text{QCD}$. Nonperturbative QCD effects generate a potential for the pseudo-Nambu--Goldstone boson of $U(1)_\text{PQ}$ that sets the effective value of $\theta$ to zero.

This mechanism requires that $U(1)_\text{PQ}$ should be classically conserved to an extremely high degree, broken only by QCD effects.
However, gravitational effects are generally expected to break global symmetries~\cite{Giddings:1987cg,Kamionkowski:1992mf,Barr:1992qq,Kallosh:1995hi,Abbott:1989jw,Coleman:1989zu}, and even relatively tiny perturbations to the axion potential can ruin the solution to the strong CP problem. A successful ``high quality'' axion model protects $U(1)_\text{PQ}$ against these gravitational intrusions by ensuring that all  PQ-charged gauge-invariant operators permitted in the Lagrangian are sufficiently suppressed~\cite{Chun:1992bn,Randall:1992ut,Cheng:2001ys,Fukuda:2017ylt,DiLuzio:2017tjx,Lillard:2017cwx,Lillard:2018fdt,Ardu:2020qmo,Nakai:2021nyf,Darme:2021cxx}.
Especially for large moose lattices, or for the ``barbershop'' arrangements of Section~\ref{sec:altcyl}, it can be relatively easy to embed $SU(3)_c$ and a high quality QCD axion within this model. 
Indeed, compared to the relative simplicity of Ref.~\cite{Lillard:2018fdt}, invoking a whole moose lattice for the sole purpose of dealing with the axion quality problem may be seen as overly aggressive.

The automatically generated hierarchy between the scales $\Lambda_N$ and $\Lambda_{3N}$ depicted in Figure~\ref{fig:cartoon} is another feature of the moose lattice that has possible model building applications. Due to the sign change in the $\beta(g_N)$ function induced by $SU(3N)$ confinement, the scale $\Lambda_N$ is suppressed by a factor of $\Lambda_{3N}^2 / M_\star$. This inverse relationship, reminiscent  of the seesaw mechanism for neutrinos,  allows for an unusually small $\Lambda_N$ even if the $SU(N)$ and $SU(3N)$ couplings are of similar size at $\mu \sim M_\star$.

\medskip

Finally, it may be worthwhile to generalize the two-dimensional triangular lattice beyond the $[SU(3N) \times SU(N)^3]^k$ paradigm to include higher dimensions and alternative lattice arrangements.

\section*{Acknowledgements}

I am grateful to Patrick Draper, Arvind Rajaraman, Yuri~Shirman, and Tim M.~P.~Tait for several conversations during the development of this paper, and to Carlos~Blanco,  Aaron~Friedman, Robert~McGehee, and Pavel Maksimov for their patience at the social occasions where I have presented the principal results.
Special thanks go to Patrick Draper for helpful feedback on this manuscript.
 This work was performed in part at Aspen Center for Physics, which is supported by National Science Foundation grant PHY-1607611. This work was partially supported by a grant from the Simons Foundation.
Some of this work was supported by NSF Grant No.~PHY-1620638 and the Chair’s Dissertation Fellowship from the Department of Physics \&~Astronomy at UC~Irvine.

\appendix

\section{Global Symmetries and 't~Hooft Anomaly Matching} \label{sec:reviewAnomaly}

For the infrared theory of $M$, $B$ and $\barB$ to be dual to the $SU(N)$ gauge theory of $Q$ and $\barQ$, it must satisfy a number of nontrivial constraints that arise from the global symmetries of the theory.
At every point on the moduli space, the 't~Hooft trace anomalies of the preserved global symmetries should match between the two theories. 
For qdms-confinement, we show that the full set of anomaly matching conditions is \emph{not} satisfied at the origin of moduli space: but, the anomaly coefficients \emph{do} match everywhere on the quantum-deformed moduli space \eqref{eq:qdmsN}, for those global symmetries which are not spontaneously broken.
This provides a separate confirmation that the origin should be excised from the qdms-confined theories. 
In the case of s-confinement, the moduli space includes the origin, so the full set of 't~Hooft anomaly coefficients must match in the infrared  theory. 

In SQCD with $N$ colors and $F$ flavors, the global symmetry group is 
\begin{align}
G_\text{global} = SU(F)_L \times SU(F)_R \times U(1)_B \times U(1)_R.
\end{align}
In addition to $G_\text{global}$, there is also an approximate symmetry $U(1)_A$, which has a nonzero $SU(N)^2$-$U(1)_A$ anomaly coefficient. It is explicitly broken by $SU(N)$ instantons, at a scale characterized by $\Lambda$.

As the $F=N$ model features prominently in the product gauge groups introduced in this paper, let us take a moment to explore its infrared effective theory in more detail. For this special case the $U(1)_R$ symmetry can be defined such that the scalar parts of the $Q$ and $\barQ$ supermultiplets have zero $R$ charge, as shown in Table~\ref{table:FisN}.
Using the canonical normalization for $U(1)_R$, the fermionic components of $Q$ and $\barQ$ (and $B$, $\barB$ and $M$) have $R$ charges $-1$.
This definition of $U(1)_R$ is not unique: as in any theory with multiple conserved $U(1)$ charges, it is possible to define a $U(1)_{R'}$ or a $U(1)_{B'}$ out of linear combinations of $U(1)_R$ and $U(1)_B$.

\begin{table}
\begin{center}
\begin{tabular}{| c | c c c | c c | c | } \hline
	& $SU(F)_\ell$	& $SU(N)_c$		& $SU(F)_r$	& $U(1)_B$	& $U(1)_R$	& $U(1)_A$	\\ \hline
\Tstrut \barQ &	\bf{F}	& $\overline{\mathbf{N}}$	 &	\bf{1}& $-1$	&	0	& $+1$ \\
$Q$ 		&	\bf{1}	& $\mathbf{N}$ 		& $\overline{\mathbf{F}}$	& $+1$	& 	0	& $+1$ \\ \hline
$\lambda$ 	&	\bf{1}	& $\mathbf{Ad}$ 	& ${\mathbf{1}}$	& $0$	& 	$+1$	& $0$ \\ 
$\Lambda^b$&	\bf{1}	& $\mathbf{1}$ 		& \bf{1}	& $0$		& 	0	& $+2N$ \\ \hline
\Tstrut \barB &	\bf{1}	&  		 &	\bf{1}& $-N$	&	0	&   \\
$B$ 		&	\bf{1}	& 		& \bf{1}	& $+N$	& 	0	&   \\ 
$M$ 		& ${\mathbf{F}}$ & 		& $\overline{\mathbf{F}}$ & $0$	& 	0	&  \\ \hline
\end{tabular}
\end{center}
\caption{Transformations of the superfields $Q$, $\barQ$, $M$, $B$ and $\barB$, under the gauged and global symmetry groups for the $F=N$ case of SQCD. 
Also shown are the $R$ charge of the gauginos ($\lambda$), and the transformation of $\Lambda^b$ under the spurious $U(1)_A$.
Here $\bf{N}$ and $\overline{\bf{N}}$ indicate the fundamental and antifundamental representations of $SU(N)$, while $\bf{Ad}$ indicates the adjoint representation. The $U(1)_R$ charges shown are those of the scalar component of the superfields.
At scales well above the scale $|\Lambda|$, $U(1)_A$ is approximately conserved, but it is broken explicitly when $SU(N)_c$ is gauged. The infrared theory is valid only for scales well below $\Lambda$, where $U(1)_A$ is badly broken, so we do not list the $U(1)_A$ charges of the gauge-invariant operators $M$, $B$, or $\barB$.}
\label{table:FisN}
 \end{table}

To demonstrate the use of the anomaly matching conditions in the presence of spontaneous symmetry breaking, consider the $U(1)_R^3$ cubic anomaly coefficient: 
\begin{align}
\mathcal A_\text{UV}(U(1)_R^3) &= F\cdot N (-1)^3  + N\cdot F (-1)^3  + (N^2 - 1)(+1)^3 = - N^2 - 1
\nonumber\\
\mathcal A_\text{IR}^{\{0\}}(U(1)_R^3) &=  1 (-1)^3 + 1 (-1)^3 + F^2(-1)^2 = - N^2 - 2,
\end{align}
where we have naively evaluated the anomaly coefficient $\mathcal A_\text{IR}$ at the origin of the moduli space. The two $\mathcal A$ do not match: this is because we have overcounted the IR degrees of freedom by neglecting the quantum modified constraint, \eqref{eq:qdmsN}.
On the $\ev{B \barB} = -\Lambda^b$ branch of the moduli space, where $SU(F)_\ell \times SU(F)_r$ is preserved and $U(1)_B$ is spontaneously broken, the only light degree of freedom between $B$ and $\barB$ is the one tangential to $B \barB = - \Lambda^b$. Excitations of $B$ and $\barB$ that change the value of $\ev{B\barB}$ acquire $\mathcal O(\Lambda)$ masses, and are not degrees of freedom of the infrared theory.

Similarly, on the $\ev{\det M} = \Lambda^b$ branch with $\ev{B} = \ev{\barB} = 0$, the $F^2 = N^2$ nominal degrees of freedom in $M$ are reduced to $N^2 - 1$. For example, at the symmetry enhanced point $\ev{M_{ij}} = \Lambda^{b/N} \delta_{ij}$, where the flavor symmetry is broken to its diagonal subgroup, $SU(F)_\ell \times SU(F)_r \rightarrow SU(F)_{d}$, the $N^2 - 1$ dimensional adjoint representation of $SU(F)_d$ remains light, while the $\text{Tr}\, M$ degree of freedom acquires an $\mathcal O(\Lambda)$ mass.
If we investigate any generic point on the moduli space, the result is the same: there are only $F^2 + 1$ degrees of freedom, and
\begin{align}
\mathcal A_\text{IR}(U(1)_R^3) &=  (2 + F^2 - 1) (-1)^3  = - N^2 - 1 = \mathcal A_\text{UV}(U(1)_R^3) .
\end{align}

Similar subtleties arise in the mixed anomalies, such as $SU(F)_\ell^2$-$U(1)_B$. In this case
\begin{align}
\mathcal A_\text{UV}(SU(F)_\ell^2 U(1)_B) &= N \cdot 1 (-1) = -N,
\nonumber\\
\mathcal A_\text{IR}^{\{0\}}(SU(F)_\ell^2 U(1)_B) &= F \cdot 1 (0) = 0.
\end{align}
Our definition of $\mathcal A$ uses the normalization of the Dynkin index $\hat\mu$ such that $\hat\mu = 1$ for the fundamental and antifundamental representations, and $\hat\mu = 2N_c$ for the adjoint of $SU(N_c)$. 
Again, we find that it is a mistake to evaluate $\mathcal A_\text{IR}$ at the origin $\{0\}$, and in fact there is no point on the moduli space where $SU(F)_\ell$ and $U(1)_B$ are simultaneously unbroken.

There is, however, the symmetry enhanced point $\ev{M_{ij}} \propto \delta_{ij}$, $\ev{B} = \ev{\barB} = 0$, where the global symmetry is $SU(F)_d \times U(1)_B \times U(1)_R$.
The degrees of freedom are $B$, $\barB$, and $M_{Ad}$, which transforms as the adjoint of $SU(F)_d$,
and the $SU(F)_d^2 U(1)_B$ anomaly coefficients do match:
\begin{align}
\mathcal A_\text{UV}(SU(F)_d^2 U(1)_B) &= N \cdot 1(-1) + N\cdot 1(+1) = 0 ,
\nonumber\\
\mathcal A_\text{IR}(SU(F)_d^2 U(1)_B) &= 1 \cdot 2F (0) = 0.
\end{align}

For other anomaly coefficients there is less need for subtlety. Evaluating $SU(F)_\ell^2 U(1)_R$ on the $\ev{M} = 0$ branch of the moduli space, for example, we find
\begin{align}
\mathcal A_\text{UV}(SU(F)_\ell^2 U(1)_R) &= N \cdot 1 (-1) ,
&
\mathcal A_\text{IR}(SU(F)_\ell^2 U(1)_R) &= F \cdot 1 (-1) = -N,
\end{align}
which match. It is similarly easy to show that the mixed $U(1)_B U(1)_R$ anomalies match when evaluated on the $\ev{B} = \ev{\barB} = 0$ branch.

\bibliography{prodsconf}

\end{document}